\documentclass{aa}
\usepackage{rotating, tabularx}
\usepackage{gensymb}
\usepackage[section]{placeins}
\usepackage{graphicx}
\usepackage{txfonts}
\usepackage[switch]{lineno}
\usepackage{amssymb}
\usepackage{url}
\usepackage{amsmath}
\usepackage[export]{adjustbox}
\usepackage{subfig}
\usepackage{bm}
\usepackage{siunitx}						    % SI units 
\usepackage{color}							    % colors
\DeclareSIUnit\gauss{G}					        % gauss

\graphicspath{{Figures/}}

\begin{document} 
%\linenumbers

   \title{Observation-based modelling of magnetised Coronal Mass Ejections with EUHFORIA}
   %\subtitle{I. Overviewing the $\kappa$-mechanism}

   \author{C. Scolini\inst{1,2},
          L. Rodriguez\inst{2},
          M. Mierla\inst{2,3},
          J. Pomoell\inst{4},
          \and
          S. Poedts\inst{1}
          }

   \institute{Centre for mathematical Plasma Astrophysics, 
             KU Leuven, 3001 Leuven, Belgium\\
             \email{camilla.scolini@kuleuven.be}
             \and
             Solar-Terrestrial Centre of Excellence -- SIDC, Royal Observatory of Belgium,
             1180 Brussels, Belgium
             \and
             Institute of Geodynamics of the Romanian Academy, 020032 Bucharest, Romania
             \and
             University of Helsinki, 
             00100 Helsinki, Finland
             }

   \date{Submitted: January 12, 2019; revised: March 25, 2019; accepted: April 15, 2019}

% \abstract{}{}{}{}{} 
% 5 {} token are mandatory
 
  \abstract
   {Coronal Mass Ejections (CMEs) are the primary source of strong space weather disturbances at Earth. 
   Their geo-effectiveness is largely determined by their dynamic pressure and internal magnetic fields, for which reliable predictions at Earth are not possible with traditional cone CME models.
   }
  % aims heading (mandatory) ============================================
   {We study two well-observed Earth-directed CMEs using the EUropean Heliospheric FORecasting Information Asset (EUHFORIA) model, testing for the first time the predictive capabilities of a linear force-free spheromak CME model initialised using parameters derived from remote-sensing observations. }
  % methods heading (mandatory) ============================================
   {Using observation-based CME input parameters, we perform magnetohydrodynamic simulations of the events with EUHFORIA, using the cone and spheromak CME models.}
  % results heading (mandatory) ============================================
   {Simulations show that spheromak CMEs propagate faster than cone CMEs when initialised with the same kinematic parameters. We interpret these differences as result of different Lorentz forces acting within cone and spheromak CMEs, which lead to different CME expansions in the heliosphere. Such discrepancies can be mitigated by initialising spheromak CMEs with a reduced speed corresponding to the radial speed only.
   Results at Earth evidence that the spheromak model improves the predictions of $B$ ($B_z$) up to 12--60 (22--40) percentage points compared to a cone model.
   Considering virtual spacecraft located within $\pm 10^\circ$ around Earth,
   $B$ ($B_z$) predictions reach 45--70$\%$ (58--78$\%$) of the observed peak values. 
   The spheromak model shows inaccurate predictions of the magnetic field parameters at Earth for CMEs propagating away from the Sun-Earth line.
   }
  % conclusions heading (optional), leave it empty if necessary ===============
   {The spheromak model successfully predicts the CME properties and arrival time in the case of strictly Earth-directed events, while modelling CMEs propagating away from the Sun-Earth line requires extra care due to limitations related to the assumed spherical shape.
   The spatial variability of modelling results and the typical uncertainties in the reconstructed CME direction advocate the need to consider predictions at Earth and at virtual spacecraft located around it.
   }


   \titlerunning{Observation-based Modelling of Magnetised CMEs}
   \authorrunning{C. Scolini et al.}
   
   \maketitle

%============================================================
%============================================================
\section{Introduction}
\label{sec:intro}

% CMEs
Coronal Mass Ejections (CMEs) are large-scale eruptions of plasma and magnetic fields from the Sun, 
and are considered to be the main drivers of strong space weather events at Earth \citep{gosling:1993, koskinen:2006}.
They are extremely common events, occurring at a rate that depends on the solar cycle and that can exceed 10 CMEs per day during solar maxima \citep{robbrecht:2009}.
CMEs mostly originate from active regions (ARs), where magnetic energy is stored in sheared and twisted magnetic field structures. Eventually these structures become unstable and erupt, releasing plasma and magnetic fields in the form of CMEs that propagate outwards in the heliosphere, subsequently affecting planetary systems and space missions in the Solar System.
From a terrestrial perspective, Earth-directed CMEs are the most important ones in terms of space weather implications and effects on our planet \citep{webb:2000, michalek:2006}, as they can cause significant damages to space missions and ground-based infrastructures, affecting a wide range of industry and service sectors \citep{schrijver:2015} as well as military operations \citep{knipp:2018}. 

% ICMEs
When observed in situ, the interplanetary counterparts of CMEs are denoted as Interplanetary CMEs (ICMEs).
The most relevant parameters assessing their potential impact on Earth, or "geo-effectiveness", are their speed, density and internal magnetic field at arrival \citep{akasofu:1973, burton:1975, dumbovic:2015, kilpua:2017}. 
The first two parameters contribute to the dynamic pressure of the impinging solar wind, which typically peaks in association with the passage of interplanetary shocks developing at the front of ICMEs. 
Although interplanetary shocks can cause significant magnetospheric compression and have been proven to be a source of geomagnetic activity \citep{tsurutani:2011, oliveira:2018}, 
strong geomagnetic storms are mainly driven by the internal magnetic structure of ICMEs \citep{gonzalez:1994, zhang:2007, lugaz:2016}.
Accurate predictions of the ICME magnetic field strength and orientation at Earth, and particularly that of its $B_z$ component, are therefore needed in order to reliably predict the geo-effectiveness of ICME structures.

% models
Over the past decades, the solar and space physics community have developed a variety of models to predict the time of arrival (ToA) of CMEs and some of their basic parameters such as the speed and density characteristics at Earth and other locations in space (see \citealt{riley:2018} for an updated list of models).
% cone models
Among them, physics-based heliospheric models that describe CMEs by means of cone models have gained an important position in space weather operations, due to their relative simplicity of use and robustness \citep[e.g. the ENLIL model,][]{odstrcil:2004}.
In cone models, CMEs are described as hydrodynamic blobs of plasma characterised by a self-similar expanding geometry \citep{xie:2004, xue:2005}, that are injected in the heliosphere with a magnetic field equal to the one of the background solar wind. 
Due to this simplified description of the CME structure, cone models are not suitable to study and predict the magnetic field structure associated to ICMEs; on the other hand, they have been successfully used to study the global evolution of CMEs and the propagation of their shock fronts in the heliosphere, to assess the CME arrival (yes/no) at Earth and other spacecraft locations, and to predict CME arrival times at a given location \citep[see for example][]{cash:2015,mays:2015,guo:2018}.
% flux-rope models
In the attempt to overcome the cone model limitations, recent efforts have focused on modelling CMEs using more realistic flux-rope models, such as spheromaks or toroidal-like structures \citep[see for example][]{shiota:2016,jin:2017}.
% EUHFORIA
In particular, EUHFORIA \citep[EUropean Heliospheric FORecasting Information Asset;][]{pomoell:2018} is a new solar wind and CME propagation model that has been recently extended to model CMEs as spheromak flux-rope structures. 
\citet{verbeke:2019} provided a detailed analysis of the spheromak model in EUHFORIA, 
highlighting promising improvements in the magnetic field predictions at Earth for one test case CME event. 
However, they initialised the spheromak CME using input parameters that were only partially derived from observations. 
In order to consistently develop a tool for predicting the ICME properties at L1 and their geo-effectiveness,
one would need to constrain all the CME input parameters from remote-sensing observations at the Sun, 
ideally reducing the number of unconstrained CME input parameters to zero. 
At the same time, a study of more than one case study CME event is necessary in order to quantify the prediction improvements in different conditions, and to asses the model limitations.

% section summary
In this work, we aim to assess how well the spheromak model can actually predict the ICME parameters at Earth, and particularly its magnetic signature, when it is initialised using observational parameters only.
The paper is structured as follows. 
In Section~\ref{sec:euhforia} we briefly describe the EUHFORIA model and compare the cone and spheromak CME models currently implemented.
In Section~\ref{sec:parameters} we discuss in the detail the determination of the CME kinematic, geometric and magnetic parameters at 0.1 Astronomical Unit (AU) from multi-spacecraft remote-sensing observations of CMEs and related source regions at the Sun.
Section~\ref{sec:case_studies} contains a detailed description of the two CME events selected as case studies.
In Section~\ref{sec:results} we present the simulation set up and we compare simulation results with observational data of the two case studies considered in this paper.
After simulating each CME event using both the cone and the spheromak model, we study the modelled CME propagation in the heliosphere and discuss similarities and differences between the two models.
Moreover, we investigate the predictions of the ICME properties at L1, discussing the spheromak capabilities and limitations in the case of well-observed CME events. 
In Section~\ref{sec:conclusions} we discuss the results and consider future improvements and applications.
In this work we investigate the solar and heliospheric evolution of the CMEs, while a detailed study of the predicted CME geo-effectiveness in terms of the induced geomagnetic activity will be addressed in a second paper.

%============================================================
%============================================================
\section{Modelling CMEs with EUHFORIA} 
\label{sec:euhforia}
%============================================================
%============================================================

EUHFORIA is a new physics-based coronal and heliospheric model designed for space weather research and prediction purposes, that models the background solar wind and CMEs in the heliosphere up to 2~AU.
The model is composed of two main parts: 
(1) the coronal model, which takes as input 
synoptic magnetograms from the Global Oscillation Network Group (GONG) and then provides the plasma quantities at 0.1~AU, corresponding to the heliospheric inner boundary, using a semi-empirical Wang-Sheeley-Arge-like model \citep[WSA;][]{arge:2004}.
(2) The heliospheric model solves three-dimensional (3D) time-dependent magnetohydrodynamics (MHD) equations to generate a self-consistent model of the background solar wind between 0.1~AU and 2~AU based on the output of the coronal model.
In addition to modelling the background solar wind, EUHFORIA can also model CMEs either using the well-established but limited cone model (Section~\ref{subsec:cone_model}) or using a linear force-free spheromak model (Section~\ref{subsec:fluxrope_model}).
CMEs are initialised as time-dependent inner boundary conditions at 0.1~AU, corresponding to the inner boundary of the heliospheric domain.

%-----------------------------------------------------------
\subsection{The cone CME model}
\label{subsec:cone_model}
%-----------------------------------------------------------
One simple approach to model CMEs in the heliosphere is by means of a cone model, 
which describes CMEs as uniformly-filled bubbles of plasma characterised by a spherical shape \citep{odstrcil:2004, scolini:2018b}. 
In cone models, CMEs are treated as dense, spherical blobs of plasma injected in the heliosphere without any internal magnetic field structure, e.g. their internal magnetic field is just the one of the background solar wind. 
Due to this simplified description, the major limitation of cone models is their inability to accurately predict the magnetic field properties of ICMEs; for this reason, they can only be used to model the propagation of CME-driven shock fronts and not that of their drivers.
In EUHFORIA, cone CMEs are initialised specifying a set of 7 input parameters 
defining the CME kinematics and geometry during the CME insertion at the heliocentric distance of 0.1~AU (=21.5~solar radii, hereafter $R_s$), 
corresponding to the inner boundary of the heliospheric model.
These parameters, namely the CME insertion time, its speed $v_\mathrm{CME}$, direction of propagation (latitude $\theta$ and longitude $\phi$), and angular half width $\omega /2$ at 0.1~AU, 
are usually derived from coronagraphic observations of the CME.
In addition, two extra parameters defining the CME mass density and temperature 
are set to be homogeneous and equal to the following default values: 
$\rho_\mathrm{CME} = 1 \cdot 10^{-18} \,\, \si{ \kg \cdot \m^{-3} } $
and $T_\mathrm{CME} = 0.8 \cdot 10^{6} \,\, \si{\kelvin}$ \citep{pomoell:2018}.

%-----------------------------------------------------------
\subsection{The linear force-free spheromak CME model}
\label{subsec:fluxrope_model}
%-----------------------------------------------------------

EUHFORIA has been recently extended to be able to model CMEs as flux-ropes structures, potentially allowing for a more realistic study of the CME propagation and evolution in the heliosphere. 
The linear force-free spheromak model \citep{chandrasekhar:1957, shiota:2016} is the first flux-rope model that has been implemented in EUHFORIA \citep{verbeke:2019}.
This model describes the flux-rope structure as a force-free magnetic field configuration characterised by a global spherical shape.
Once completely inserted in the heliosphere, a spheromak CME will therefore be completely disconnected from the Sun.
It is important to note that studies on the global shape of ICMEs at 1~AU based on in-situ and remote-sensing observations, have provided evidence that the axes of magnetic flux-rope structures in ICMEs can be described as having ellipsoidal shapes often still connected to the Sun \citep{janvier:2013}.
As such, the spheromak flux-rope model is able to approximate the structure of ICME flux-ropes only locally, while it is not able to reproduce their global, large-scale geometry.

When simulating CMEs in EUHFORIA using the spheromak model, 
three additional input parameters are needed compared to that required by the cone model.
These parameters, that determine the CME internal magnetic field, are 
the helicity sign (chirality), the tilt, and the toroidal magnetic flux
at 0.1~AU. 
In the current implementation, the mass density and temperature 
inside the CME are set to be uniform, and prescribed according to the same default values as used in cone CMEs.

%-----------------------------------------------------------
\subsection{Role of the Lorentz force on CME propagation}
\label{subsec:cone_vs_fluxrope}
%-----------------------------------------------------------

In the ideal MHD description, Newton's second law assumes the form of a momentum equation which in an Eulerian frame can be written as
\begin{equation}
    \frac{\partial (\rho \mathbf{v}) }{\partial t} = 
    - \rho \vec{v} \cdot  \nabla \vec{v} 
    + \vec{j} \times \vec{B}
    - \nabla P \label{eqn:momentum1} 
\end{equation}
where
$\rho$ is the mass density, 
$\vec{v}$ is the fluid velocity vector,
$\vec{B}$ is the magnetic field, 
$P = \rho T \frac{k_B}{m_p} $ is the plasma (thermal) pressure, 
and $\vec{j} \times \vec{B}$ is the Lorentz force.
The Lorentz force can also be expressed as the sum of a magnetic pressure and magnetic tension term, as:
\begin{equation}
    \frac{\partial (\rho \mathbf{v}) }{\partial t} =
    - \rho \vec{v} \cdot  \nabla \vec{v}
    + \frac{ (\vec{B} \cdot \nabla) \vec{B} }{\mu_0}
    - \nabla (P+P_{mag}), \label{eqn:momentum2}
\end{equation}
where
$\mu_0$ is the magnetic permeability of vacuum,
and $P_{mag} = \frac{B^2}{2 \mu_0}$ is the magnetic pressure.
% pressures
In Equation \ref{eqn:momentum2} a positive pressure gradient $\nabla (P+P_{mag}$) acts as an expanding force on a parcel of fluid, while a negative pressure gradient generates a compression. 
% magnetic tension
On the other hand, the magnetic tension $\frac{(\vec{B} \cdot \nabla) \vec{B}}{\mu_0}$ acts as a restoring force against the bending of magnetic field lines.
In an MHD description, the evolution of any plasma structure, particularly that of CMEs, is therefore governed by the interplay between the two terms, as well as by the plasma inertia.

\medskip
In general, it can be envisaged that, for a given background solar wind,
the plasma characterising a CME will evolve differently depending on the particular CME model used.
% cone CMEs
Cone CMEs have very weak internal magnetic fields, hence their internal pressure is simply
\begin{equation}
P_c = P + P_{mag} \simeq P.
\end{equation}
No significant magnetic pressure gradient or tension terms are present, due to the fact that the CME only has the background solar wind magnetic field. 
In heliospheric simulations, prescriptions of $T_{CME}$ and $\rho_{CME}$ in cone CMEs are usually such that $P > P_{sw}$, so that the positive pressure gradient at the CME-solar wind interface generates an expansion of the CME body.

The evolution of flux-rope CMEs is more heavily affected by Lorentz forces acting within and around their bodies.
% non-force-free flux-rope
In general, flux-rope configurations are non-force-free ($\vec{j} \times \vec{B} \neq 0 $), and in this case internal electric currents $\vec{j}$ non-parallel to $\vec{B}$ are responsible for the occurrence of the so-called Lorentz self-force acting within CME bodies \citep[][]{subramanian:2014}. 
%
% force-free flux-rope
In the case of force-free flux-ropes ($\vec{j} \times \vec{B} = 0 $) such as the spheromak model employed in this work, internal electric currents $\vec{j}$ are, by construction, parallel to $\vec{B}$. 
Although within these flux-ropes the Lorentz force vanishes as long as the force-free condition holds, 
a non-zero Lorentz force can develop at the CME-solar wind interface due to local force imbalances mainly associated to pressure gradients. 
Within spheromak CMEs, the internal pressure is
\begin{equation}
P_\mathrm{FR} = P + P_{mag}.
\end{equation}
where $P_{mag} \gg P$, i.e. spheromak CMEs are generally low-$\beta$, magnetically-dominated objects.
As $P_\mathrm{FR} > P_c$, spheromak CMEs are subject to higher (positive) pressure gradients at the CME-solar wind interface than cone ones, suggesting a stronger expansion according to Equation~\ref{eqn:momentum2}. At the same time, magnetic tension terms can become significant in response to strong bendings of the flux-rope magnetic field lines.
In heliospheric simulations, the Lorentz force can therefore be expected to play a major role in CME evolution 
even when the internal magnetic field structure of the CME is defined as a force-free configuration.

\medskip
Lorentz forces acting on CMEs are interpreted to be at the origin of two major global effects that are often observed in relation to CME/ICME evolution: CME acceleration and CME expansion. The two effects are discussed below.

% hoop force ====================
\textit{CME acceleration/propagation.}
CME accelerating behaviours in the corona have often been explained in terms of Lorentz self-forces \citep[][and references therein]{subramanian:2014}.
From Equation~\ref{eqn:momentum1}, the Lorentz force can manifest in the form of self-force due to misaligned currents and magnetic fields within evolving flux-rope structures. This is expected to occur particularly in the case of traditional, loop-like flux-ropes connected at both end to the Sun, where the curvature of their toroidal (axial) magnetic field induces an asymmetry between the leading and the trailing parts of the loop. This asymmetry is associated to a magnetic pressure gradient that results in an outwardly-directed force that accelerates the flux-rope \citep{subramanian:2009}.
\citet{subramanian:2007} investigated the impact of the magnetic pressure on CME kinematics in the range 2-30~$R_s$, studying the energetics of CMEs in terms of the evolution of their kinetic and magnetic energy reservoirs from coronal observations. Their study provided observational evidence that the CME kinematics in such range of distances cannot be explained in terms of the drag force alone, and that the Lorentz self-force needs to be taken into account to explain CME kinematics, {i.e.} their speed behaviour.
Due to its symmetrical magnetic structure, the linear force-free spheromak model employed in this work should in principle be weakly associated to the Lorentz self-force, as its magnetic structure is defined as symmetric in its leading and trailing portions. 

% CME expansion ====================
\textit{CME expansion.}
The second major observational consequence of Lorentz forces acting on CMEs/ICMEs, is their expansion.
In relation to Equation~\ref{eqn:momentum2}, indications that the internal pressure in CME bodies in the corona and in the heliosphere is dominant compared to the magnetic tension acting against the bending of magnetic field lines are provided by evidences of CME  expansion in both remote-sensing and in-situ observations of CMEs/ICMEs.
% solar corona
For example, in the solar corona starting from a height of about 2.5-3~$R_s$, 
CMEs are often observed to evolve with a self-similar expanding behaviour \citep{cremades:2004, kilpua:2012}, which has been interpreted as evidence of non-force-free conditions \citep{subramanian:2014}.
% IP space
An evidence that CME expansion is still occurring within MCs at 1~AU is provided by 
the plasma velocity profiles, which often show a linear variation along the spacecraft trajectory, 
with a higher velocity at the ICME front than at its back \citep{burlaga:1982,lepping:2008}.
As discussed by \cite{demoulin:2009b}, heliospheric ICME expansion is primarily due to the drop in the solar wind pressure at increasing radial distance from the Sun.
This effect is expected to affect the propagation of spheromak CMEs in our simulations as consequence of force imbalances (via magnetic pressure gradient) developing at the CME-solar wind interface.

\medskip
% conclusions
In summary, the CME evolution is related to both the internal CME properties, and the (external) solar wind plasma properties, and it originates from force unbalances within CMEs and at their interaction surface with the solar wind. 
When modelling CMEs in EUHFORIA, which forces dominate depends on the particular CME model chosen.
In Section~\ref{sec:results} we will discuss how this affects the heliospheric propagation of CMEs in EUHFORIA, and how these differences can be mitigated by adjusting the CME input parameters in the simulations, based on observational parameters in the corona.

%===============================================================================
%===============================================================================
\section{Deriving the CME parameters from source region and coronal observations} 
\label{sec:parameters}
%===============================================================================
%===============================================================================

In this Section, we discuss how to constrain the CME geometric, kinematic, and magnetic parameters that are needed as input parameters at 0.1~AU, from remote-sensing observations of CMEs and their source regions.

%-----------------------------------------------------------
\subsection{Kinematic and geometric parameters}
\label{subsec:gcs}
%-----------------------------------------------------------

To derive the CME geometric and kinematic parameters, we fit each CME with a croissant-like 3D shape using the Graduated Cylindrical Shell model \citep[GCS;][]{thernisien:2009,thernisien:2011}.
We use contemporaneous observations of CMEs in the solar corona from the \textit{Large Angle and Spectrometric COronagraph} (LASCO) instrument on board the \textit{Solar and Heliospheric Observatory} \citep[SOHO][]{brueckner:1995}, and from the \textit{Sun Earth Connection Coronal and Heliospheric Investigation} (SECCHI) instrument on board the \textit{Solar TErrestrial RElations Observatory} \citep[STEREO][]{howard:2008}. 
\begin{figure}[h]
\centering
{\includegraphics[width=\hsize,trim={0mm 0mm 0mm 0mm},clip]{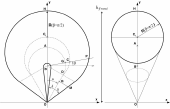} } 
\caption{Schematic of the GCS model, adapted from \citep{thernisien:2011}:
face-on (left) and edge-on (right) representations.
In the case $\alpha=0$, the face-on and edge-on views coincide.}
\label{fig:gcs} 
\end{figure}
The results of the fitting using the GCS model allow to estimate the following instantaneous quantities:
the CME direction of propagation in terms of its longitude $\phi$ and latitude $\theta$ (in Stonyhurst heliographic coordinates), 
the height of the CME apex $h_{front}$, 
the tilt angle around the axis of symmetry $\gamma$ (with respect to the solar equator), 
the half angle between the legs $\alpha$, 
and the half angle of the cone $\delta$, related to the "aspect ratio" $\kappa$ by the relation $\kappa = \sin \delta$.
The geometrical meaning of all the parameters is shown in Figure~\ref{fig:gcs}.
By applying the GCS model to a sequence of images, 
one can extract the 3D speed at the CME apex from the derivative of the CME apex height over time:
\begin{equation}
v_{3D} = \frac{d h_{front}}{d t}.
\label{eqn:gcs_v3d}
\end{equation}
Following the discussion in Section~\ref{subsec:cone_vs_fluxrope}, here we are interested in estimating the contributions to the total 3D speed coming from both the expansion and radial speed terms, starting from observations.
As pointed out by several previous studies \citep{dallago:2003,schwenn:2005,gopal:2012}, 
separating the expansion and radial speed contributing to the total speed of a CME is non-trivial.
In the case of single-spacecraft coronagraphic observations, 
the expansion speed can be directly quantified only for CMEs that are observed as limb events.
In the case of multi-spacecraft observations, however, one can estimate the expansion term fitting the CME body with a geometrical shape. In this work, we propose an approach based on employing the CME parameters obtained from the GCS model, as described below. 
At the CME apex, i.e. along the CME axis of propagation, for a self-similarly propagating CME in the corona, the 3D speed at any time can be expressed as sum of two contributions, the radial speed $v_{rad}$ and the expansion speed $v_{exp}$:
\begin{equation}
v_{3D} = v_{rad} + v_{exp}.
\label{eqn:3Dspeed}
\end{equation}
Using the same notation as \citet{thernisien:2011}, and for $\kappa$ constant in time (i.e. a self-similarly expanding CME),  one can express the radial and expansion contribution as
\begin{align}
v_{rad} &= \frac{1}{1+\kappa} \frac{d h_{front}}{dt}
\label{eqn:gcs_vrad}
\end{align}
and 
\begin{align}
v_{exp} &= \frac{\kappa}{1+\kappa} \frac{d h_{front}}{dt}.
\label{eqn:gcs_vexp}
\end{align}
We redirect the reader to Appendix \ref{app:appendix_b} for the analytical derivation of Equations \ref{eqn:gcs_vrad} and \ref{eqn:gcs_vexp}, including the general case when $\kappa = \kappa(t)$.
These relations provide a geometrically-based method that allows to quantify the expansion and radial speeds associated to a CME directly from the parameters obtained from the GCS reconstruction.
This approach represents an alternative to using empirical relations to derive the CME expansion speed. Since empirical relations only apply to statistically relevant set of events, they may provide inaccurate results for a specific CME event \citep[see][]{dallago:2003,schwenn:2005,gopal:2012}.
At the same time, the methodology described above implicitly assumes that a 3D reconstruction of the CME event under study is possible, 
i.e., that at least two coronagraphs observe the CME from different view points.
Should this not be the case, {e.g.} only single spacecraft observations are available, the use of empirical relations would still be needed.
For this reason, in Section~\ref{sec:parameters} we compare the method above with empirical relations based on single-spacecraft observations of CME events.
In particular, we consider the empirical relation proposed by \cite{dallago:2003} and \cite{schwenn:2005}, 
linking the CME (3D) front speed and the CME expansion speed as:
\begin{equation}
{v_{3D} = 2 \cdot 0.88 \, v_{exp}} = 1.76 \, v_{exp},
\label{eqn:v3d_dallago2003}
\end{equation}
where $v_{exp}$ is the variation of the CME radius over time.
This relation provides an estimate of the 3D speed starting from observations of what we think is the expansion speed of a CME.
Although more sophisticated relations have been developed to better capture the plethora of expansion/radial speed combinations observed \citep{gopal:2012}, in this work we limit our attention to Equation \ref{eqn:v3d_dallago2003}, as this is the one relying on the smallest number of parameters.
Equation \ref{eqn:v3d_dallago2003} has been fine-tuned for a set of CME events that have been observed as full-halo events from Earth (but the same could apply to any other spacecraft in space), and for which no observations from other directions were available. 
In such cases, one can assume that $v_{2D} \simeq v_{exp}$.
If the 3D reconstruction of the CME is possible, one can invert Equations \ref{eqn:v3d_dallago2003} and \ref{eqn:3Dspeed}
to estimate the expansion and radial speed from the 3D speed, as 
\begin{equation}
\begin{cases} 
v_{exp} = 0.57 \, v_{3D} \\
v_{rad} = v_{3D} - v_{exp} = 0.43 \, v_{3D}.
\end{cases}
\label{eqn:vexp_dallago2003_vexp}
\end{equation}

%-----------------------------------------------------------
\subsection{Magnetic parameters}
\label{subsec:magnetic_parameters}
%-----------------------------------------------------------

With the aim of developing a fully predictive methodology, 
in this work we constrain the flux-rope magnetic parameters needed to initialise spheromak CMEs in EUHFORIA directly from remote-sensing observations of the corona available at the time of the observed eruptions.
The magnetic input parameters needed in the case of spheromak CMEs are the flux-rope tilt, the flux-rope chirality, and the flux-rope toroidal magnetic flux at 0.1~AU.
Estimating each of those CME parameters at 0.1~AU from observations is extremely challenging.
In general, strong approximations combined with photospheric and low-coronal observations of the source active region before and after the eruption are needed.
To derive each of those parameters for the case studies analysed in this work, 
we use the approaches described below.

\medskip
\textit{Flux-rope chirality.}
Magnetic helicity provides a quantification of how much the magnetic field is sheared and twisted compared to the lowest-energy state, i.e. the potential field. It exhibits the unique property of being almost completely conserved over time, even in presence of magnetic reconnection events \citep{berger:2005}.
As a consequence, also its sign, commonly referred to as \textit{handedness} or \textit{chirality}, is a quantity conserved over time.

Observationally, the chirality of active regions and erupting filaments can be inferred from different morphological features 
\citep[see][and references therein]{demoulin:2009a,palmerio:2017}.
Indeed, several studies have found that the chirality of most MCs matches with the one inferred from the morphological features of the associated source regions and erupting filaments \citep{bothmer:1998,palmerio:2018}.
Nevertheless, examples of inconsistency between the chirality of the source active region and the one of the associated filament have been observed and interpreted as due to local phenomena of helicity injections in active regions prior and during the eruption \citep{chandra:2010, romano:2011, zuccarello:2011}. 

Assuming that the large-scale magnetic field of the active region has the same chirality of the one of the associated MC, in this work we determine the chirality of erupting flux-rope from pre-eruption EUV observations of the CME source regions, considering in particular EUV sigmoids as its main proxy.
We make use of images obtained by the \textit{Atmospheric Imaging Assembly} (AIA) instrument on board the \textit{Solar Dynamics Observatory} \citep[SDO;][]{lemen:2012}.
Being aware that a robust determination of the chirality should be based on more than one indicator/proxy,
we then compare our estimation with that reported by \citet{palmerio:2018},
who already performed a detailed analysis of both the events considered in this work.

\medskip
\textit{Flux-rope tilt angle/orientation.}
In general, the orientation of a CME/flux-rope axis can be altered by rotation phenomena over a wide range of distances from the Sun. 
By comparing remote-sensing and in-situ observations at 1~AU in the case of 20 CME events, \cite{palmerio:2018} found rotations of the flux-rope axis ranging between $0^\circ$ and $>150^\circ$. 
Several studies suggest that rotations tend to occur within 4~$R_s$ from the Sun \citep[see][and references therein]{kay:2015}, although cases of extreme rotations ($>60^\circ$) taking place in the middle corona and in the heliosphere have also been reported \citep[e.g.][]{vourlidas:2011,isavnin:2014}.
As the amount of these rotations largely depends on the magnetic configuration of the surroundings of the CME source regions \citep[e.g.][]{kay:2015}, estimating the orientation of CMEs at 0.1~AU requires a detailed analysis of the source region and its surrounding magnetic fields, which is difficult to carry out solely on the basis of EUV and magnetic observations of the source region.

As first approximation, we assume that no CME rotation occurs in the corona, 
e.g. its orientation at 0.1~AU matches the one of the filament prior to the eruption.
In this case, we can infer the orientation of the CME flux-rope from the orientation of the source region polarity inversion line \citep[PIL;][]{marubashi:2015} and/or from the orientation of the Post-Eruption Arcades \citep[PEAs;][]{yurchyshyn:2008}. 
To determine the orientation of the PIL we make use of images obtained by the \textit{Helioseismic and Magnetic Imager} (HMI) instrument on board SDO \citep{schou:2012}.

\medskip
\textit{Flux-rope toroidal magnetic flux.}
To derive the flux-rope toroidal magnetic flux in the corona, we apply a modified version of the FRED method described by \cite{gopal:2017}.
The FRED method uses the PEA area as primary signature indicating the position of flare ribbons, which in turn can be used to mark the area of a source region where magnetic reconnection has occurred. 
Under this assumption, one can compute the reconnected flux 
during an erupting event by computing the total (unsigned) magnetic flux over the PEA area from line-of-sight magnetic field data.
Dividing it by 2 in order to recover the (signed) reconnected flux $\phi_\mathrm{RC}$, one has:
\begin{equation}
\phi_\mathrm{RC} = \frac{1}{2} \int_\mathrm{PEA} \mid B_{los} \mid dA =  \frac{1}{2}  \mid B_{los} \mid_{tot} A_\mathrm{PEA}.
\label{eqn:reconnected_flux}
\end{equation}
We emphasise that the determination of the reconnected flux is subject to large uncertainties.
For example, \citet{gopal:2017} found a difference of about $38\%$ between the value obtained using the PEA method and the one from a similar method based on flare ribbon observations, due to the difficulties in the identification of the ribbon edges. 
In another case, \citet{pal:2017} found a difference of $25\%$ in the $\phi_{RC}$ obtained from EUV and X-ray observations of the PEA. 
This was interpreted as consequence of the fact that the area of the PEA appeared smaller in EUV than in X-ray images.

Assuming that all reconnected flux $\phi_\mathrm{RC}$ goes into the poloidal magnetic flux $\phi_{p}$ of the erupted flux-rope \citep{qiu:2007}, one can estimate the axial field strength $B_0$ for a spheromak flux-rope as (see Appendix \ref{app:appendix_a})
\begin{equation}
B_0 =   \frac{\alpha^3}{ 2 \pi}
        \frac{\phi_p (r_*) \, r_* }{\Big(\sin(\alpha r_* ) - \alpha r_*  \cos(\alpha r_* ) \Big)},
\label{eqn:axial_field}
\end{equation}
with $r_*$ being the distance from the center of the spheromak, 
on the plane $\theta = \pi/2$, where the magnetic field becomes completely axial ($B_r = 0, B_\theta = 0$).
The flux-rope toroidal magnetic flux can be calculated as
\begin{equation}
\phi_{t} = \frac{2 B_0}{\alpha^2} \, 
\Big[ - \sin(x_{01}) + \int_{0}^{x_{01}} \frac{\sin x}{x} dx \Big],
\label{eqn:toroidal_flux}
\end{equation}
where $x_{01} = \alpha r_0 = 4.4934$ is the 1st zero of $J_1$
and $r_0$ is the spheromak radius.

%========================================================================
%========================================================================
\section{Case studies: CMEs on 12 July 2012 and on 14 June 2012} 
\label{sec:case_studies}
%========================================================================
%========================================================================

In this Section we present an analysis of the observations of the two CME events 
that were selected as case studies in this work, according to the following criteria: 
\begin{enumerate}
\item 
They were observed as Earth-directed, fast halo CMEs by the SOHO/LASCO C2 and/or C3 coronagraphs, 
and they were unambiguously associated with ICMEs at Earth. 
\item 
The in-situ ICME signatures were characterised by a shock followed by a turbulent sheath region and a magnetic cloud (MC) structure. 
This condition was verified from visual inspection and by consulting the Richardson and Cane ICME list 
\citep[][\url{http://www.srl.caltech.edu/ACE/ASC/DATA/level3/icmetable2.htm}]{richardson:cane:2003, richardson:cane:2010}.
\item 
The source regions from where the eruptions originated were within $40^\circ$ from the solar disk center to limit projection effects \citep{gopal:2017}, 
and both SDO/HMI as well as SDO/AIA remote observations of the source regions were available to determine the magnetic parameter of the erupted flux-rope. 
\item 
SOHO/LASCO and STEREO/SECCHI images of the CMEs in the corona 
were available from favourable vantage points in order to perform a 3D multi-spacecraft reconstruction of the CME kinematic and geometric parameters. 
%The Earth and the two STEREO spacecraft were almost evenly spaced on the ecliptic plane, with relative separations ranging between $115^\circ$ and $125^\circ$ for both events. These conditions ensure a good spacecraft configuration to perform the 3D reconstruction of the CME in the corona based on multi-spacecraft observations.
\end{enumerate}
Throughout this work, latitudinal and longitudinal coordinates are given in Stonyhurst/Heliocentric Earth Equatorial (HEEQ) coordinates, unless specified otherwise.

%===========================================
\subsection{Event 1: CME on 12 July 2012} 
\label{subsec:20120712}
%===========================================

The first event studied in this work is the halo CME that erupted on 12 July 2012 from NOAA AR 11520.
On the day of the eruption, the AR was classified as having $\beta\gamma\delta$ magnetic topology, according to the Mount Wilson classification \citep{hale:1919, kunzel:1965}, and it was located at coordinates S17E06 on the solar disk.
AIA images of the source region show a sigmoid brightening in the 94 \r{A} filter starting around 15:00~UT, which was closely followed by an intense GOES X1.4 flare (onset: 15:37 - peak: 16:49 - end: 17:30). 
The associated CME was first observed in the LASCO C2 coronagraph at 16:48~UT, appearing as a fast halo CME propagating towards the Earth with an average projected speed of 885~\si{ \km \,\, \s^{-1} }.
This event has been already extensively investigated in multiple studies \citep[see for example][]{hu:2016, gopal:2018, marubashi:2017}.
In terms of CME geo-effectiveness forecasting, it is worth noting that the impact of this event was originally underestimated by the space weather community, which was not expecting the ICME signature at Earth to be characterised by a such long-lasting, steady and intense southward $B_z$ as was eventually observed \citep{webb:2017}.

% Source region and coronal observations
\subsubsection{Source region and coronal observations} 
\begin{figure*}
\centering
\subfloat[]
{\includegraphics[width=.25\hsize, trim={5mm 20mm 5mm 20mm},clip]{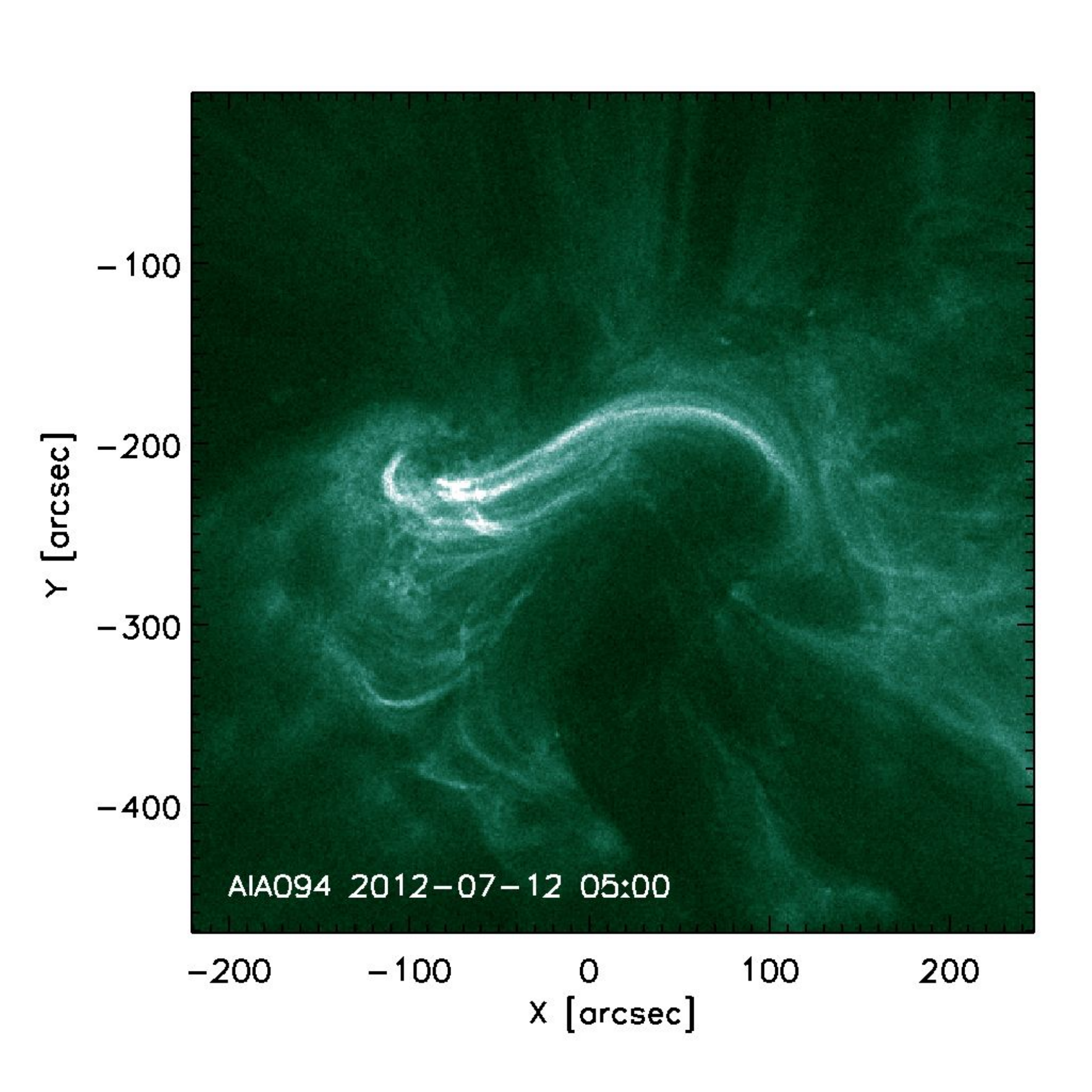} } \quad
\subfloat[]
{\includegraphics[width=.25\hsize, trim={5mm 20mm 5mm 20mm},clip]{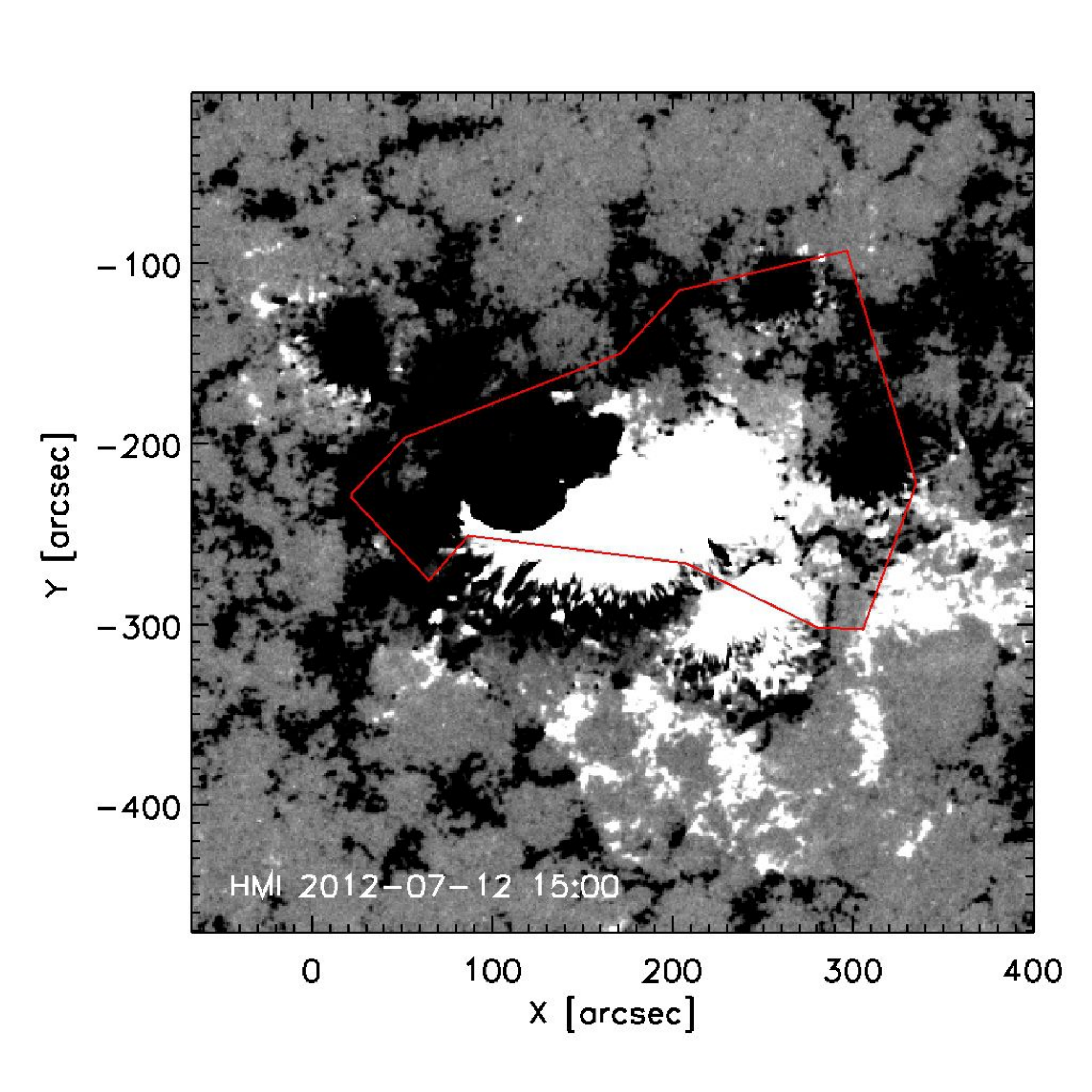}} \quad
\subfloat[]
{\includegraphics[width=.25\hsize, trim={5mm 20mm 5mm 20mm},clip]{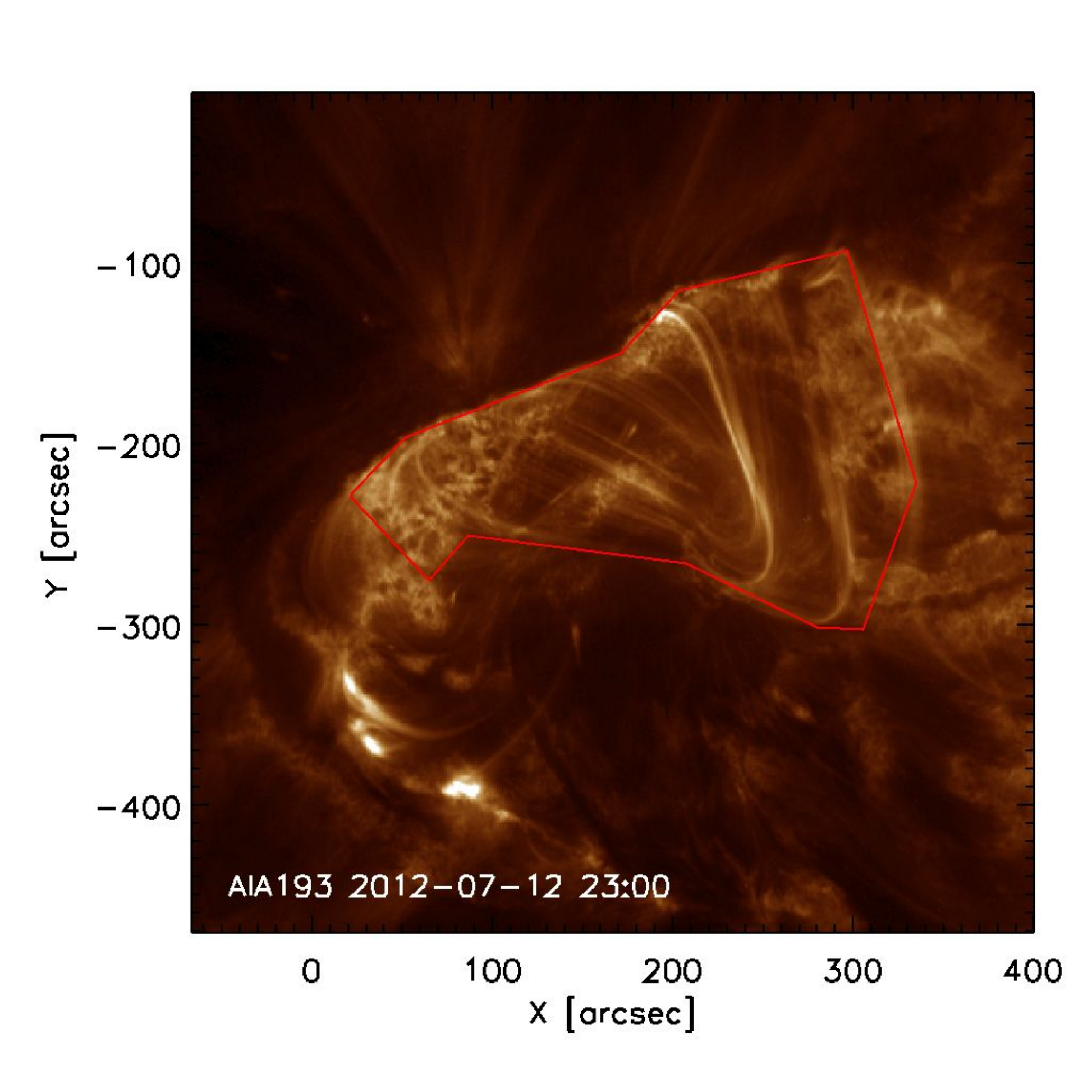}}
\caption{Event 1: AR 11520 on 12 July 2012.
        (a) AIA 94 \r{A} image of the pre-eruptive sigmoid.
        (b) HMI magnetogram with PEA area overlaid (saturated at -100~gauss and +100~gauss).
        (c) PEA from AIA 193 \r{A} with the area outlined by a polygon.
		The dates and times are shown as YYYY-MM-DD hh:mm in all panels.}
\label{fig:20120712_source} 
\end{figure*}

\noindent
% chirality, tilt and reconnected flux =====================
\textit{Source region observations.} 
 Figure~\ref{fig:20120712_source} shows AR 11520 on 12 July 2012 
 as observed by HMI, and by AIA in different EUV channels.
%
% chirality
The source region in AIA 94 \r{A} is characterised by the presence of a forward-S sigmoid, 
suggesting a positive chirality for the erupting flux-rope.
% orientation
From a visual inspection of the HMI magnetogram, the PIL appears inclined by about $45^\circ$ with respect to the solar equator. 
Assuming a positive chirality and having a positive magnetic polarity west of the PIL, we conclude that the flux-rope that formed in the AR is expected to be a low- to mid-inclination flux-rope exhibiting a north-east-south (NES) or east-south-west (ESW) magnetic field rotation \citep[see flux-rope type classification as described by][]{palmerio:2017}, with an axial field pointing towards south-east.
These results are consistent with those reported by \citet{hu:2016}, \citet{gopal:2018}, and \citet{palmerio:2018}.
\citet{kay:2016} studied in detail the magnetic environment surrounding the CME source active region with the ForeCAT model, 
concluding that this particular event underwent almost no deflection or rotation during its early evolution, 
probably due to its very rapid propagation. Therefore, it is reasonable to expect the orientation of the flux-rope structure at 0.1~AU to be consistent with the one at the source region.

% reconnected flux 
Over the 12 hours following the eruption, a long-lasting, stable PEA developed in the active region.
Applying the method described by \citet{gopal:2017} (see also Section~\ref{subsec:magnetic_parameters}) to AIA 193 \r{A} images of the PEA between 12 July at 18:00~UT and 13 July at 00:00~UT, we derive a PEA area of $A_{PEA} = 5.5 - 7.7 \cdot 10^{15}$~m$^2$.
Over-plotting the PEA area on the HMI magnetogram at 15:00~UT on 12 July 2012, 
we estimate the reconnected magnetic flux in the PEA region to be 
$\phi_{RC} = 1.1 - 1.4 \cdot 10^{14}$~Wb. 

% RIBBONDB
We also compare the results obtained above with the ones listed in the {\tt RIBBONDB} catalog \citep[][\url{solarmuri.ssl.berkeley.edu/~kazachenko/RibbonDB/}]{kazachenko:2017}.
The catalog contains properties of ARs and flare ribbons associated with well-observed GOES solar flares of class C1.0+.
The properties of flare ribbons are obtained using AIA observations in the 1600 \r{A} filter, and the estimate of the reconnected flux 
is computed using HMI vector magnetograms.
For AR 11520, the catalogue gives a ribbon area of $A_{R} = 1.3 \cdot 10^{15}$~m$^2$, 
associated with an uncertainty of $\pm 2.8 \cdot 10^{14}$~m$^2$, corresponding to about $20 \%$ of the total value.
The estimated reconnected magnetic flux in the ribbon region is reported as 
$\phi_{RC, R} = 4.3 \pm 0.7 \cdot 10^{13}$~Wb, 
with the uncertainty corresponding to about $15 \%$ of the total value. 

As $A_{PEA}$ is more than a factor four larger than $A_{R}$, the resulting $\phi_{RC,R}$ calculated from the ribbon observations is about a factor two smaller than $\phi_{RC}$ obtained using the PEA observations.
A similar discrepancy in magnitude between the two estimates was also reported by \citet{gopal:2017} comparing two analogous methods.
Although a detailed comparison of the two methods for a large number of events would certainly be extremely valuable to clarify the relationship between PEA and ribbons, in this work we limit ourselves to the use of the results obtained from the PEA-based method as it provides the highest $\phi_{RC}$ estimate. As shown in Section~\ref{sec:results}, this maximises CME magnetic field signals at 1~AU and provides a better match with in-situ observations.

\begin{figure}[t]
\centering
{\includegraphics[width=0.9\hsize]{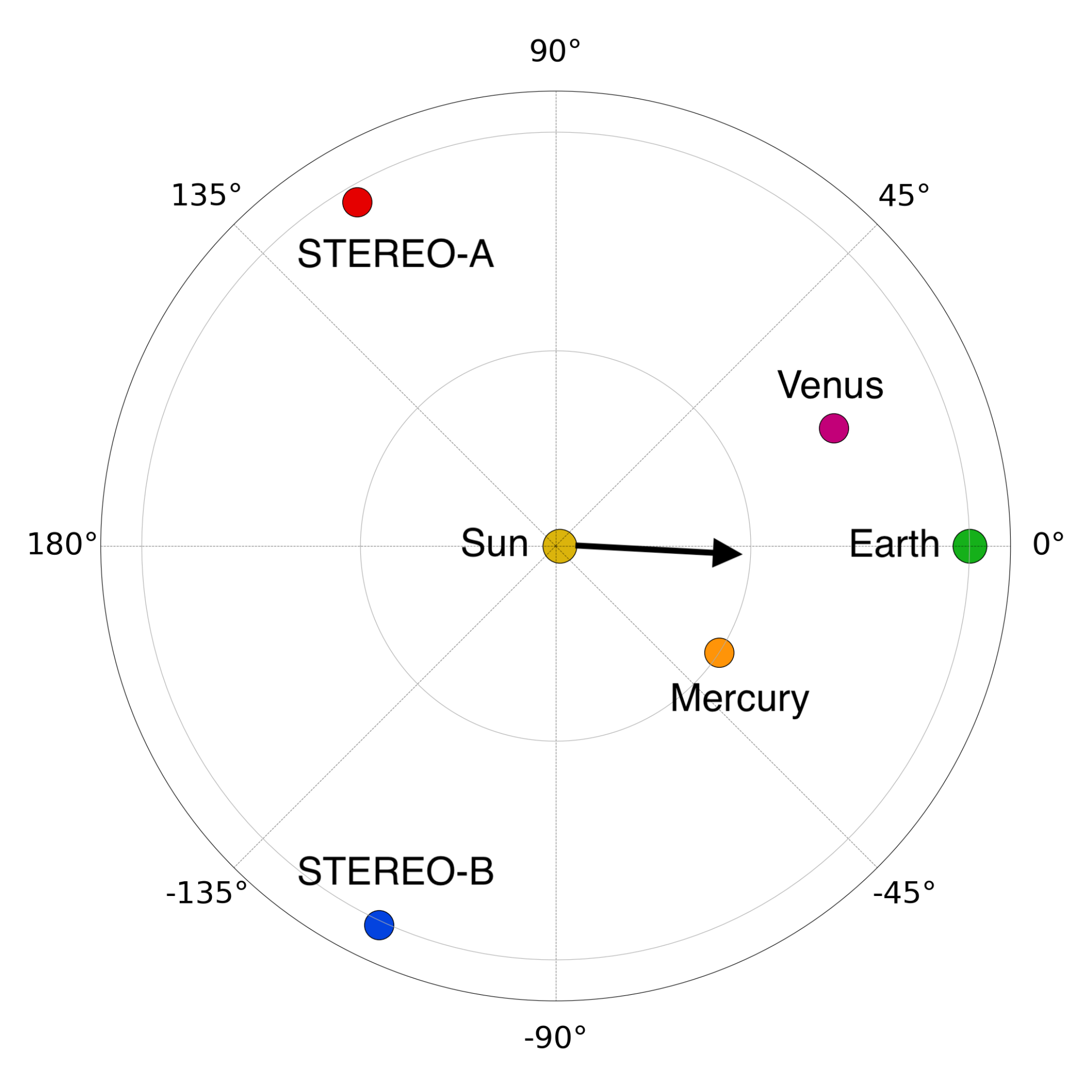} } 
\caption{Event 1: projected position of Earth, STEREO-A, STEREO-B, Mercury and Venus on the ecliptic plane on 12 July 2012 00:00~UT. The black arrow shows the reconstructed longitude of the CME from the GCS fitting.
Longitude is in HEEQ coordinates. }
\label{fig:20120712_ecliptic} 
\end{figure}

\begin{figure}[h]
\centering
{ \includegraphics[width=.48\hsize]{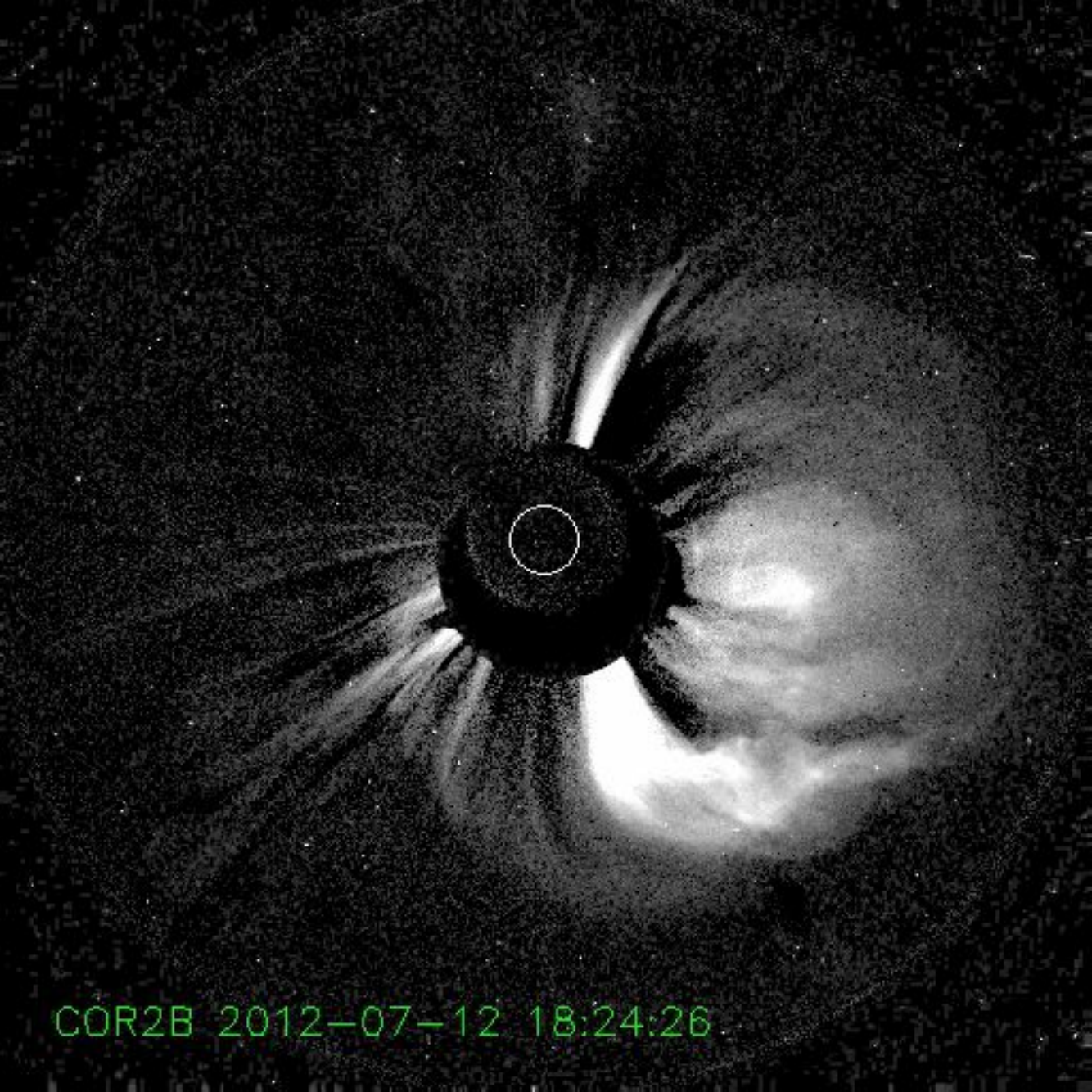} 
\includegraphics[width=.48\hsize]{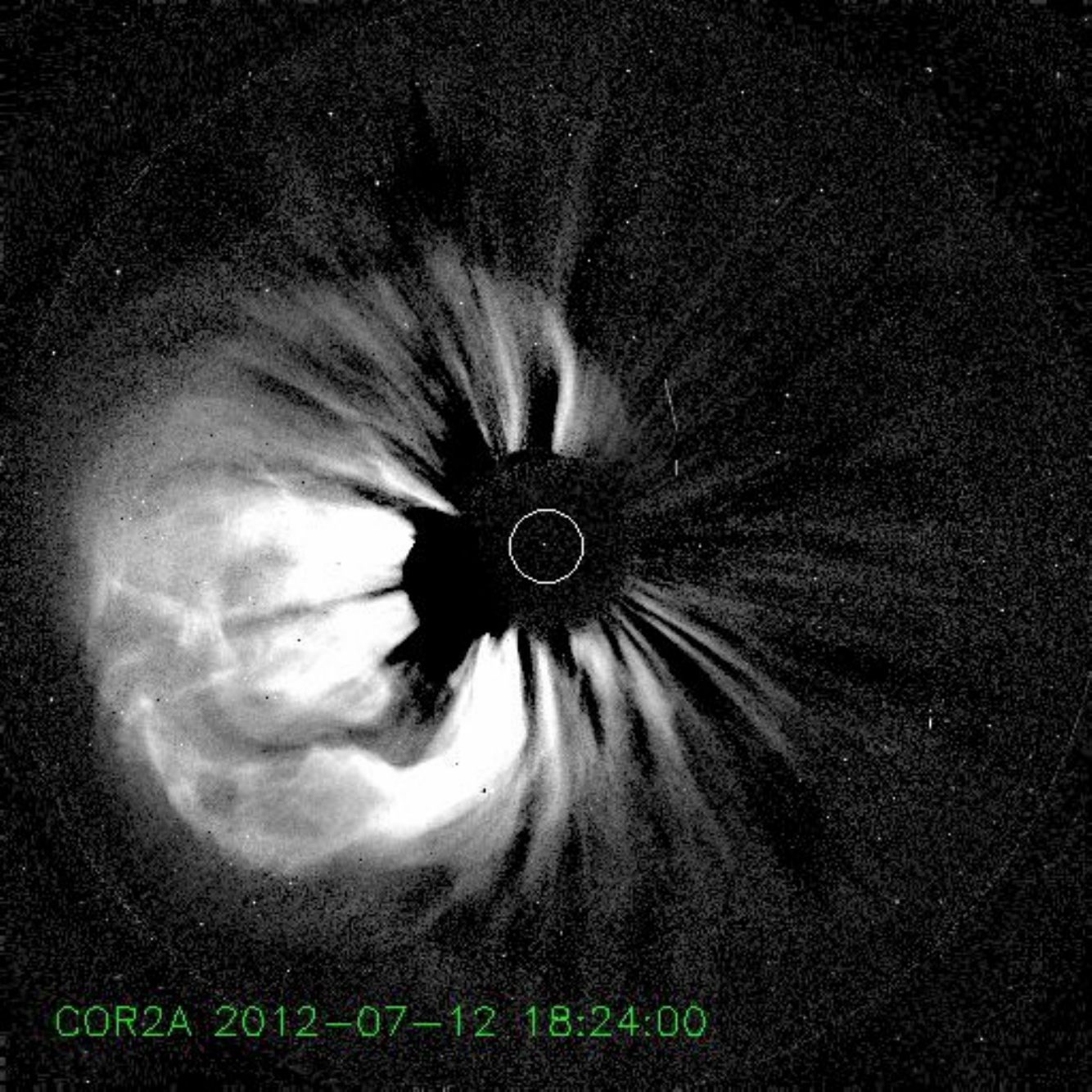} \\
\includegraphics[width=.48\hsize]{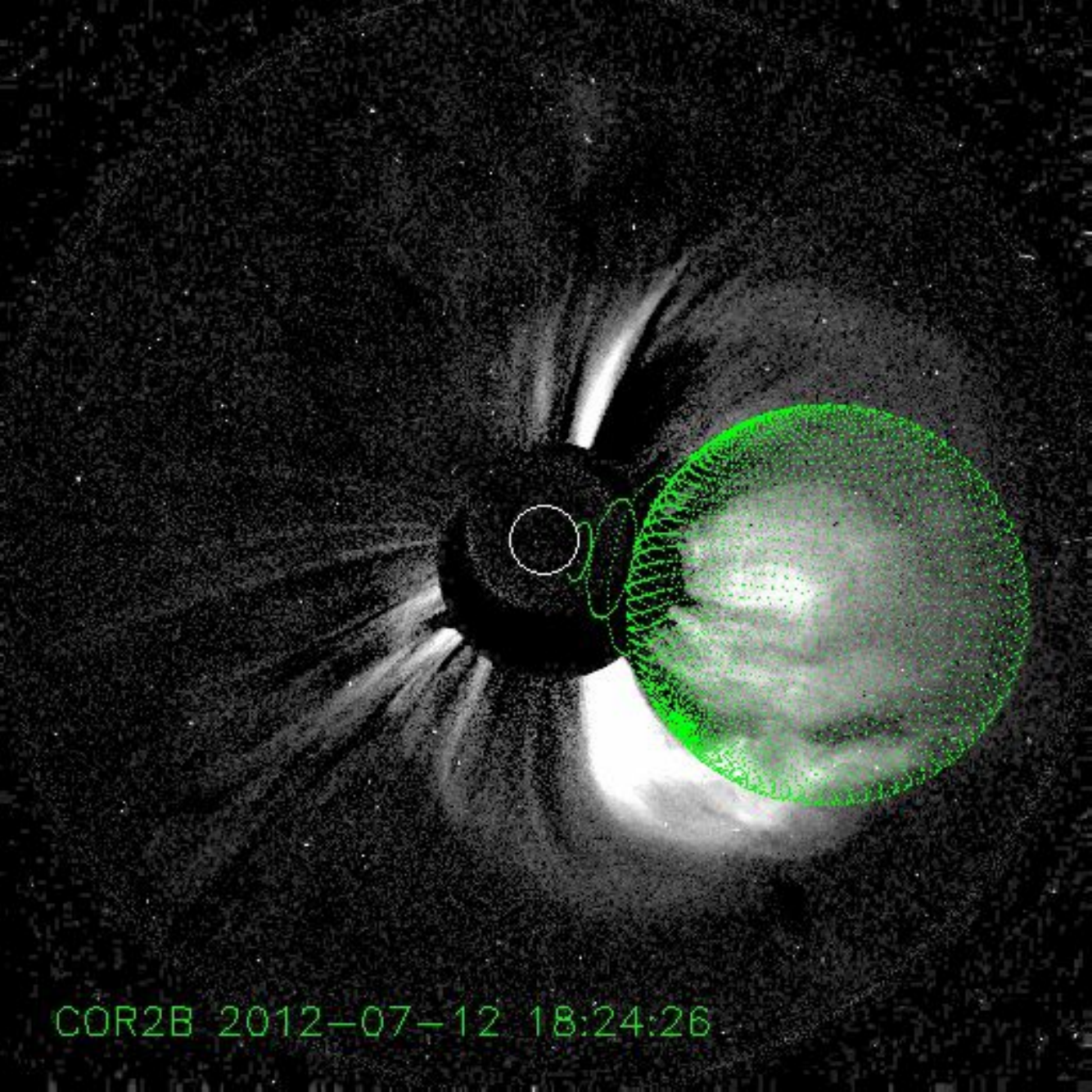}
\includegraphics[width=.48\hsize]{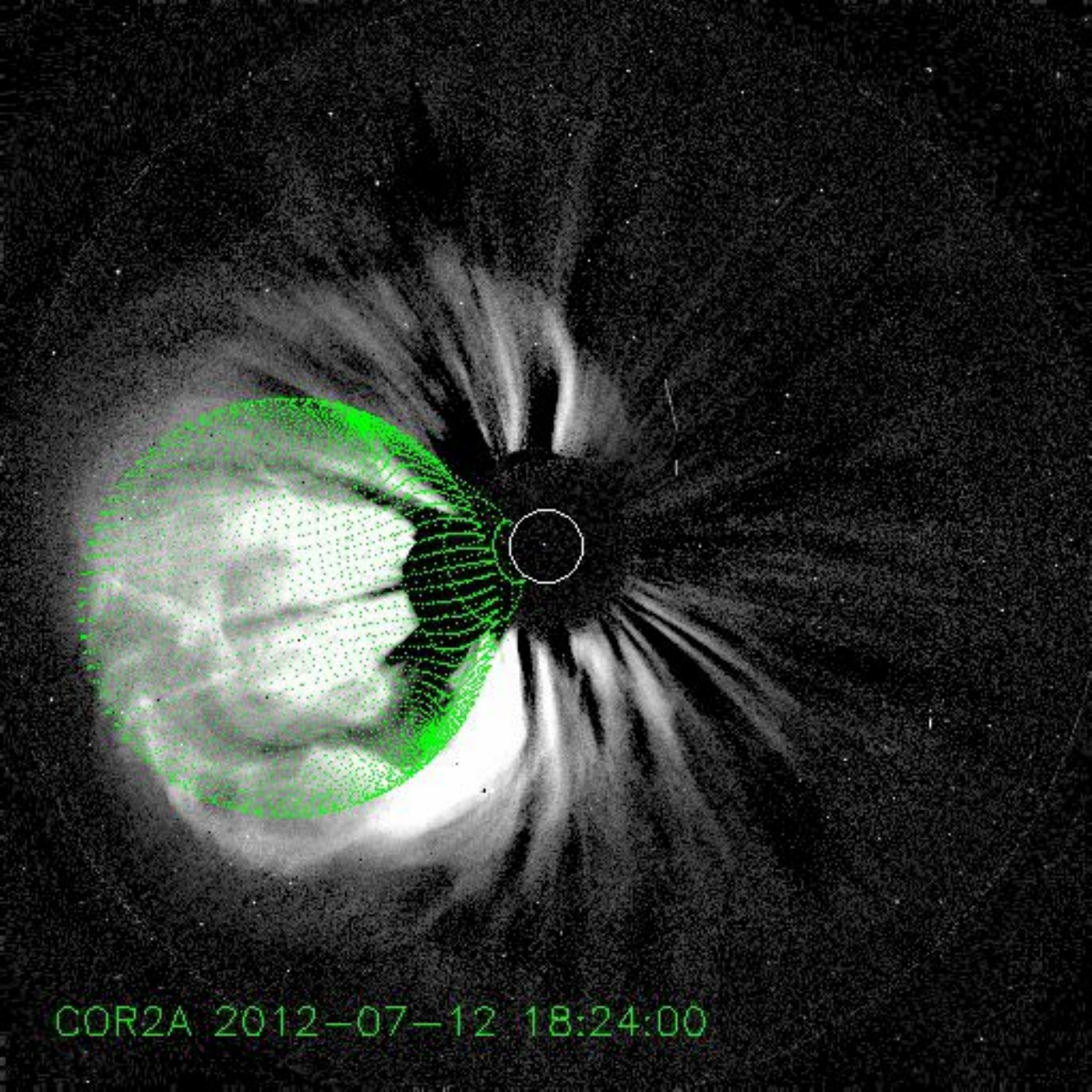} \\ }
\caption{Event 1: SECCHI/COR2B (left) and SECCHI/COR2A (left) pre-event background-subtracted intensity images on 12 July 2012 at 18:24~UT, 
with and without the GCS model wireframe (in green). }
\label{fig:20120712_gcs} 
\end{figure}

\medskip
% coronal observations and GCS =====================
\textit{Coronal observations and GCS reconstruction.} 
As shown in Figure~\ref{fig:20120712_ecliptic}, on the date of the CME eruption the separation of the STEREO spacecraft relative to Earth was $120^\circ$ for STEREO-A and $115^\circ$ for STEREO-B. 
The separation between the two STEREO spacecraft was $125^\circ$.
Due to a data gap, the CME was observed only very early on by LASCO C2 (between 16:48~UT and 17:24~UT), 
and it was not observed at all by the LASCO C3 instrument.
For this reason, we apply the GCS fitting to contemporaneous images of the CMEs from SECCHI/COR2B and SECCHI/COR2A only,
available in the time interval 16:54~UT - 18:24~UT.
Figure~\ref{fig:20120712_gcs} shows the GCS fitting of the CME as observed by COR2A and COR2B on 12 July 2012 at the last available frame (18:24~UT) when the CME was still fully contained within the field of view of the instruments.
We fit the CME with a spherical geometry ($\alpha=0$) in order to be consistent with the spherical shapes characterizing the CME models in EUHFORIA. 
The results obtained from the GCS fitting for the last frames available ({i.e.} closest to 0.1~AU) are listed in Table~\ref{tab:obs_event_1}.
% Table
\begin{table*}
\centering
\begin{tabular}{l|llll}
\hline
\hline
 Parameter      & \multicolumn{4}{c}{Method} \\
 \hline
                & GCS fitting & GCS fitting & & \\
 \hline
 Date           & 2012-07-12      & 2012-07-12  & & \\
 Time           & 18:24~UT        & 17:12~UT  & & \\
 $h_{front}$    & 14.9 $R_s$      &  5.6 $R_s$ & & \\
 $\phi$         & $-4^\circ$      &  $3^\circ$ & & \\
 $\theta$       & $-8^\circ$      &  $-5^\circ$ & & \\
 $\kappa$       & 0.66            &  0.60 & & \\
 $\omega/2$     & $38^\circ$      &  $34^\circ$ & & \\
 $r_0$ at 0.1 AU         & 16.8 $R_s$      & 14.5 $R_s$ & & \\
 \hline
                & Geometrical     & Geometrical  & Empirical-3D   & Empirical-2D \\
 \hline
 $v_{3D}$       & 1266~\si{ \km \,\, \s^{-1} }       &  1352~\si{ \km \,\, \s^{-1} } & 1352~\si{ \km \,\, \s^{-1} } & 1922~\si{ \km \,\, \s^{-1} } \\
 $v_{rad}$      & 763~\si{ \km \,\, \s^{-1} }        &  845~\si{ \km \,\, \s^{-1} } & 582~\si{ \km \,\, \s^{-1} } & 827~\si{ \km \,\, \s^{-1} } \\
 $v_{exp}$      & 503~\si{ \km \,\, \s^{-1} }        &  507~\si{ \km \,\, \s^{-1} } & 770~\si{ \km \,\, \s^{-1} } & 1092~\si{ \km \,\, \s^{-1} } \\
 \hline
\end{tabular}
\caption{Event 1: CME kinematic parameters derived from the GCS fitting and from the application of the geometrical, empirical-3D and empirical-2D approaches to derive total (3D), expansion and radial speeds.}
\label{tab:obs_event_1}
\end{table*}
% geometrical approach
In the geometrical approach, we estimate the CME 3D speed from Equation~\ref{eqn:gcs_v3d},
and the radial and expansion speeds from Equations~\ref{eqn:gcs_vrad} and \ref{eqn:gcs_vexp}.
Extrapolating the CME height in time assuming a constant CME speed,
the CME passage at 0.1~AU is estimated to occur on 12 July 2012 at 19:24~UT.
% comparison with empirical relations
We then compare the geometrical approach with two empirical approaches based on Equations~\ref{eqn:v3d_dallago2003} and \ref{eqn:vexp_dallago2003_vexp}.
% geometrical
As LASCO C2 images were only available before 17:24~UT,
we first perform the GCS fitting to LASCO C2, SECCHI/COR2B and SECCHI/COR2A images taken at 17:00~UT and 17:12~UT (column 2 in Table~\ref{tab:obs_event_1}), 
estimating the radial and expansion speeds from Equation \ref{eqn:gcs_vrad} and \ref{eqn:gcs_vexp}.
% semi-empirical
We then apply an empirical approach (hereafter "empirical-3D") based on the CME 3D reconstruction performed above.
In this case, we use $v_{3D}$ derived from the GCS fitting to derive $v_{rad}$ and $v_{rad}$ from Equation \ref{eqn:vexp_dallago2003_vexp} (column~3 in Table~\ref{tab:obs_event_1}).
% fully-empirical
Finally, we test a completely empirical approach (hereafter "empirical-2D")  based on single-spacecraft observations of the CME from LASCO C2. Using projected speed data provided by the CDAWeb CME Catalog (\url{https://cdaw.gsfc.nasa.gov/CME_list/}), we apply Equations~\ref{eqn:v3d_dallago2003} and \ref{eqn:vexp_dallago2003_vexp} to derive $v_{rad}$ and $v_{rad}$ from observations at 17:24~UT (column~4 in Table~\ref{tab:obs_event_1}).

% discussion
Both the empirical-3D and the empirical-2D approaches give different results than the geometrical method.
On one hand, the total (3D) speed of the CME obtained from the empirical-2D approach is almost a factor 2 higher than the one obtained from the geometrical approach. On the other hand the empirical-3D approach estimates the expansion speed to be higher than the radial speed, while the geometrical approach finds the opposite condition.

\medskip
% FR magnetic parameters =====================
\textit{Derived magnetic parameters.} 
From PEA observations at 23:00~UT, we derive the flux-rope axial magnetic field and toroidal magnetic flux at 0.1~AU from Equations \ref{eqn:axial_field} and \ref{eqn:toroidal_flux}, using the spheromak radius calculated from the half width derived from the GCS fitting of the CME (see Table~\ref{tab:obs_event_1}), assuming that the CME evolved self-similarly in the corona up to 0.1~AU.
The resulting values from $A_{PEA} = 7.7 \cdot 10^{15}$~m$^2$ and $\phi_{RC} \simeq 1.4 \cdot 10^{14}$~Wb are $B_0 \simeq 2.9 \cdot 10^{-6}$~T and $\phi_{t} = 1.0 \cdot 10^{14}$~Wb, consistent with those reported by \citet{gopal:2018}, who analysed the same event assuming a Lundquist flux-rope structure.

% CME propagation in the heliosphere
\subsubsection{CME propagation in the heliosphere}

\begin{figure}[h]
\centering
{\includegraphics[width=\hsize, trim={30mm 30mm 30mm, 30mm},clip]{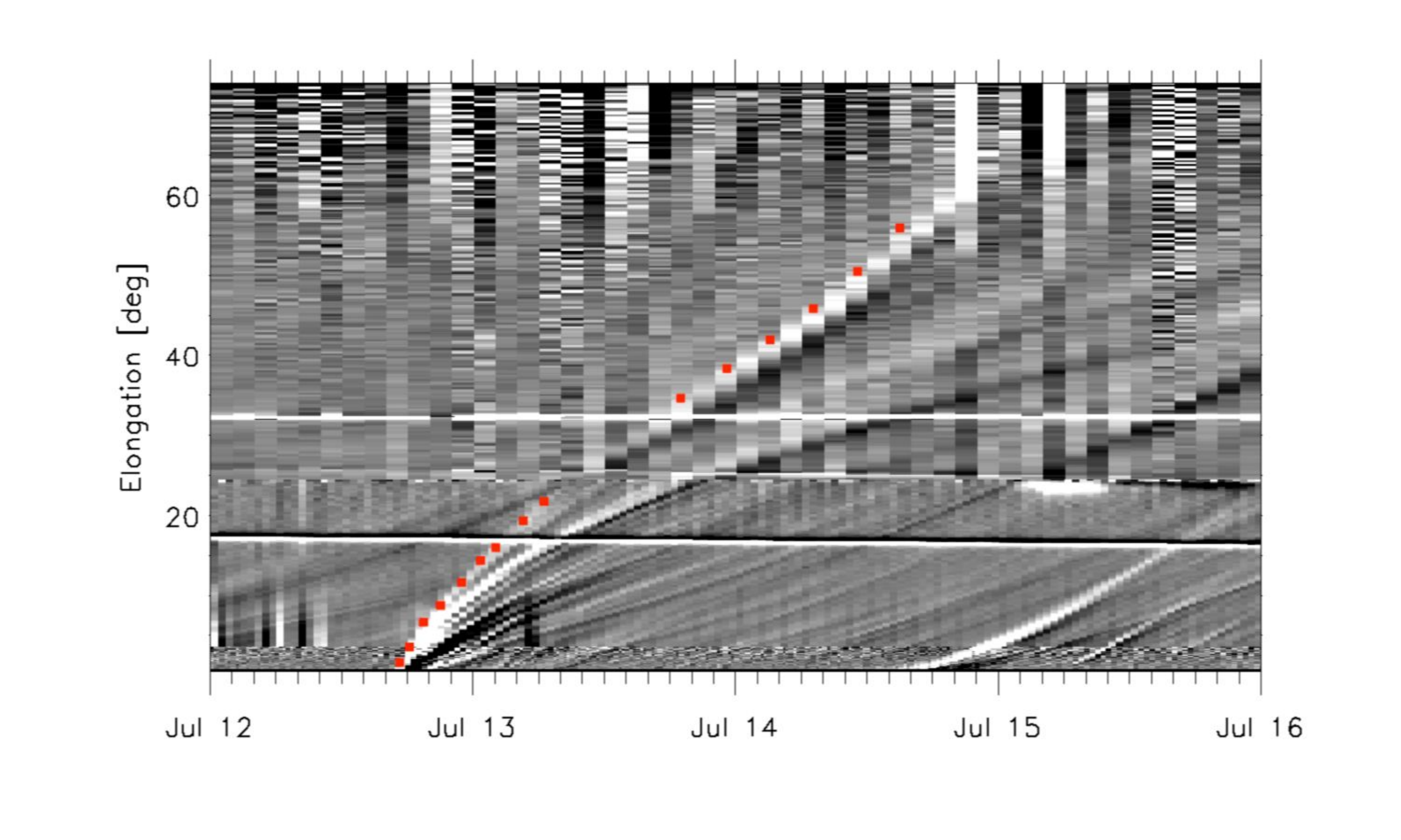} } 
\caption{Event 1: STEREO-B running-difference J-maps at PA=$267^\circ$.
The red dots mark the leading edge of the CME.}
\label{fig:20120712_jmaps} 
\end{figure}

\textit{STEREO-B time-elongation maps.}
To constrain the CME propagation in the heliosphere we extract the position over time of the CME leading edge from STEREO-B time-elongation maps \citep[J-maps;][]{sheeley:2009,lugaz:2009}, obtained by stacking SECCHI/COR2B-HI1B-HI2B images at a position angle (PA) of $267^\circ$, i.e. the ecliptic plane.
As shown in Figure~\ref{fig:20120712_jmaps}, the CME leading edge in the STEREO-B J-map is clearly visible from $2^\circ$ up to $56^\circ$ in elongation.

\medskip
\textit{SSE and iSSE techniques.}
% SSE technique
In order to construct time-height profile of the leading edge based on tracking the edge in the time-elongation map, we apply the Self-Similar Expansion (SSE) technique proposed by \citet{davies:2012}.
In the SSE technique any solar transient propagating away from the Sun is assumed to be characterised by a circular cross-section, with a radius that increases in a self-similar way as it propagates anti-sunward. 
In order to generate time-height profiles from time-elongation profiles of the CME apex, 
we apply the following relation \citep{davies:2012}:
\begin{equation}
h_\mathrm{SSE} = \frac{d_0 \cdot \sin \epsilon \cdot (1+ \sin{\omega/2} ) }{\sin(\epsilon + \phi) + \sin{\omega/2}},
\label{eqn:sse}
\end{equation}
where $d_0$ is the heliocentric distance of the STEREO spacecraft used in the J-maps analysis,
$\epsilon$ is the elongation of the CME leading edge recovered from J-maps,
$\omega/2$ is the CME half width, 
and $\phi$ is the and the angle between the observer, the Sun and the CME propagation direction.
In our analysis, we use the $\omega/2$ parameter derived from the GCS fitting. 
The angle $\phi$ between the observer and the CME propagation direction is also calculated based on the spacecraft location and the CME direction estimated from the GCS fitting, 
as $\phi = \arccos(\cos{\phi_\mathrm{HGRTN}} \cos{\theta_\mathrm{HGRTN}})$, 
where ($\phi_\mathrm{HGRTN},\theta_\mathrm{HGRTN}$) are the CME longitude and latitude in the reference system centered at the Sun and for which the $\theta_\mathrm{HGRTN}=0^\circ$ points towards the observing spacecraft.

The SSE method as originally proposed by \citet{davies:2012} 
was formulated to describe the propagation of the CME apex only.
Therefore, strictly speaking this method can only be used to estimate the arrival time of CMEs at Earth in the case of central encounters. 
As in general this is not the case, we also apply the most recent version proposed by \citet{moestl:2013}, who extended the model to account for the geometrical correction in the case of a spacecraft crossing a CME off axis.
Hereafter, we refer to this approach as the in-situ SSE (iSSE) method.
The heliocentric distance of the CME portion that is propagating along the Sun-Earth line is then described by the following equation \citep{moestl:2013}:
\begin{equation}
h_\mathrm{iSSE} = \frac{\cos \Delta + \sqrt{\sin^2{\omega/2} - \sin^2{\Delta}}}{1 + \sin \omega/2 } \cdot h_\mathrm{SSE},
\label{eqn:isse}
\end{equation}
where $\Delta$ is the angle between the CME propagation direction, the Sun and the spacecraft where one wants to predict the impact (in our case, the Earth), 
and $h_\mathrm{SSE}$ is the height of the CME apex as recovered from Equation~\ref{eqn:sse}.
The angle $\Delta $ is calculated based on the spacecraft location and the CME direction estimated from the GCS fitting, as $\Delta = \arccos(\cos{\phi_\mathrm{HEEQ}} \cos{\theta_\mathrm{HEEQ}})$ (in HEEQ coordinates).

In this case the CME was propagating very close to the Sun-Earth line ($\Delta = 9^\circ$), 
so the results from the SSE and iSSE techniques almost coincide - as visible from Figure~\ref{fig:20120712_propagation}.

\medskip
\textit{MESSENGER data.}
To further constrain the CME propagation in the inner heliosphere, in addition to remote-sensing tracking of the CME in the corona and heliosphere we make use of data from the ICME Catalog at Mercury 
\citep[][\url{http://c-swepa.sr.unh.edu/icmecatalogatmercury.html}]{winslow:2015}, based on data from the magnetometer on board the MESSENGER mission \citep[MAG;][]{anderson:2007}.
On the day of the eruption (12 July 2012) the MESSENGER spacecraft was orbiting around Mercury, which was located at a position of $\theta_M= -3^\circ$, $\phi_M = -35^\circ$ in HEEQ coordinates, and at a distance of 0.466~AU (=100~$R_s$) from the Sun (Figure~\ref{fig:20120712_ecliptic}). 
The spacecraft angular separation from Earth was $\Delta \theta \simeq 3^\circ$ in latitude and $\Delta \phi  \simeq 35^\circ$ in longitude.
From the relative position of the MESSENGER spacecraft and the CME direction of propagation, 
it is therefore reasonable to expect that the CME encountered MESSENGER with its eastern flank.
The ICME Catalog at Mercury reports that the ICME-driven shock arrived at MESSENGER on 13 July 2012 at 10:53~UT. 
The flux-rope signature started at 13:44~UT of the same day, and ended at 02:46~UT of the following day.

% in-situ observations
\subsubsection{ICME signatures at Earth}

\begin{figure}[t]
\centering
{
   \includegraphics[width=\hsize]{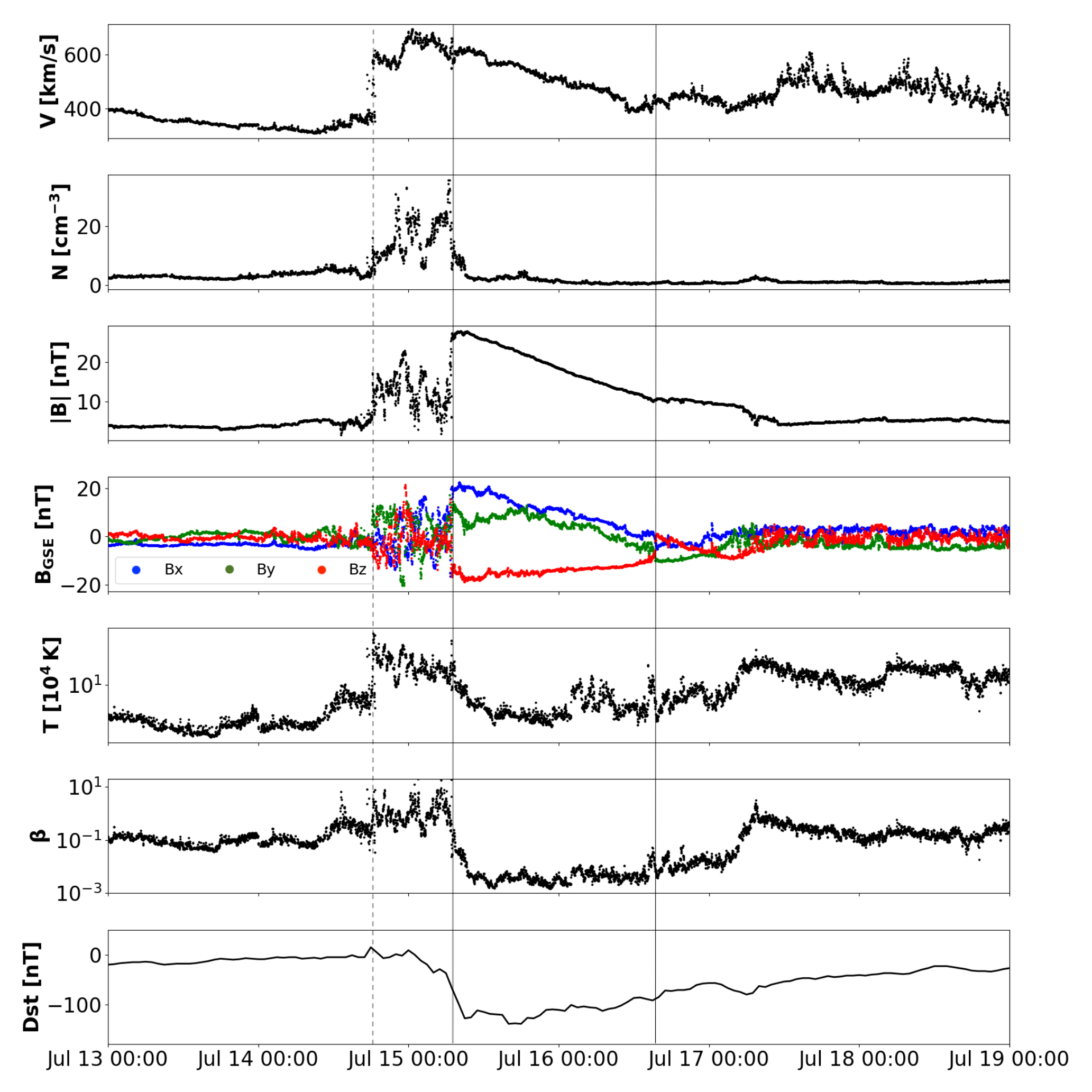}  } 
\caption{Event 1: in-situ observations for the 12 July 2012 event from OMNI 1-min data.
From  top  to  bottom: speed, number density, magnetic field strength, 
$B_x$, $B_y$, $B_z$ components in GSE coordinates, temperature, plasma $\beta$ and \textit{Dst} index.
The vertical dashed black line marks the interplanetary shock associated to the ICME, 
whilst the region delimited by the continuous black lines marks the MC period.}
\label{fig:20120712_omni} 
\end{figure}

 Figure~\ref{fig:20120712_omni} shows in-situ data from the OMNI database (\url{https://omniweb.gsfc.nasa.gov/ow_min.html}) on the days following the eruptions of CME event 1.
The Wind spacecraft \citep{ogilvie:1995} orbiting L1 detected the interplanetary shock associated with the ICME on 14 July 2012 at 17:39~UT
\citep[from the Heliospheric Shock Database, \url{www.ipshocks.fi};][]{kilpua:2015}.
The shock was followed by a turbulent sheath region of the duration of about 12 hours.
As reported by the Richardson and Cane ICME list, clear MC signatures can be identified starting from 06:00~UT on 15 July 2012 up to 05:00~UT on 17 July 2012. The MC duration is about 23 hours, and it is characterised by enhanced magnetic field, smooth rotation of the interplanetary magnetic field (IMF) components, low density and temperature, that resulted in a low plasma $\beta$.
The maximum magnetic field in the magnetic cloud is 27~nT, while the average $B$ is 16~nT. The observed minimum $B_z$ is $-18$~nT.
The MC also exhibits a decelerating plasma velocity profile indicating significant expansion, with a maximum speed of $694$~\si{ \km \,\, \s^{-1} } at the front
and a speed difference of $220$~\si{ \km \,\, \s^{-1} } between the front and the back.
The presence of the long-lasting southward $B_z$ region in the MC also triggered an intense geomagnetic storm, as indicated by the $Dst$ index reaching a minimum value of -139~nT on 15 July 2012.

From a visual inspection of the magnetic field signatures,
we observe that $B_y$ rotates from positive (east) to negative (west), while $B_z$ shows a prolonged long-lasting southward component, compatible with a ESW flux-rope type at Earth.
On the other hand, \citet{palmerio:2018} fitted the in-situ flux-rope using the Minimum Variance Analysis technique \citep[MVA;][]{sonnerup:1967}, and found an orientation of the ICME axis equal to $(\theta_\mathrm{MVA,1}, \phi_\mathrm{MVA,1})=(-4^\circ, 305^\circ)$.
This result suggests that the flux-rope structure at Earth was characterised by a very low inclination on the equatorial plane (as indicated by the low $\theta_\mathrm{MVA,1}$) 
and hence that the magnetic structure underwent a clockwise rotation of about $30^\circ - 40^\circ$ as it propagated from the Sun to 1 AU. 
Based on the same reconstruction technique, the HELCATS ICME catalog ({\tt ICMECAT}; \url{https://www.helcats-fp7.eu/catalogues/wp4_icmecat.html}) reports an orientation of the ICME axis equal to $(\theta_\mathrm{MVA,2}, \phi_\mathrm{MVA,2})=(-22^\circ, 315^\circ)$.
The value of $\theta_\mathrm{MVA,2}$ suggests that the flux-rope structure arrived at Earth with an inclination similar to the one of the source region PIL at the Sun.
Despite the slight difference between $\theta_\mathrm{MVA,1}$ and $\theta_\mathrm{MVA,2}$ ($ \simeq 20^\circ$, reflecting the uncertainties affecting the determination of the 3D flux-rope geometry from single-spacecraft in-situ observations), such results are both consistent with a NES flux-rope at Earth.
\citet{palmerio:2018} also considered the location angle $L$, defined by \citet{janvier:2013} as 
\begin{equation}
\sin L = \cos \theta_\mathrm{MVA} \cos \phi_\mathrm{MVA},
\label{eqn:location_angle}
\end{equation}
giving an indication of distance of the spacecraft crossing from the ICME flux-rope nose. In this case, they found $L = 35^\circ $, suggesting that the flux-rope impacted on Earth in between its nose and its leg.
Using the MVA results listed in the HELCATS catalog, we obtain a similar result of $L = 40^\circ $.

%===========================================
\subsection{Event 2: CME on 14 June 2012}\label{subsec:20120614}
%===========================================

The second event considered in this work is the halo CME that erupted on 14 June 2012, first discussed in detail by \cite{palmerio:2017}.
As discussed recently by \citet{srivastava:2018}, this event was composed by a sequence of two CMEs that were launched from NOAA AR 11504.
% CMEs
The first CME (CME1) erupted on 13 June 2012 and it was observed by LASCO as a partial halo, entering the C2 field of view at 13:25~UT and propagating with an average projected speed of 632~\si{ \km \,\, \s^{-1} }.
On the following day, a second CME (CME2) entered the C2 coronagraph at 14:12~UT, appearing as a fast halo CME propagating towards the Earth with an average projected speed of 987~\si{ \km \,\, \s^{-1} }.
% active region
On the day of the first eruption (13 June 2012), the active region was located at coordinates S17E28 on the solar disk and was classified as a $\beta$ region.
On 14 June 2012 the AR rotated to S17E14 and showed an increased level of magnetic complexity, being classified as $\beta \gamma \delta$.
% flaring
EUV images of the source region show that the eruption of CME1 took place around 11:30~UT on 13 June 2012, as indicated by an M1.2 flare (onset: 11:29 - peak: 13:17 - end: 14:31) detected by the GOES satellites. 
The second, moderate GOES M1.9 flare was detected at 12:52~UT on 14 June 2012 (onset: 12:52 - peak: 14:35 - end: 15:56) in association with the eruption of CME2. 

% source region and coronal observations
\subsubsection{Source region and coronal observations}

\begin{figure*}[t]
\centering
\subfloat[]
{\includegraphics[width=.25\hsize]{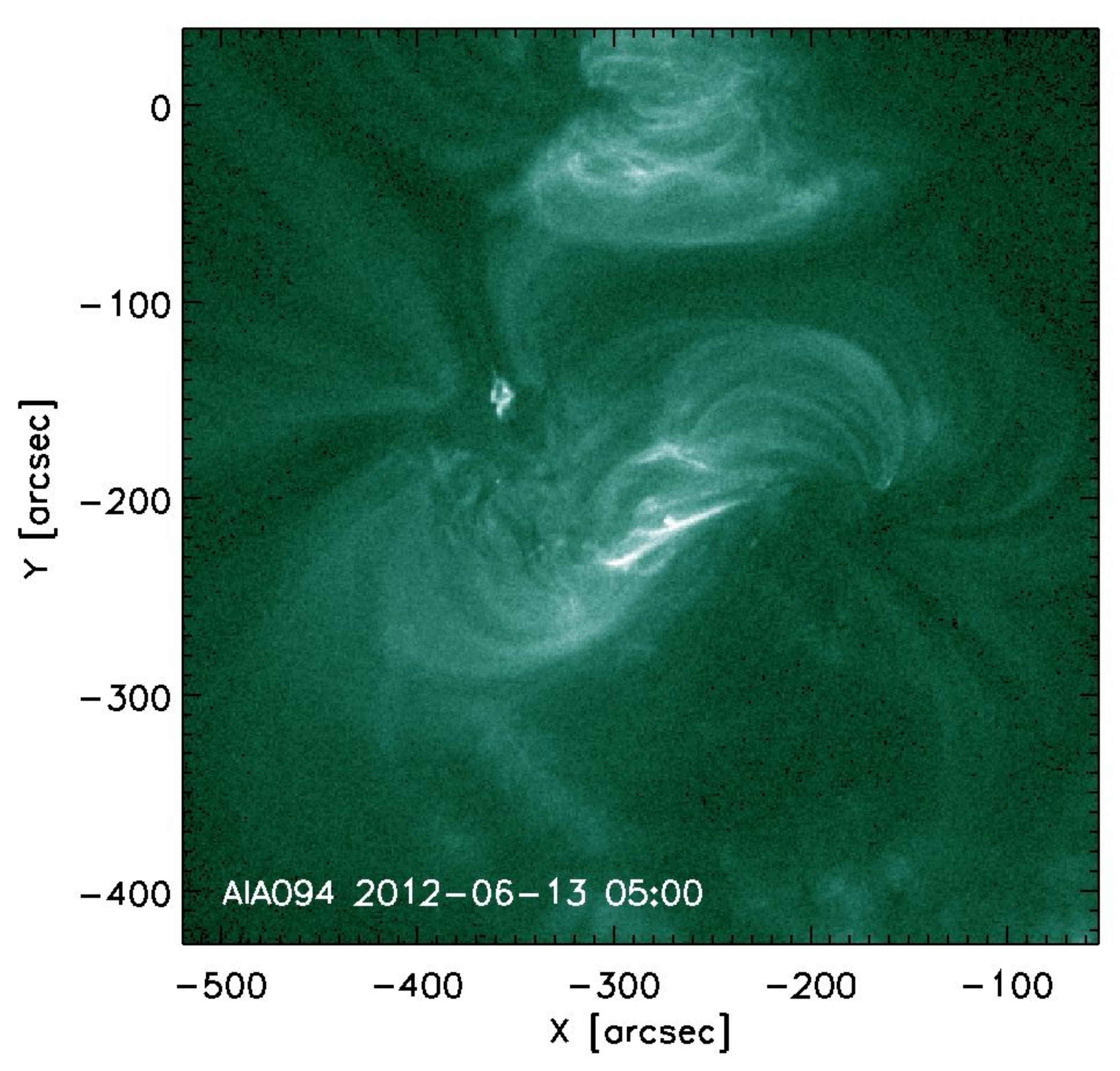} \quad
 \includegraphics[width=.25\hsize]{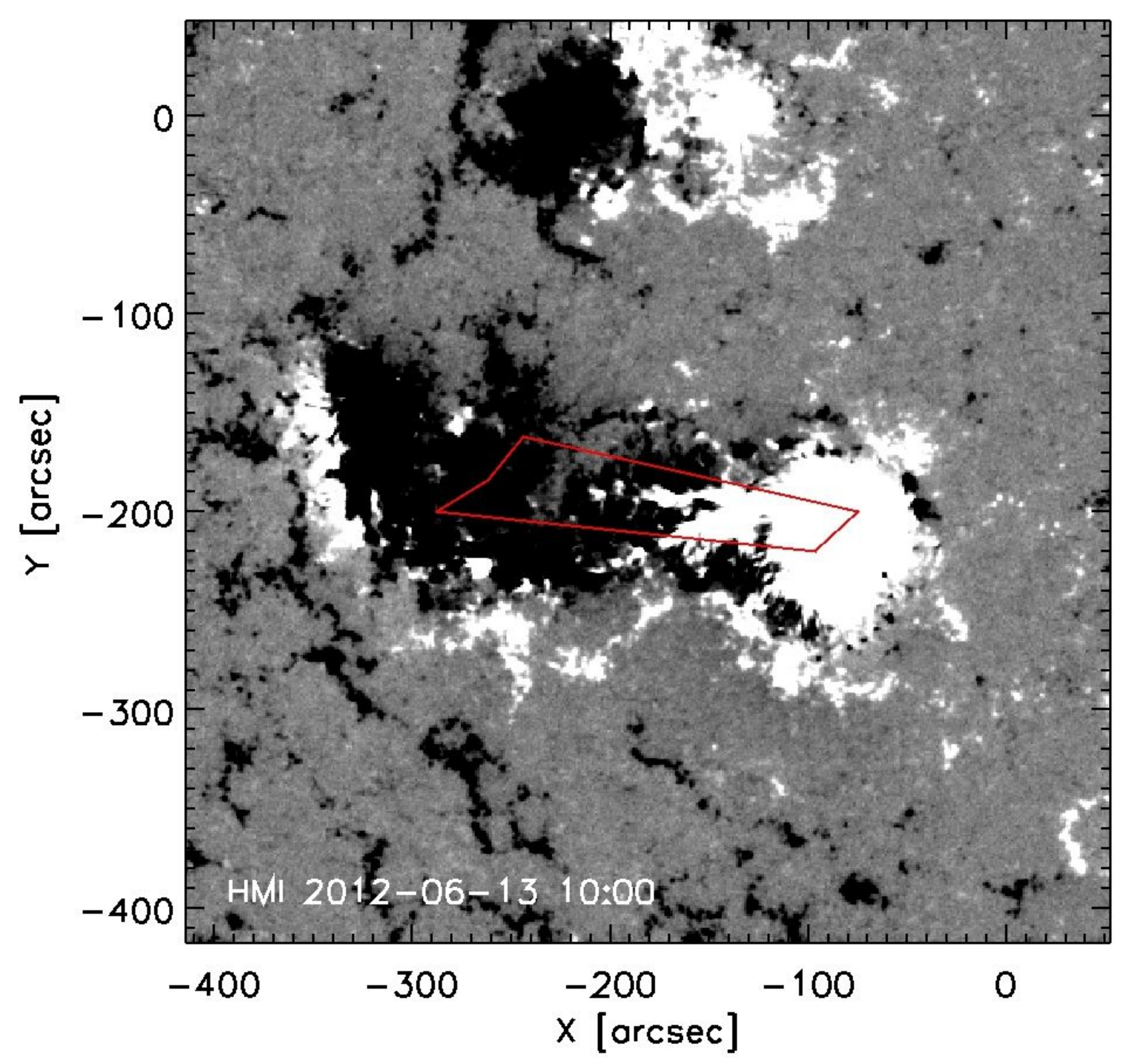} \quad
 \includegraphics[width=.25\hsize]{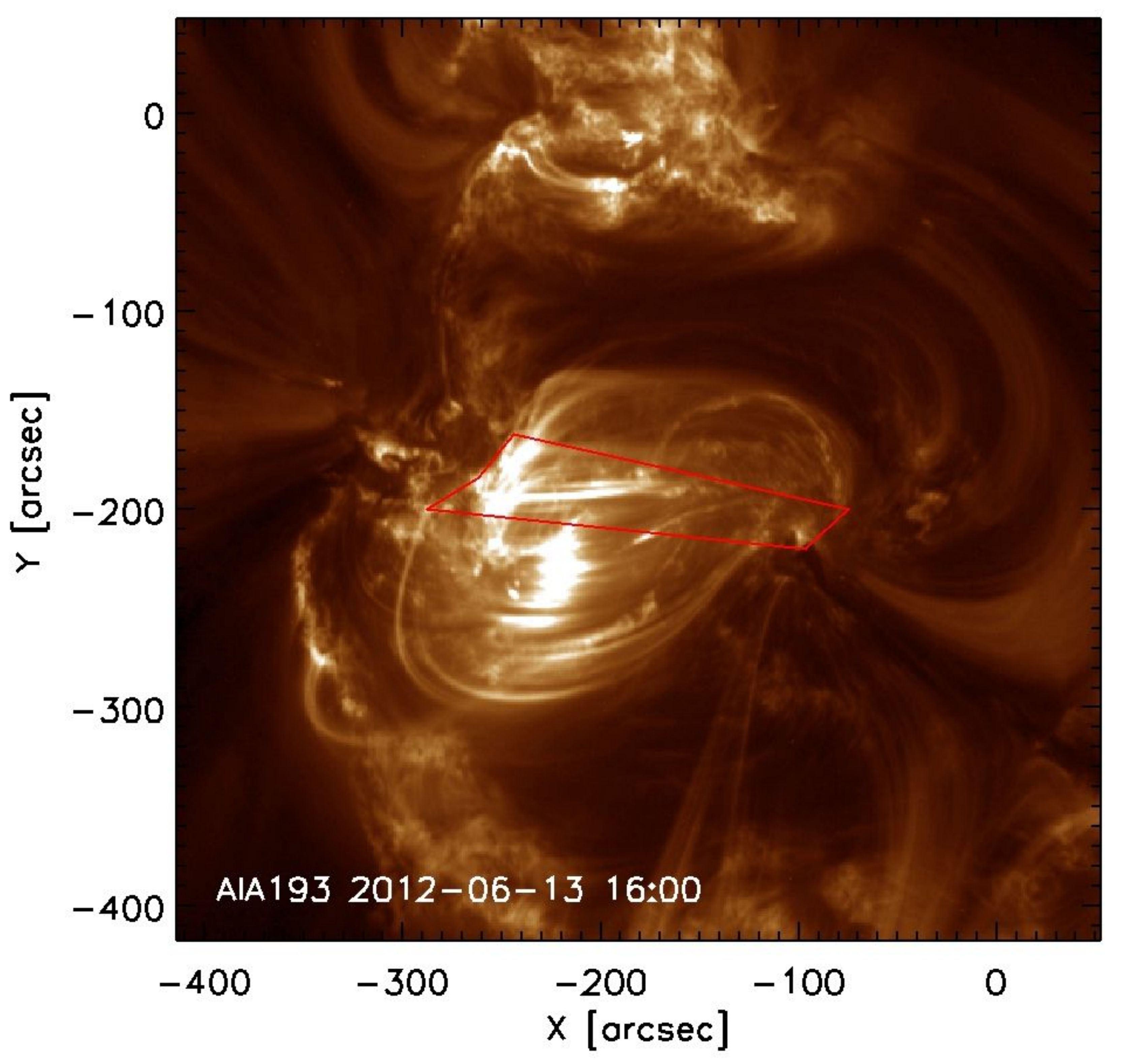}} \\
\subfloat[]
{\includegraphics[width=.25\hsize]{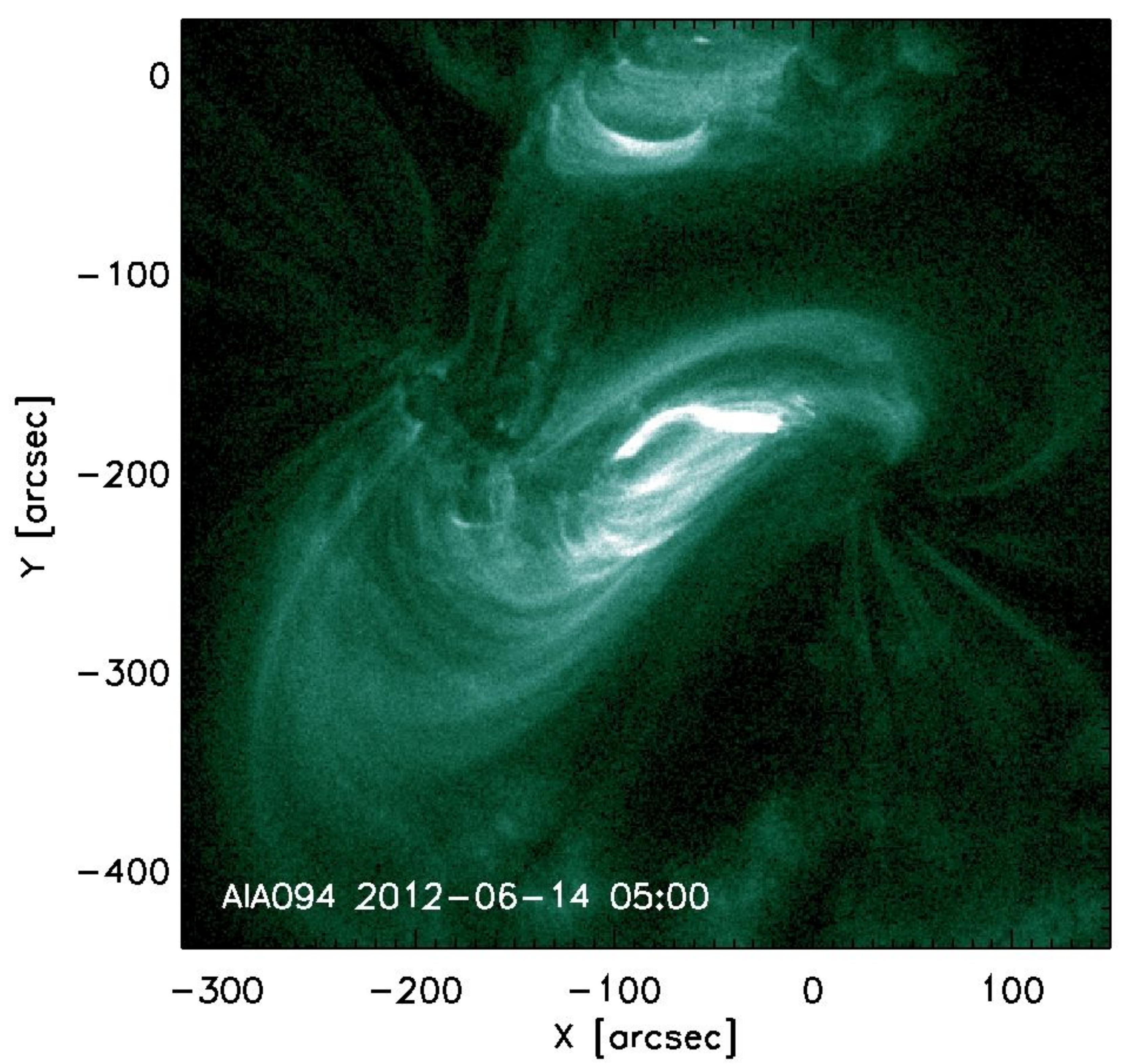} \quad
 \includegraphics[width=.25\hsize]{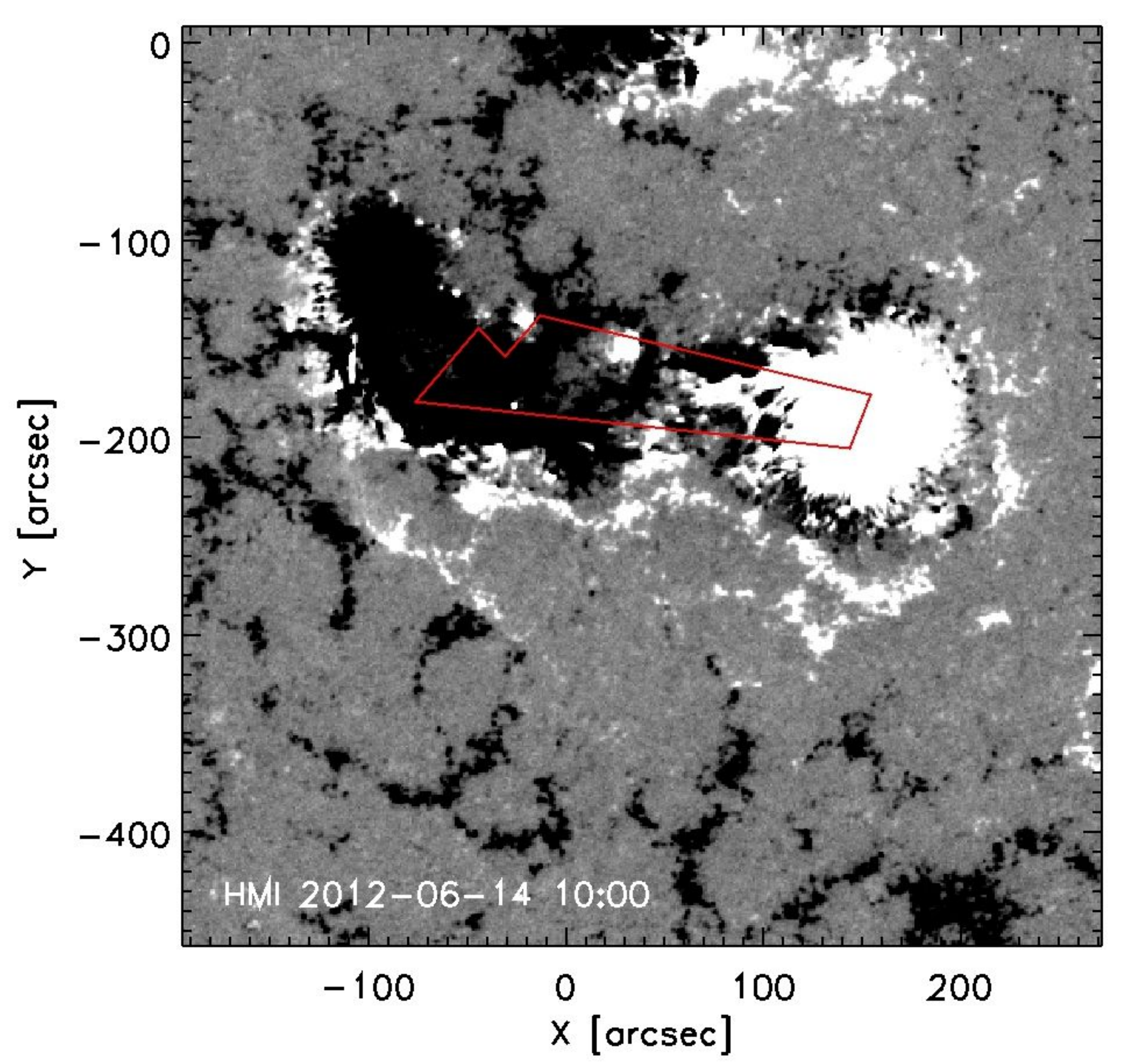} \quad
   \includegraphics[width=.25\hsize]{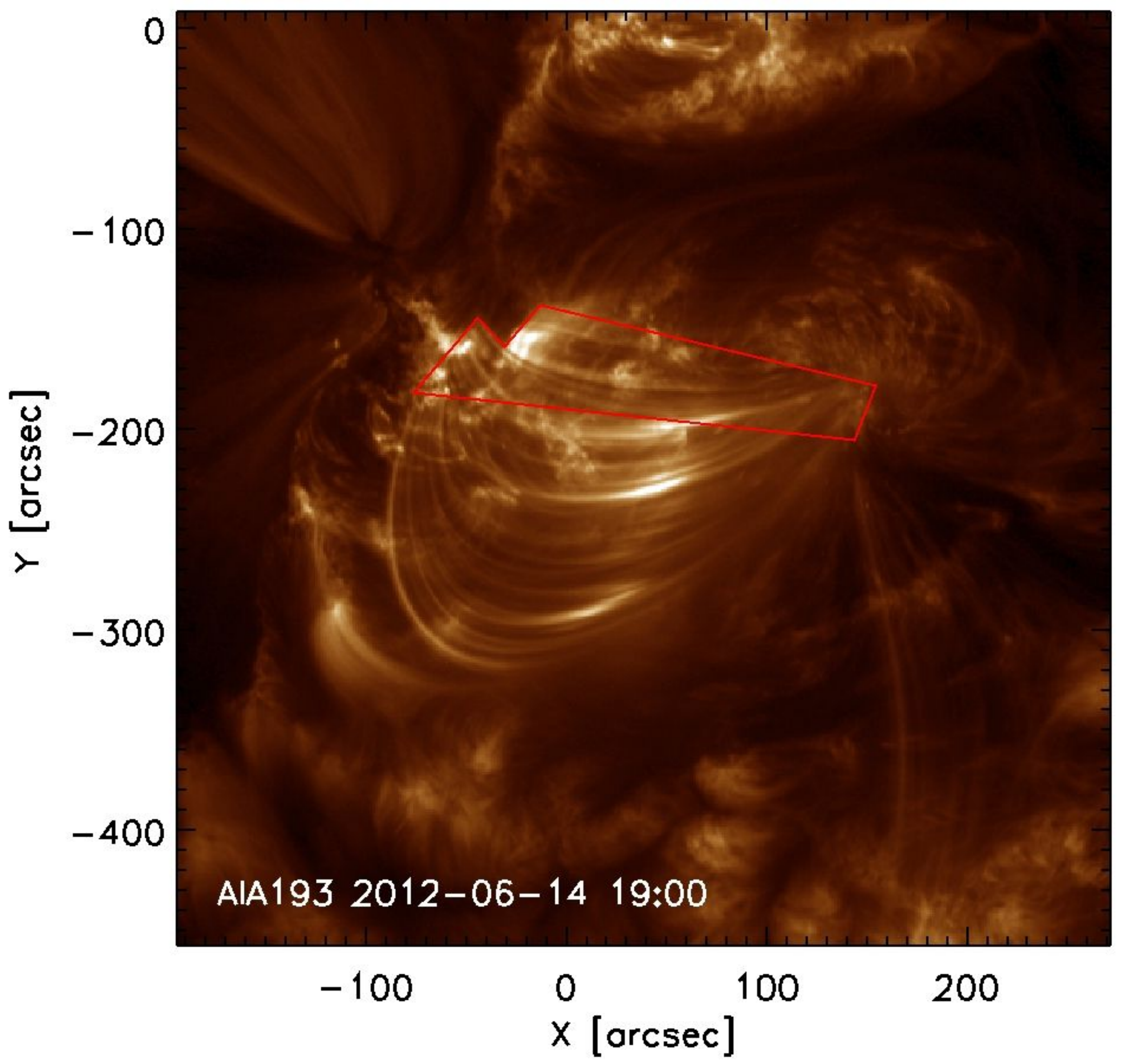}} 
\caption{Event 2: AR 11504 on 13 June 2012 (a) and 14 June 2012 (b).
		Left: AIA 94~\r{A} image of the pre-eruptive sigmoids.
		Center: HMI magnetogram with PEA areas overlaid (saturated at -100~gauss and +100~gauss).
		Right: PEAs from AIA 193~\r{A} with the area outlined by a polygon.
		The dates and times are shown as YYYY-MM-DD hh:mm in all panels.} 
\label{fig:20120614_source} 
\end{figure*}

\medskip
% source region ============================
\textit{Source region observations.} 
 Figure~\ref{fig:20120614_source} shows AR 11504 as observed on 13 and 14 June 2012 by HMI, and by AIA in different EUV channels.
The source region in AIA 94 \r{A} shows the presence of a forward-S sigmoid, indicating a positive  chirality. From a visual inspection of the HMI image, the PIL appears inclined by about $30^\circ$ with respect to the solar equator. Assuming a positive chirality and having a positive polarity west of the PIL, we conclude that flux-ropes formed in the region are expected to be a low-inclination flux-rope of NES-type, with an axial field pointing towards the south-east.
These results are consistent with those found by \cite{palmerio:2017}.
\citet{kay:2017} performed a statistical analysis of the rotations and deflections
in the solar corona and interplanetary space of 45 CMEs between 2007 and 2014, including the CME2
here considered. The result of the analysis with the ForeCAT and FIDO models for this event indicates that the flux-rope axis rotated by $<8^\circ$ between the low corona and 1~AU. Therefore, also in this case the orientation of the flux-rope structure at 0.1~AU can be expected to be consistent with that at the source region, here assumed to coincide with the orientation of the PIL.

After the eruption of CME1, a first PEA (PEA1) was observed 
and reached its maximum extension around 16:00~UT of 13 July 2012. 
After the eruption of CME2, a second PEA (PEA2) was observed to develop, 
peaking around 20:00~UT on 14 July 2012.
The two PEAs are shown in Figure~\ref{fig:20120614_source}. 
Both PEAs were characterised by very dynamic structures that made the identification of their extent over time very difficult.
For this reason, we calculate the area of PEA1 from AIA 193 \r{A} images at 16:00~UT on 13 June 2012 only. 
The estimated area results to be $ A_{PEA1} \simeq 1.0 \cdot 10^{15}$ m$^2$.
Overplotting its area with the HMI pre-eruptive magnetogram, we estimate the reconnected magnetic flux in the PEA region to be $\phi_{RC1} \simeq 2.2 \cdot 10^{13}$~Wb. 
Applying the same method to AIA 193 \r{A} images of PEA2 between 17:00~UT and 21:00~UT of 14 June 2012, we estimate its area to be $A_{PEA2}=1.0-2.1 \cdot 10^{15}$ m$^2$, and the reconnected magnetic flux in the PEA region to be in the range $\phi_{RC2} = 2.1-4.0 \cdot 10^{13}$~Wb. 

% RIBBONDB
The {\tt RIBBONDB} catalog reports a ribbon area equal to $A_{R1}=3.8 \pm 1.6 \cdot 10^{14}$~m$^2$ in association to the 13 June 2012 event.
The estimated reconnected magnetic flux in the ribbon region is 
$\phi_{RC,R1} = 1.1 \pm 0.45 \cdot 10^{13}$~Wb.
For the ribbon developing in AR 11504 in association to the flare class M1.9 on 14 June 2012, 
they found a ribbon area corresponding to $A_{R2} = 4.5 \pm 1.8 \cdot 10^{14}$~m$^2$.
The estimated the reconnected magnetic flux in the ribbon region is 
$\phi_{RC,R2} = 1.9 \pm 0.6 \cdot 10^{13}$~Wb.
The high uncertainties reported in the case of these two events reflect the more complex evolution of AR 11504 after the two eruptive flares associated to CME1 and CME2.
Similarly to Event 1, in this case we find PEA areas about a factor 2 larger than the ribbon areas, leading to $\phi_{RC,R}$ calculated from ribbon observations that are about a factor 2 smaller than $\phi_{RC}$ obtained from PEA observations.

\begin{figure}[h]
\centering
{\includegraphics[width=0.8\hsize]{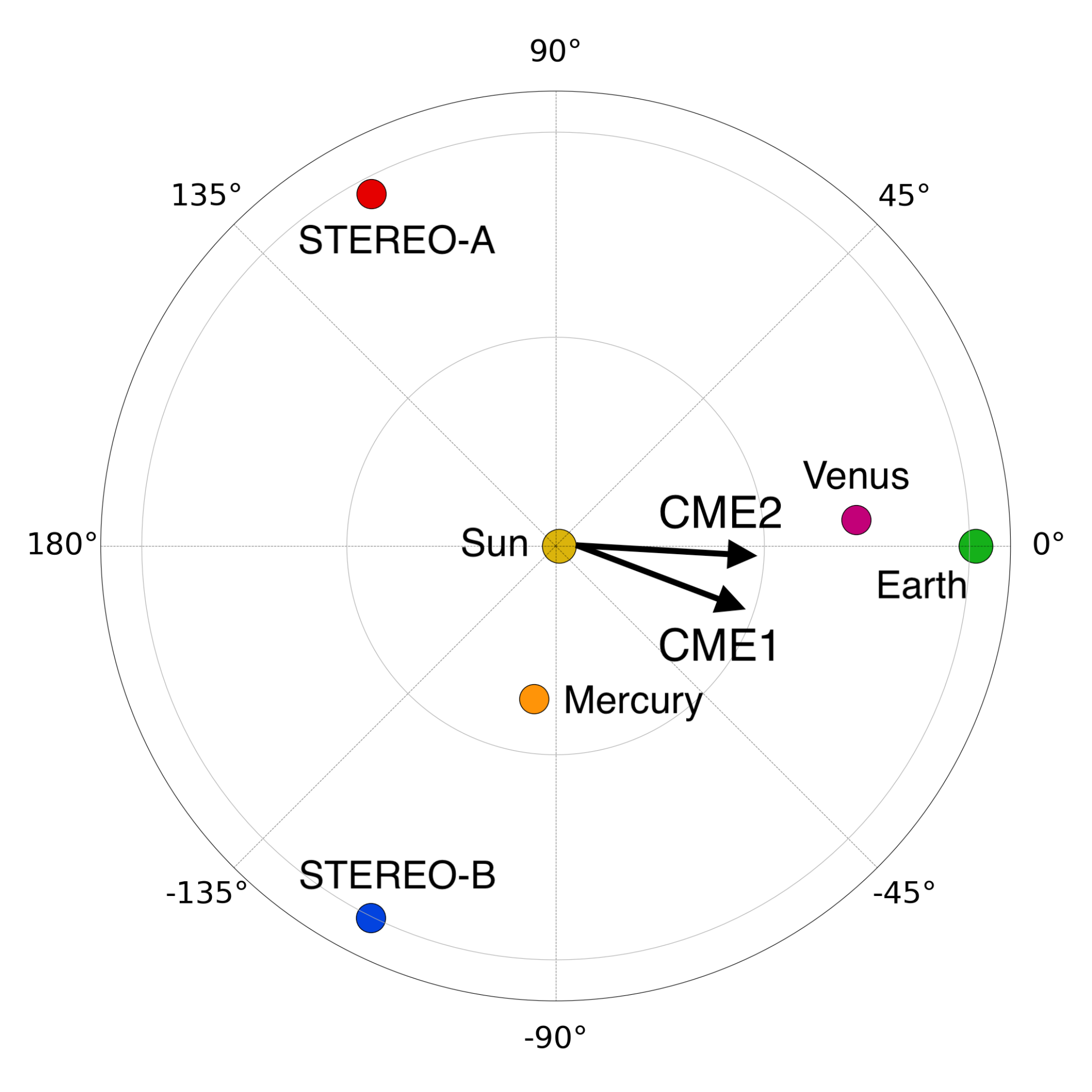} } 
\caption{Event 2: position of Earth, STEREO-A, STEREO-B, Mercury and Venus on the ecliptic plane on 14 June 2012 00:00~UT. The black arrows show the reconstructed longitude of the CME from the GCS fitting. Angles are in HEEQ coordinates. }
\label{fig:20120614_ecliptic} 
\end{figure}

\begin{figure*}[t]
\centering
\subfloat
{\includegraphics[width=.25\textwidth]{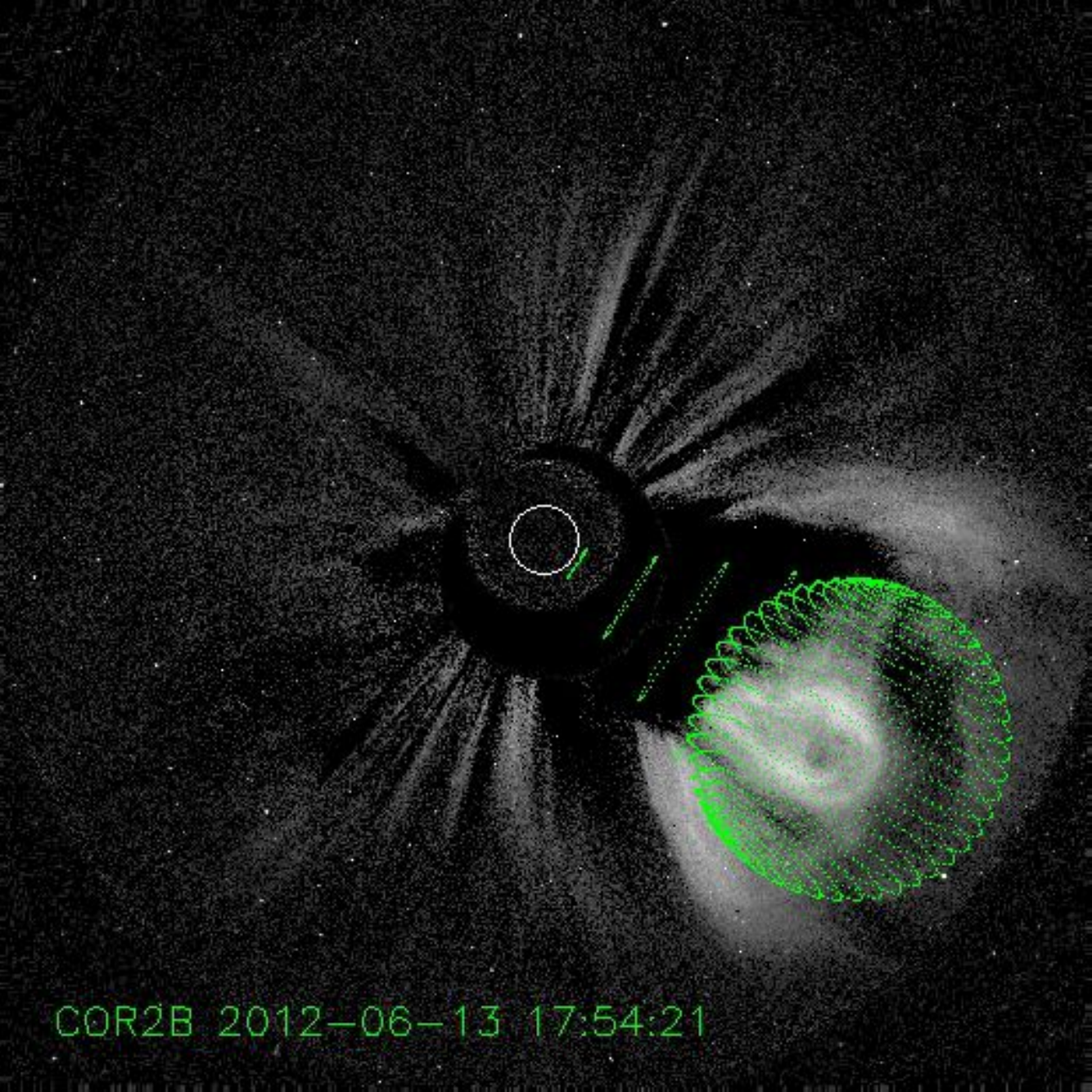} 
\includegraphics[width=.25\textwidth]{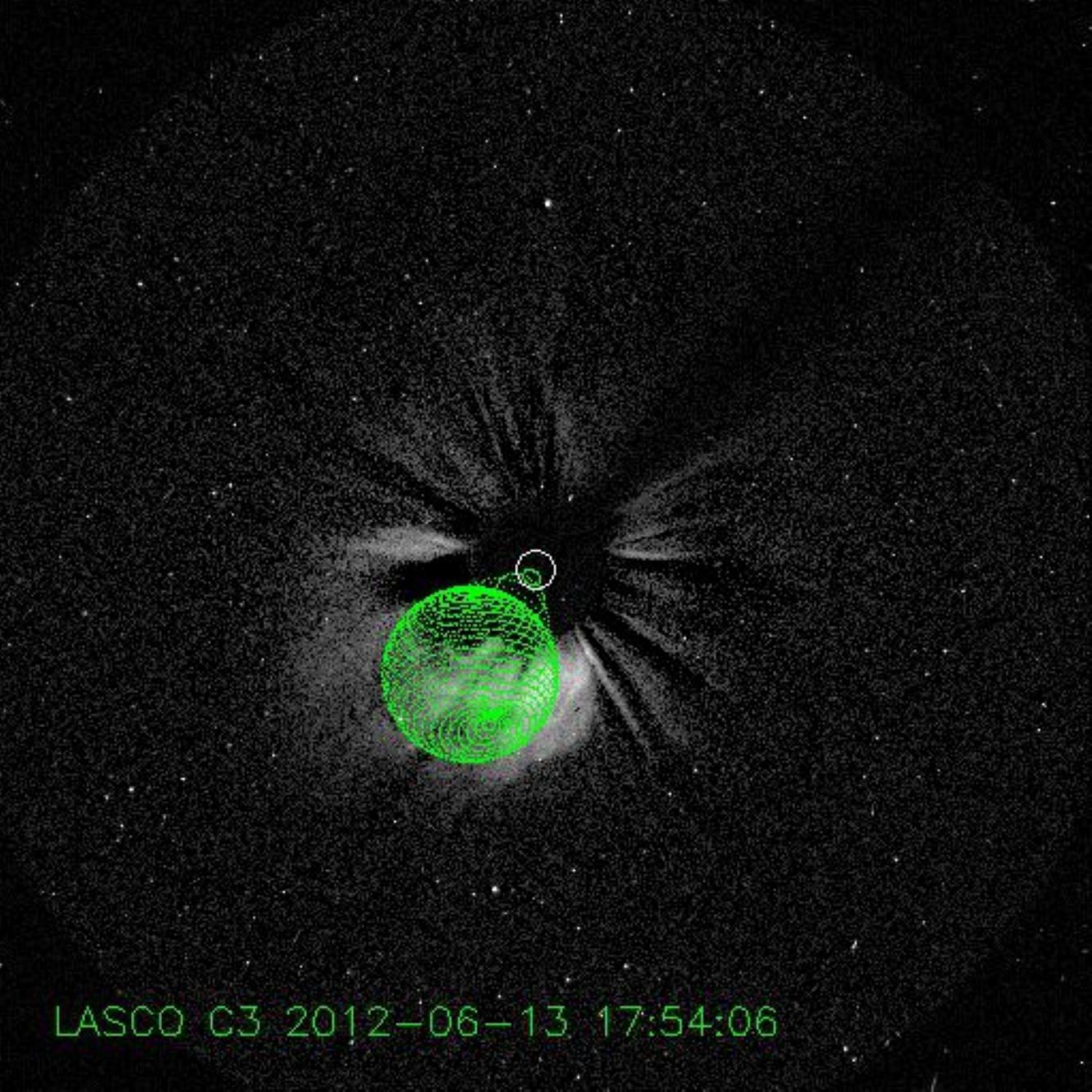} 
\includegraphics[width=.25\textwidth]{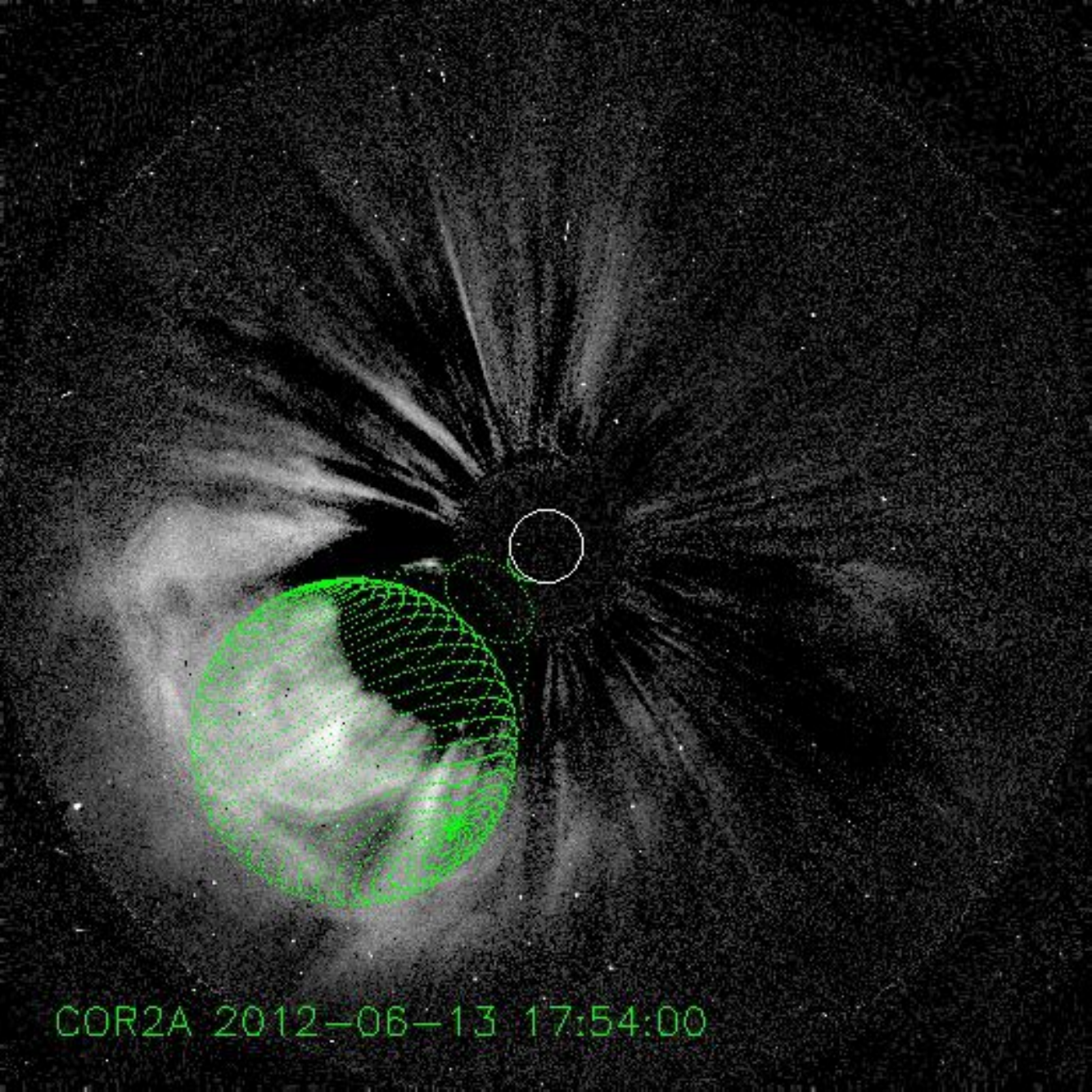}} \\
\setcounter{subfigure}{0}%
\subfloat[]
{\includegraphics[width=.25\textwidth]{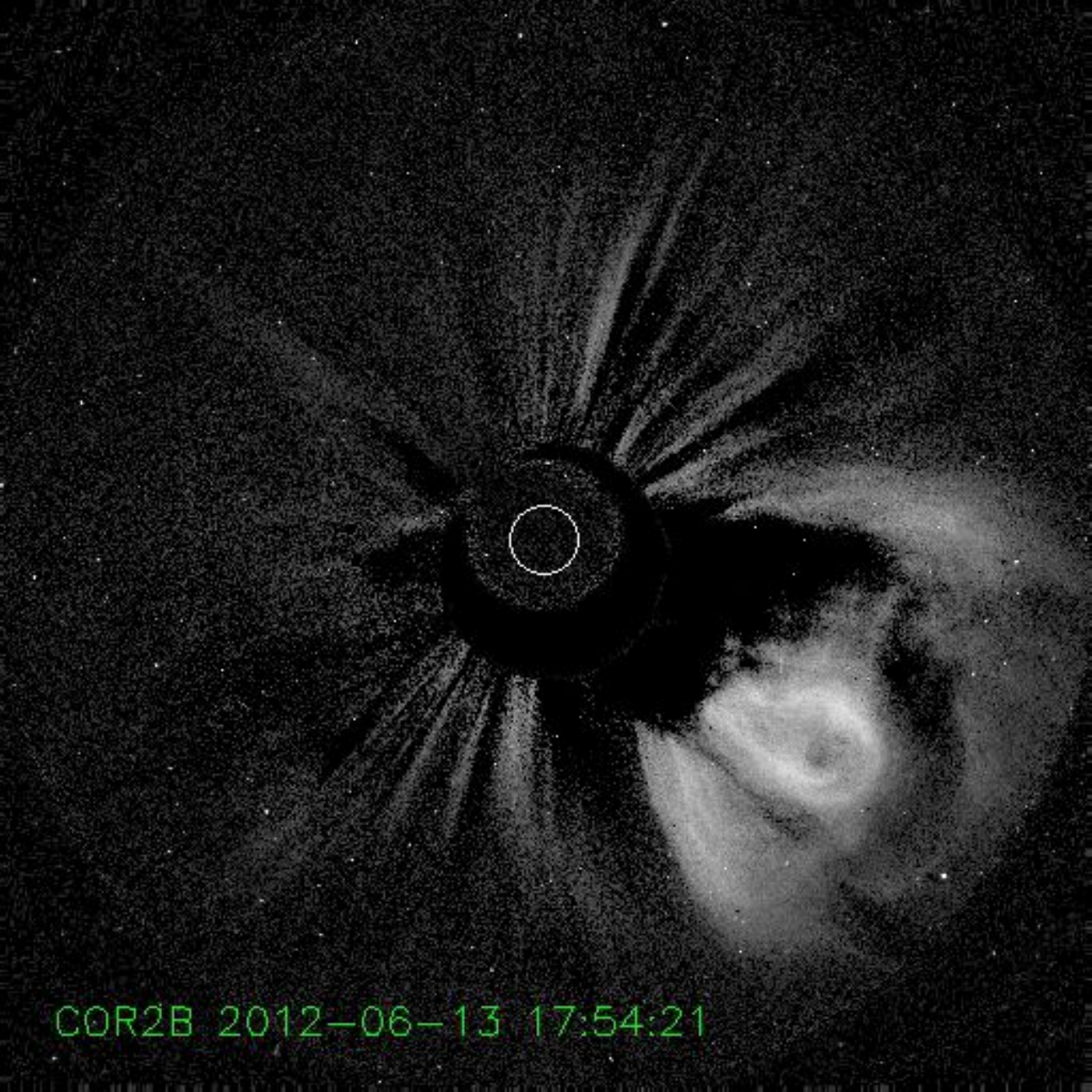} 
\includegraphics[width=.25\textwidth]{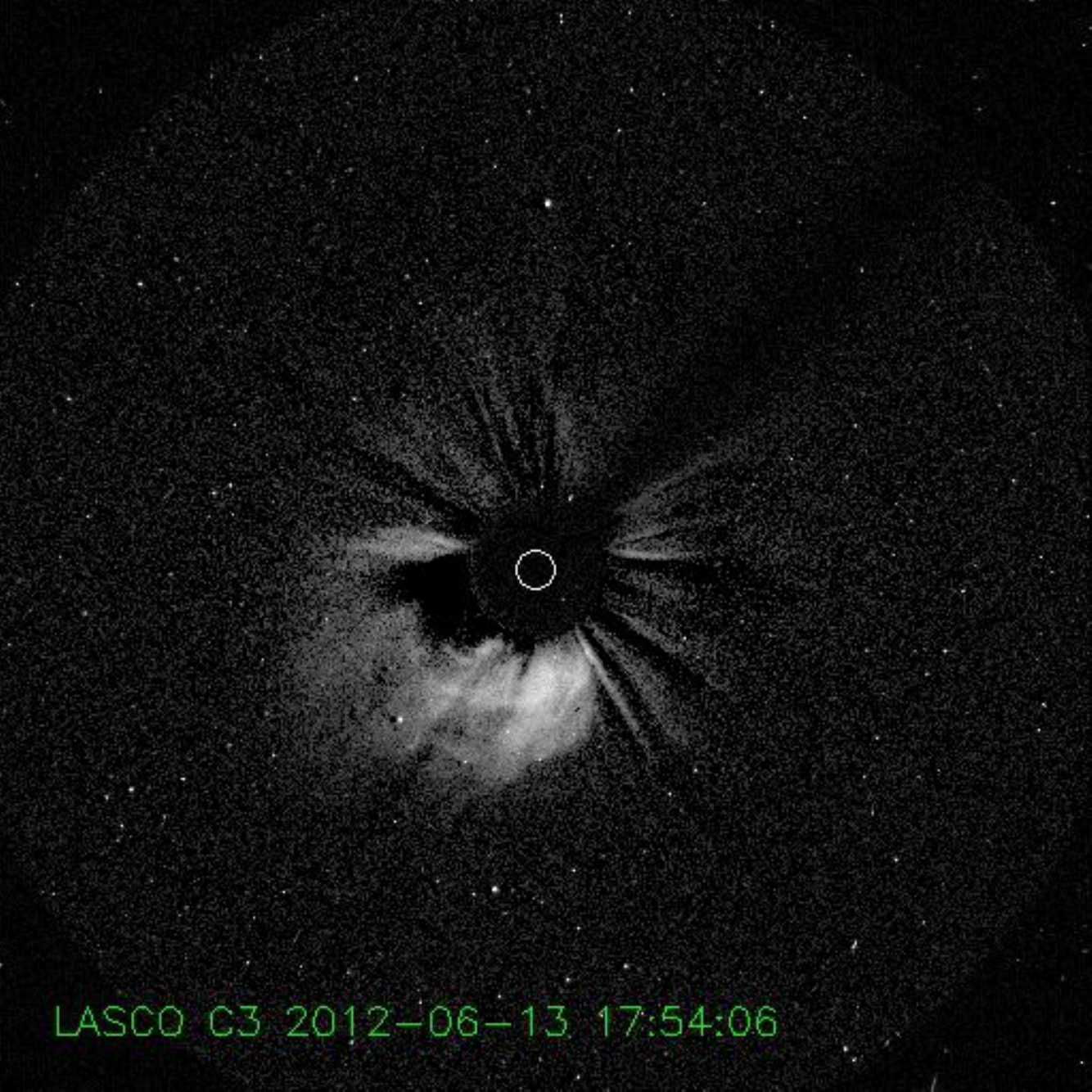} 
\includegraphics[width=.25\textwidth]{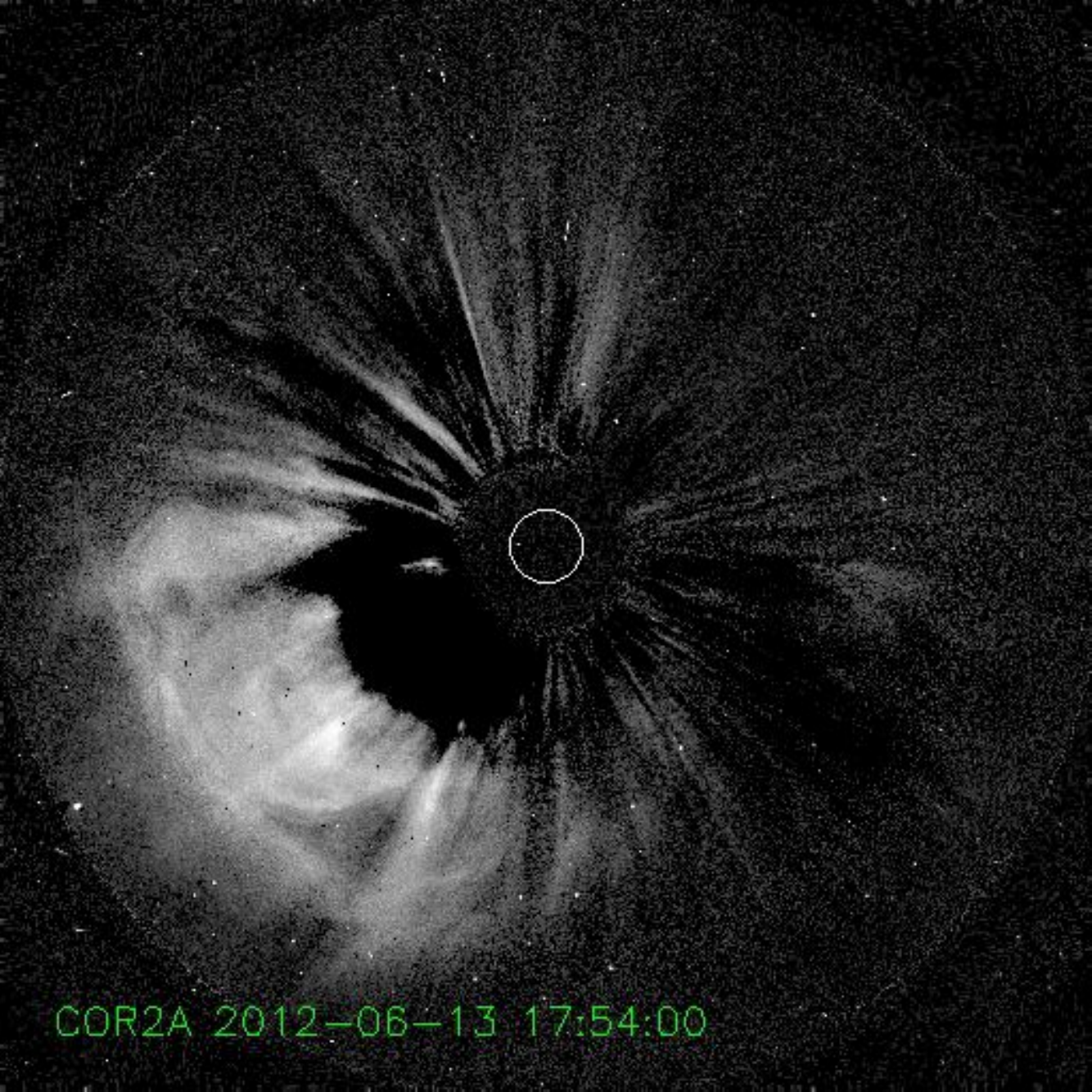}} \\
\subfloat
{\includegraphics[width=.25\textwidth]{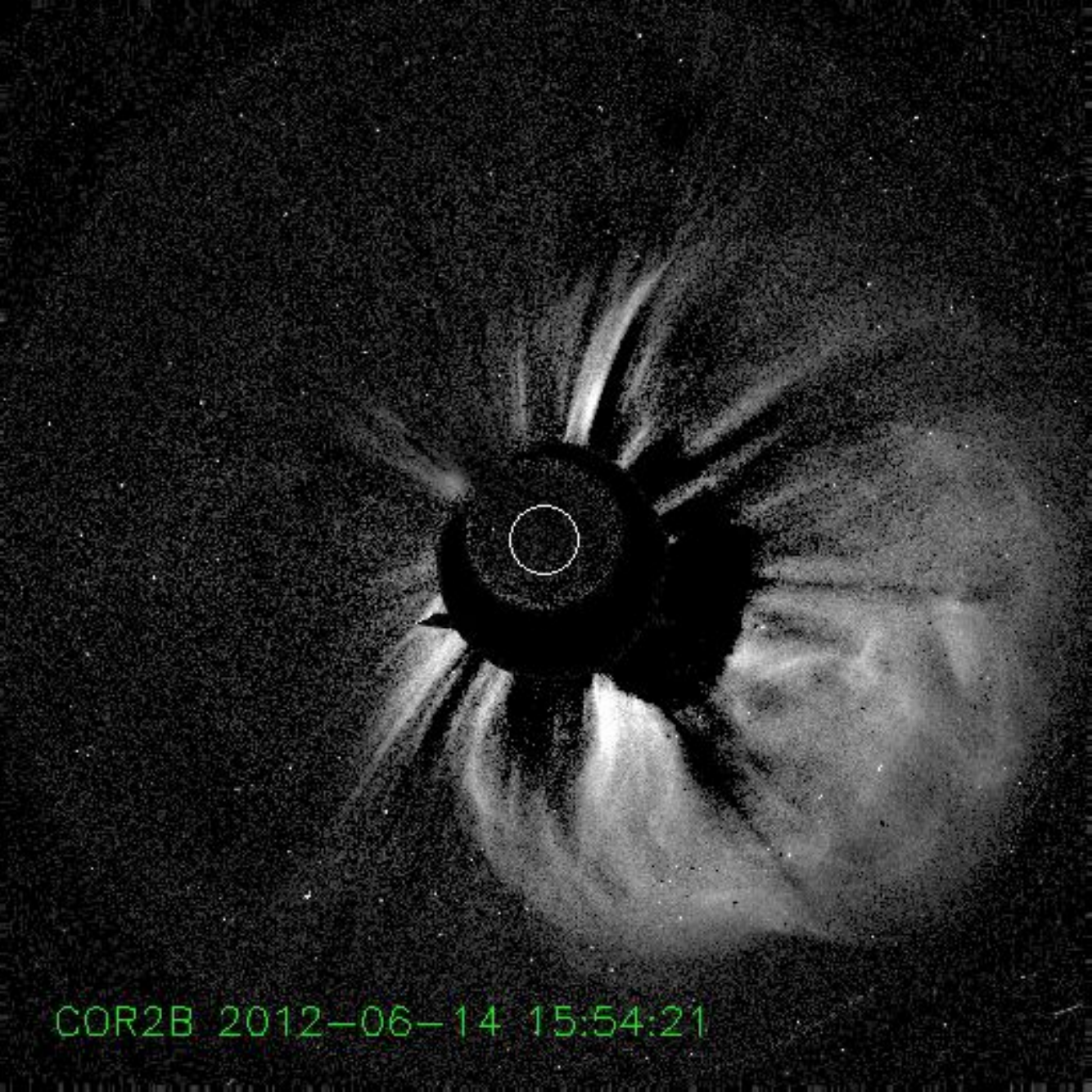} 
\includegraphics[width=.25\textwidth]{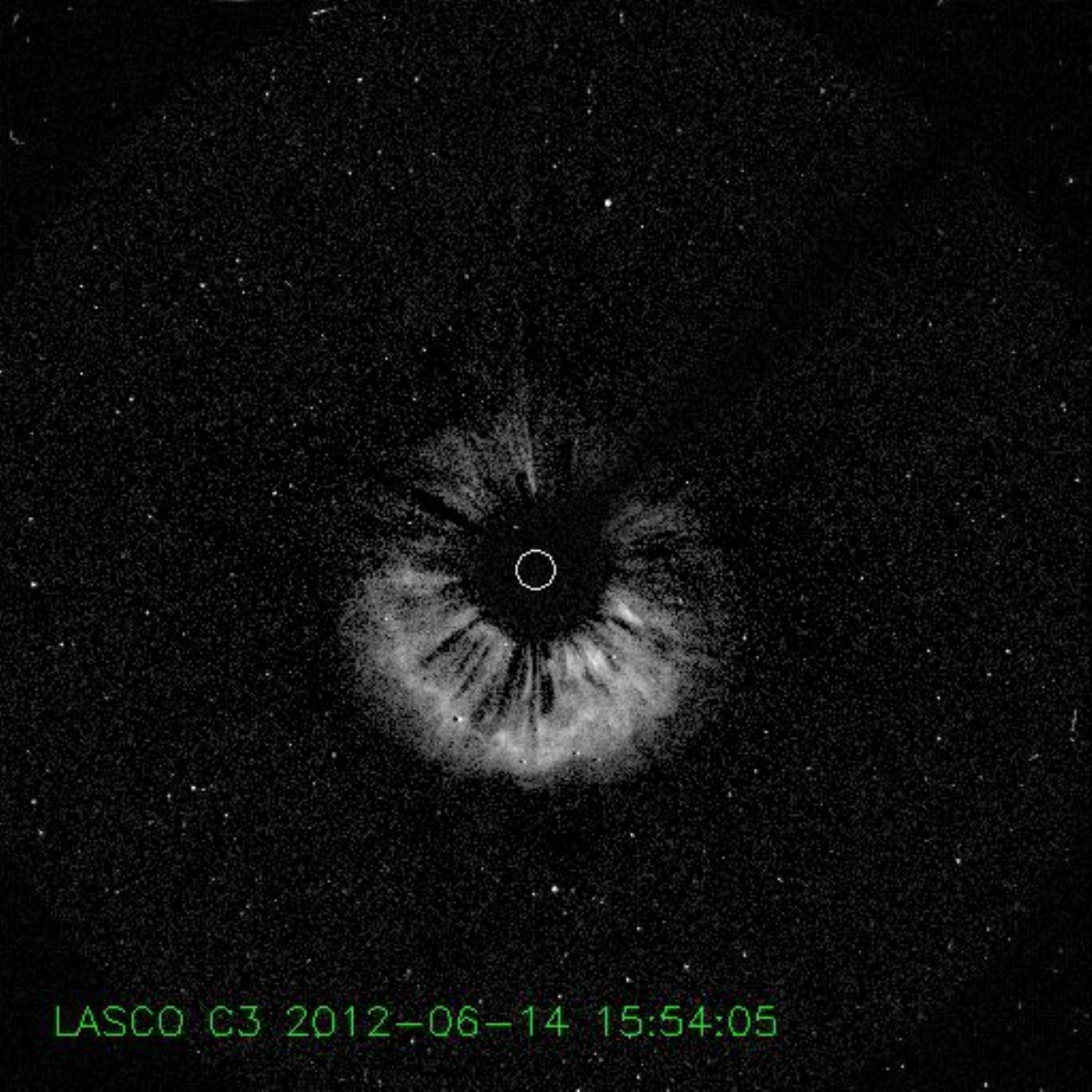} 
\includegraphics[width=.25\textwidth]{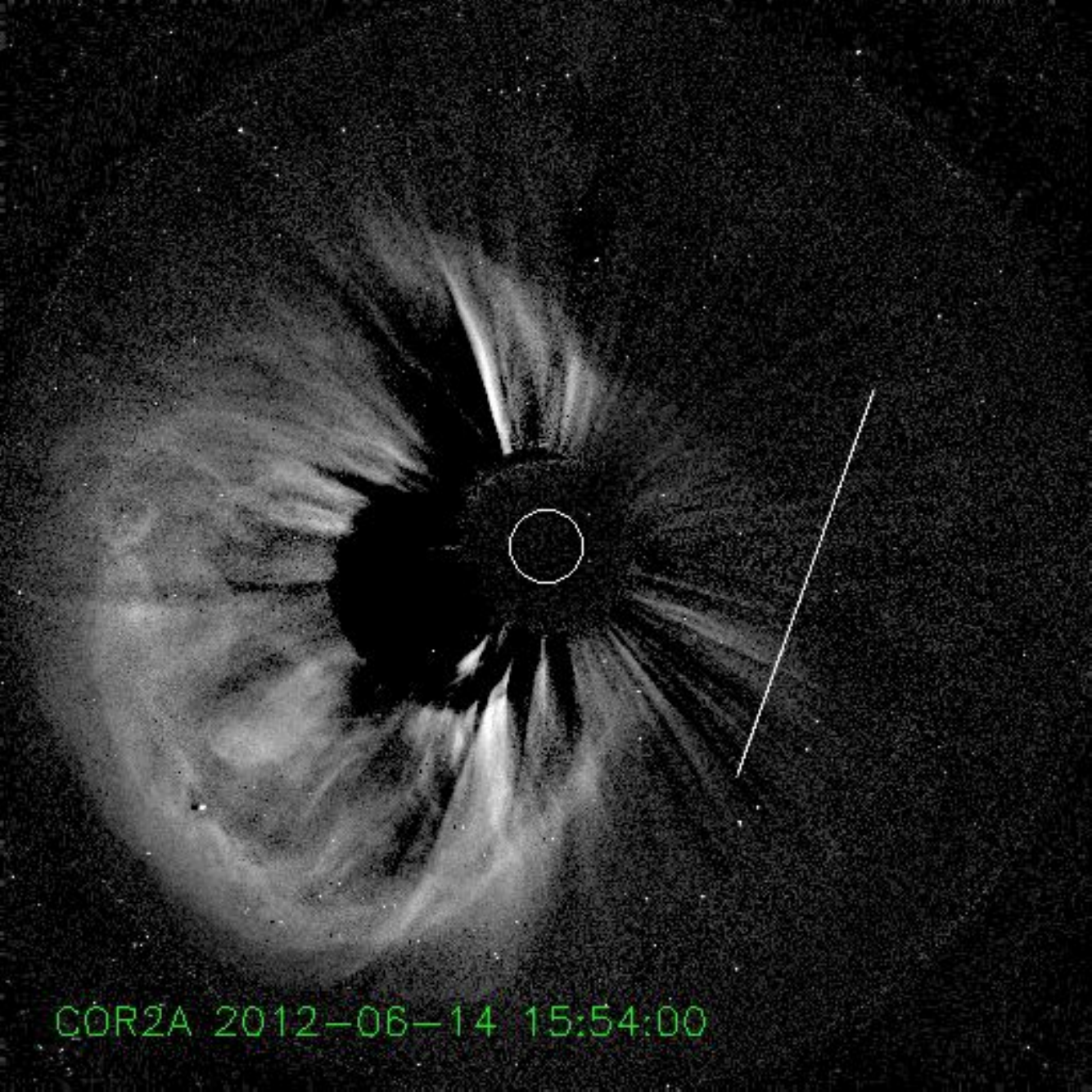} } \\
\setcounter{subfigure}{1}%
\subfloat[]
{\includegraphics[width=.25\textwidth]{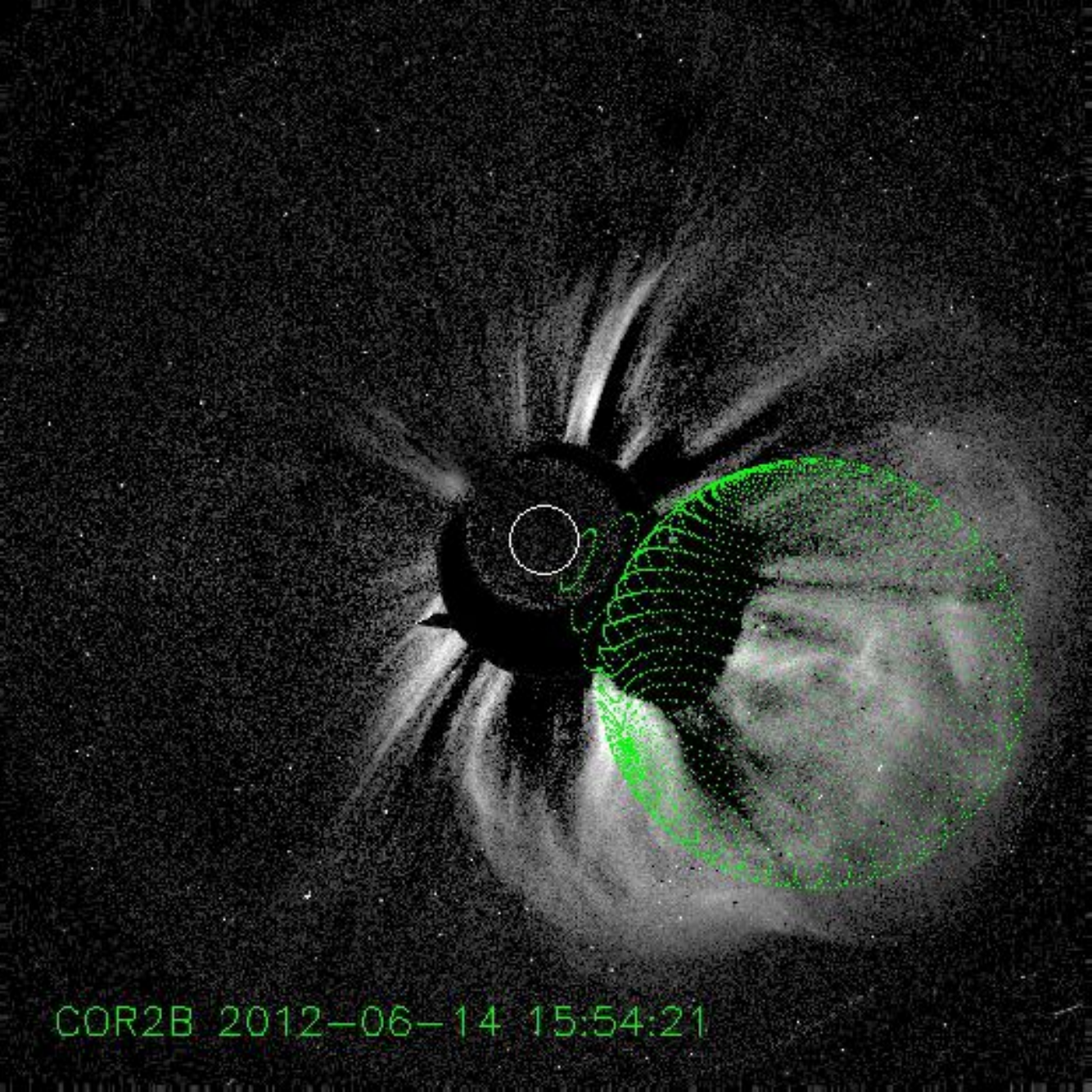} 
\includegraphics[width=.25\textwidth]{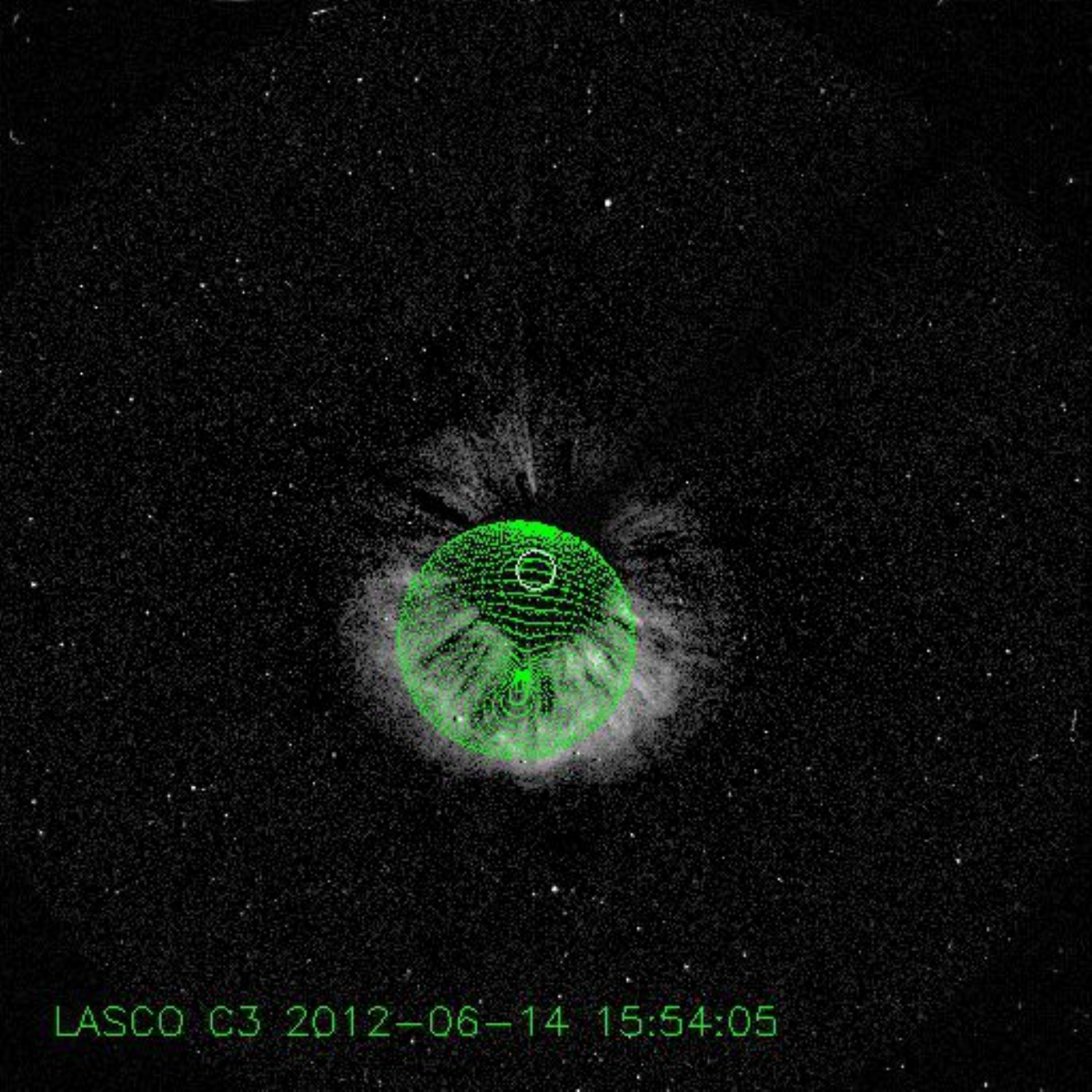} 
\includegraphics[width=.25\textwidth]{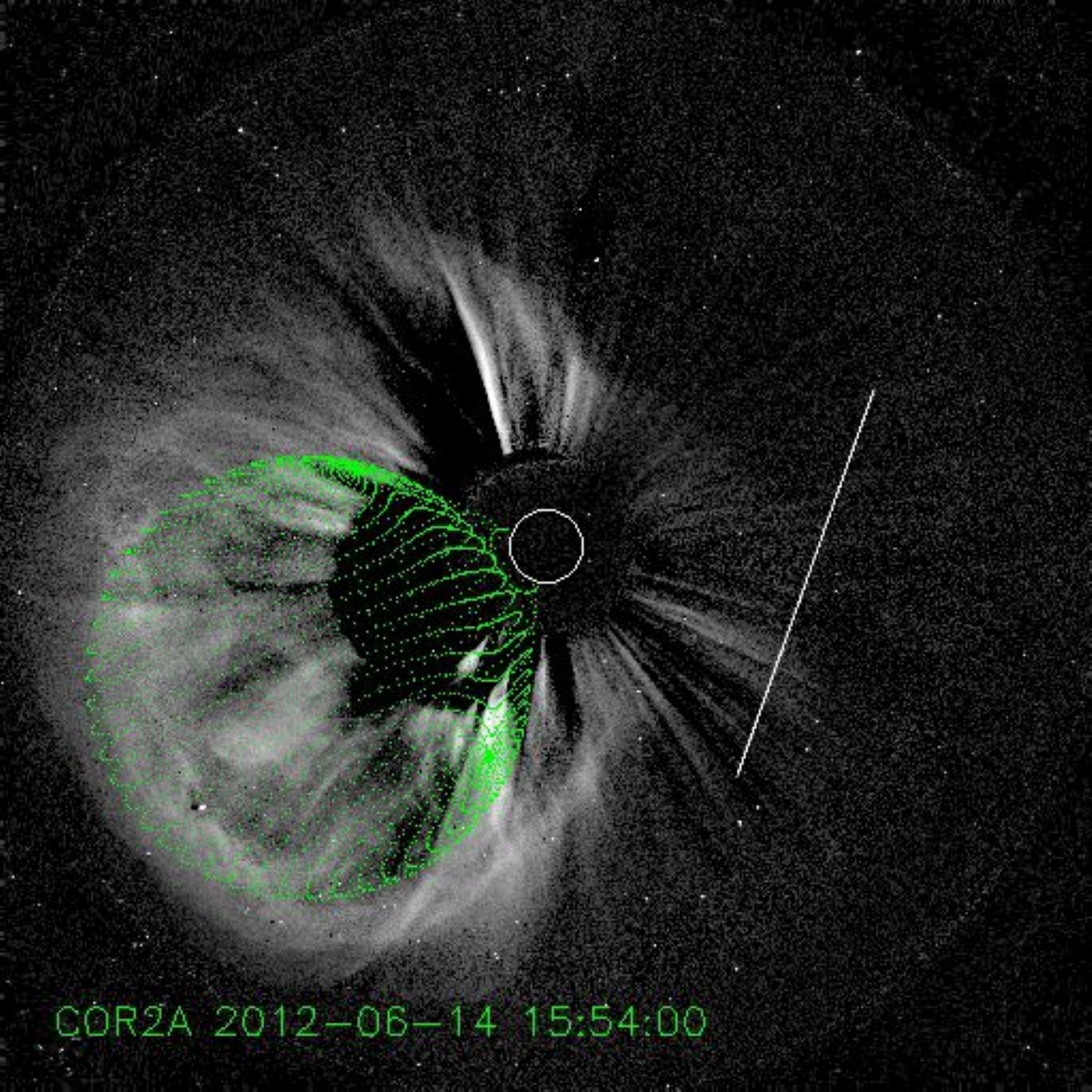}} 
\caption{Event 2: GCS fitting of CME1 (a) and CME2 (b).
SECCHI/COR2B (left column), LASCO C3 (cental column) and SECCHI/COR2A (left column) pre-event background-subtracted intensity images of the two CME events with and without the GCS model wireframe (in green).
}
\label{fig:20120614_gcs} 
\end{figure*}

\medskip
% coronal observations ============================
\textit{Coronal observations and GCS reconstruction.} 
As shown in Figure~\ref{fig:20120614_ecliptic}, on the dates of the CME eruptions the separation of the STEREO spacecraft relative to Earth was about $117^\circ$ for STEREO-A and about $116^\circ$ for STEREO-B. The separation between the two STEREO spacecraft was about $127^\circ$.
Both CMEs were observed by three spacecraft (SOHO and the two STEREO) in the corona, so that the GCS fitting using three view points could be performed. 
We apply the GCS fitting to contemporaneous images of the CMEs from SECCHI/COR2B, LASCO C2 and C3, and SECCHI/COR2A, in the following time intervals: CME1:  15:45~UT -- 17:54~UT on 13 June 2012; CME2: 15:24~UT -- 15:54~UT on 14 June 2012.
As for Event 1, we fit the CMEs with a spherical geometry ($\alpha=0$). The results of the GCS fitting for the latest time frames available ({i.e.} closest to 0.1~AU) are shown in Figure~\ref{fig:20120614_gcs}. 
The results are listed in Table~\ref{tab:obs_event_2}.
% Table
\begin{table*}
\centering
 \begin{tabular}{l|l|lll}
 \hline
 \hline
 Parameter      &  \multicolumn{4}{c}{Method} \\
 \hline
                & \multicolumn{1}{c|}{CME1} & \multicolumn{3}{c}{CME2} \\
 \hline
               & GCS fitting &  GCS fitting & \\
 \hline
 Date           & 2012-06-13      & 2012-06-14 & \\
 Time           & 17:54~UT        & 15:54~UT & \\
 $h_{front}$    & 15.0 $R_s$      & 15.2 $R_s$ & \\
 $\phi$         & $-20^\circ$     & $-5^\circ$ & \\
 $\theta$       & $-35^\circ$     & $-25^\circ$ & \\
 $\kappa$       & 0.45            & 0.70 & \\
 $\omega/2$     & $26^\circ$      & $40^\circ$ & \\
 $r_0$ at 0.1 AU        & 10.5 $R_s$      & 18.0 $R_s$ & & \\
 \hline
                & Geometrical     & Geometrical  & Empirical-3D   & Empirical-2D \\
 \hline
 $v_{3D}$       & 719~\si{ \km \,\, \s^{-1} }        &  1213~\si{ \km \,\, \s^{-1} }   & 1213~\si{ \km \,\, \s^{-1} }  & 1737~\si{ \km \,\, \s^{-1} } \\
 $v_{rad}$      & 496~\si{ \km \,\, \s^{-1} }        &  713~\si{ \km \,\, \s^{-1} }    & 523~\si{ \km \,\, \s^{-1} }   & 747~\si{ \km \,\, \s^{-1} } \\
 $v_{exp}$      & 223~\si{ \km \,\, \s^{-1} }        &  500~\si{ \km \,\, \s^{-1} }    & 690~\si{ \km \,\, \s^{-1} }   & 990~\si{ \km \,\, \s^{-1} } \\
 \hline
\end{tabular}
\caption{Event 2: CME kinematic parameters derived from the GCS fitting and from the application of the geometrical, empirical-3D and empirical-2D approaches to derive total (3D), expansion and radial speeds.}
\label{tab:obs_event_2}
\end{table*}
%
% geometrical approach
Extrapolating the CME height in time assuming a constant CME speed,
the passage of CME1 at 0.1~AU is estimated to occur on 13 June 2012 at 19:38~UT,
while the passage of CME2 at 0.1~AU is estimated to occur on 14 June 2012 at 16:55~UT.
% comparison with empirical relations
For CME2 (the only full halo one), we compare the geometrical approach proposed in Section~\ref{subsec:gcs} with the empirical approaches presented in Equations~\ref{eqn:v3d_dallago2003} and \ref{eqn:vexp_dallago2003_vexp}, similarly to what we have done for Event~1.
As the CDAWeb CME Catalog reports a projected CME speed $v_{2D}$ very steady during the CME propagation in the instruments FoV from 3.3~$R_s$ to 28.1~$R_s$, we can just compare the results therein with the results from the GCS fitting at the latest time available.
All results are listed in columns 4 and 5 of Table~\ref{tab:obs_event_2}.
As for the previous case, the results from the empirical-2D approach significantly overestimate the total (3D) speed of the CME. On the other hand, results obtained using the empirical-3D approach overestimate the expansion speed and underestimate the radial one.

\medskip
% derived magnetic parameters ============================
\textit{Derived magnetic parameters.} 
From observations of PEA1 at 16:00~UT and of PEA2 at 19:00~UT, we derive the flux-rope axial magnetic fields and toroidal magnetic fluxes. 
In the case of PEA1, starting from $A_{PEA} = 1.0 \cdot 10^{15}$~m$^2$ and $\phi_{RC} \simeq 2.2 \cdot 10^{13}$~Wb, the resulting values are $B_0 \simeq 1.4 \cdot 10^{-6}$~T and $\phi_{t} \simeq 1.9 \cdot 10^{13}$~Wb. 
In the case of PEA2, $A_{PEA} = 1.4 \cdot 10^{15}$~m$^2$ and $\phi_{RC} \simeq 2.7 \cdot 10^{13}$~Wb give as result $B_0 \simeq 8.6 \cdot 10^{-7}$~T and $\phi_{t} \simeq 4.0 \cdot 10^{13}$~Wb.

% CME propagation in the heliosphere
\subsubsection{CME propagation in the heliosphere}

\begin{figure}[h]
\centering
{\includegraphics[width=\hsize, trim={30mm 30mm 30mm, 30mm},clip]{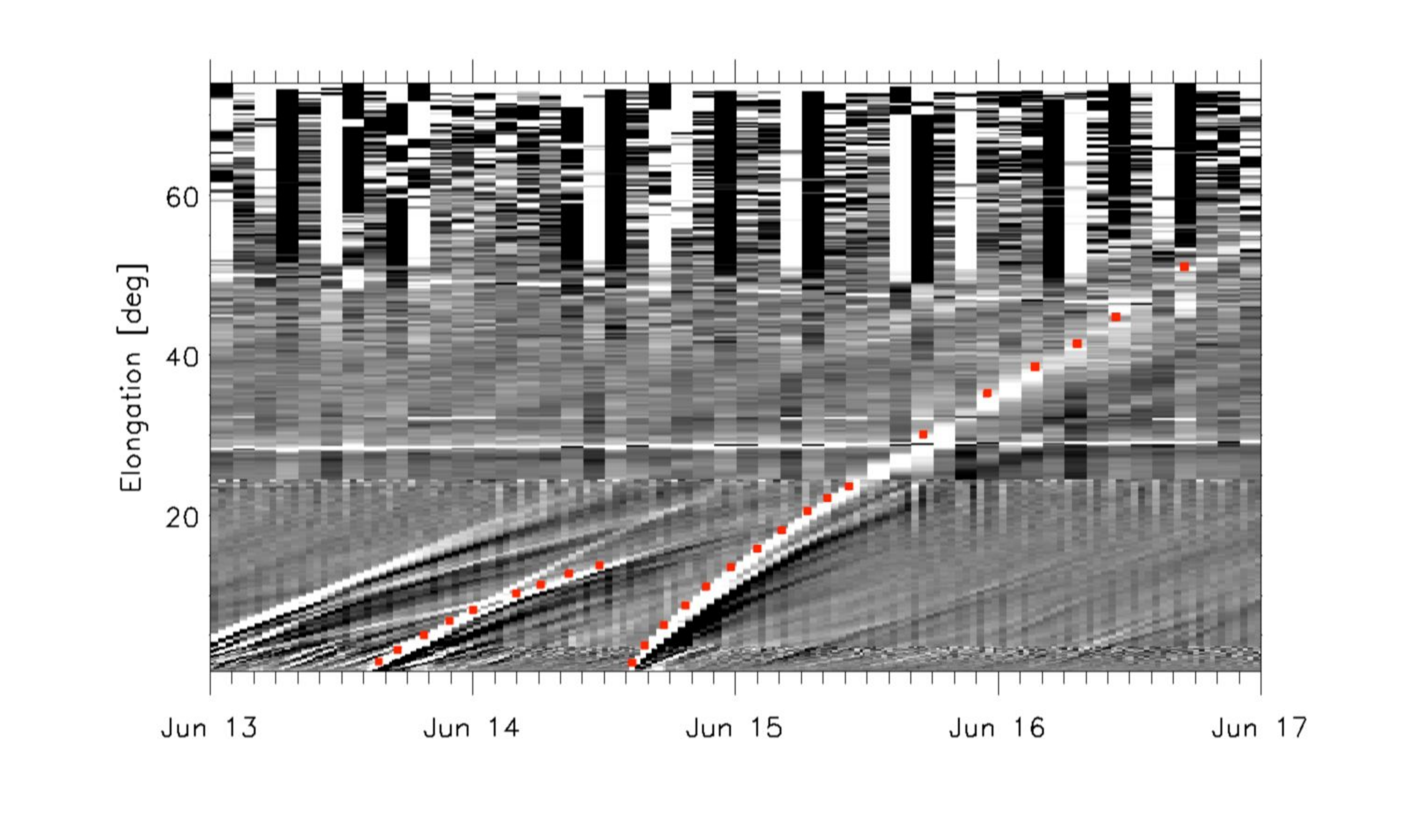} } 
\caption{Event 2: STEREO-A running-difference J-maps at PA=$90^\circ$.
The red dots mark the leading edge of CME1 and CME2.}
\label{fig:20120614_jmaps} 
\end{figure}

\textit{STEREO-A time-elongation maps.}
As shown in Figure~\ref{fig:20120614_jmaps}, to constrain the CME propagation in the heliosphere we track the position over time of the CME leading edges as extracted from STEREO-A J-maps, obtained by stacking SECCHI/COR2A-HI1A-HI2A images at PA=$90^\circ$, e.g. tracking the CME leading edges on the ecliptic plane.
The leading edge of CME1 in STEREO-A images could be tracked between $2^\circ$ and $14^\circ$ in elongation, and that of CME2 between $2^\circ$ and $51^\circ$ in elongation. 

\medskip
\textit{SSE and iSSE techniques.}
In order to recover the time-height profiles of the CME apex, we first apply the SSE model (Equation~\ref{eqn:sse}) to the time-elongation profiles of CME1 and CME2,
using the CME half widths derived from the GCS fitting. 
The angles $\phi_1, \phi_2$ between the observer and the propagation directions of CME1 and CME2 were also calculated based on the directions estimated from the GCS fitting.

In this case the angles between the CME propagation directions and the Sun-Earth line 
were $\Delta_1 = 40^\circ$ and $\Delta_2 = 25^\circ$ for the two CMEs.
Therefore, the application of the iSSE technique was needed in order to recover the actual propagation of the portion of the CME leading edge that travelled towards the Earth.
In the case of CME1, $\Delta_1 > \omega/2$, i.e. according to this model, CME1 does not intersect the Sun-Earth line, and hence the iSSE method predicts that CME1 does not arrive at Earth at all.
On the other hand, as visible from Figure~\ref{fig:20120614_propagation}, the iSSE and SSE techniques in the case of CME2 gave significantly different results.

\medskip
\textit{Venus Express data.}
In addition to remote-sensing tracking of the CME, we make use of in-situ data from Venus Express \citep[VEX;][]{zhang:2006} to better constrain the CME propagation in the heliosphere. 
At the time of the eruptions, VEX was orbiting Venus and it was located at $\theta_V=0.88^\circ$ and $\phi_V=5^\circ$ in HEEQ coordinates in the heliosphere, at $0.726$~$R_s$ (=156~$R_s$) from the Sun (Figure~\ref{fig:20120614_ecliptic}).
The spacecraft was separated from Earth by $\Delta \theta  < 1^\circ $ in latitude and $\Delta \phi = 5^\circ $ in longitude from Earth. Therefore, VEX and Earth were in quasi-alignment.
In a previous study, \citet{good:2016} reported an ICME flux-rope leading edge to arrive at VEX on 15 June 2012 at 19:26~UT, while the trailing edge was reported to pass at 08:28~UT on the following day. 
From an inspection of coronal and low-coronal images on the days prior the eruption of CME2, we consider this CME as the most promising candidate to be associated with the ICME observed at VEX, as no other suitable CME candidates were identified. A similar conclusion was also reached by \cite{kubicka:2016}.
The flux-rope configuration at VEX was identified to be a NES type, with a positive handedness/chirality, 
i.e. a configuration that is consistent with the one recovered from the analysis of the source region. 
This would provide an additional indication that the flux-rope underwent only slight rotation between the low corona and 0.7~AU.

% in-situ observations
\subsubsection{ICME signatures at Earth}
\begin{figure}[t]
\centering
{
   \includegraphics[width=\hsize]{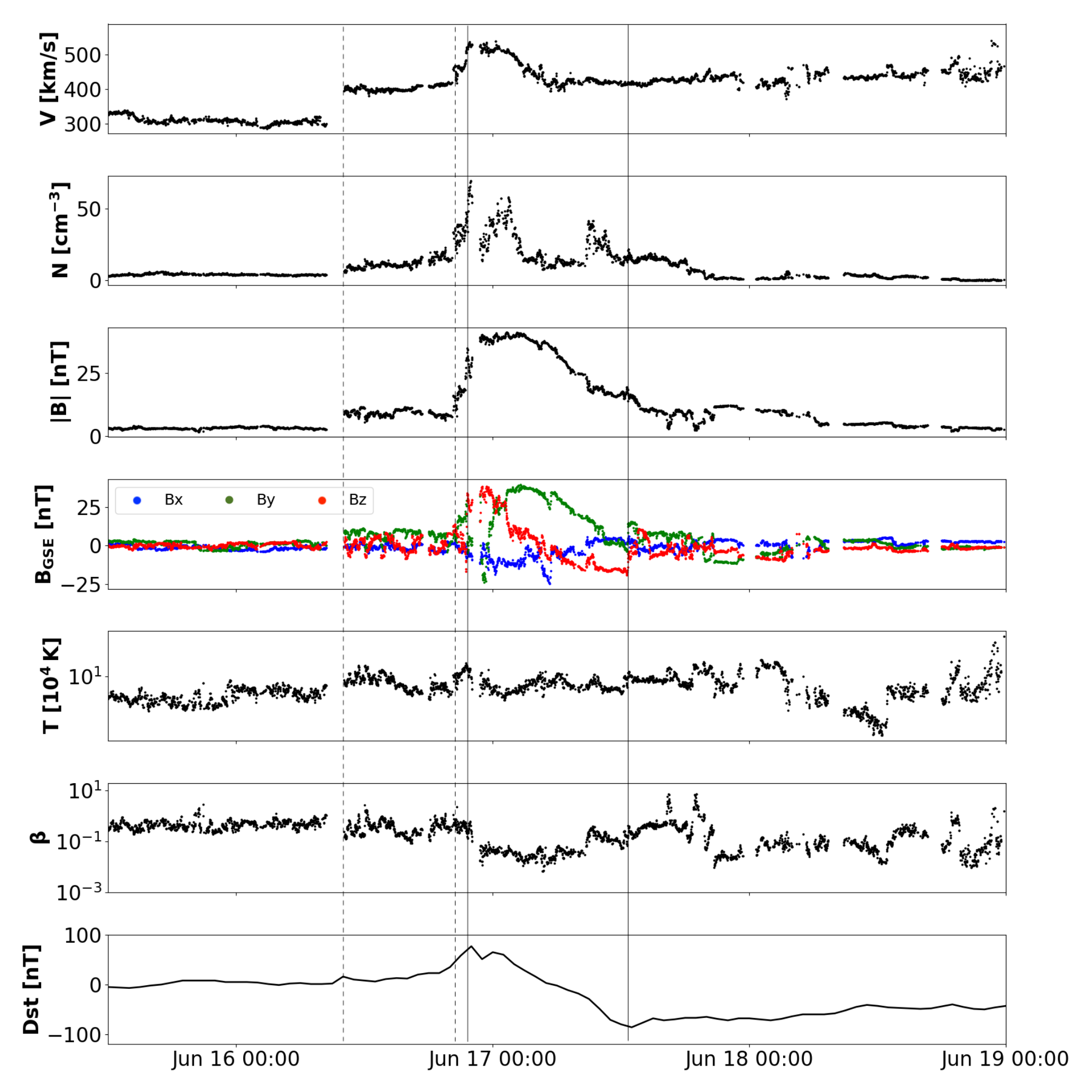}  } 
\caption{Event 2: in-situ observations of the 13-14 June 2012 event from OMNI 1-min data. 
From  top  to  bottom: speed, number density, magnetic field strength, 
$B_x$, $B_y$, $B_z$ components in GSE coordinates, temperature, plasma $\beta$ and \textit{Dst} index.
The dashed black lines mark the shocks associated to ICME1 (S2) and ICME2 (S2), 
whilst the region delimited by the continuous black lines marks the MC associated to ICME2.
}
\label{fig:20120614_omni} 
\end{figure}

Figure~\ref{fig:20120614_omni} shows in-situ magnetic field and plasma measurements from the OMNI database, on the days following the eruptions of the two interacting CMEs. 
A first forward shock (S1), associated to the interplanetary signature of CME1 (ICME1), was detected by the $Wind$ spacecraft on 16 June 2012 at 08:42~UT (from the Heliospheric Shock Database), as indicated by sudden increases in plasma speed and magnetic field.
The shock was followed by a region of enhanced speed, increasing density and fluctuating magnetic fields that lasted approximately 12 hours.
Such region does not show any coherent magnetic field rotation, and it is characterised by $\beta \sim 1$, compatible with a long-lasting sheath region that suggests a flank encounter of ICME1 at Earth. 
A second forward shock (S2), associated to the interplanetary signature of CME2 (ICME2), was detected by $Wind$ on 16 June at 19:34~UT.
As reported by the Richardson and Cane ICME list, MC signatures can be identified in in-situ data starting from 23:00~UT on 16 June, up to 12:00~UT on 17 June.
The MC duration is about 13 hours, and it is characterised by enhanced magnetic field and $\beta \ll 1$, while the presence of density peaks suggests some compression inside and near the trailing edge of the MC.
The maximum magnetic field in the MC is 40~nT, while the average $B$ is 28~nT.
The observed minimum $B_z$ is -19~nT.
The MC also exhibits a moderate expansion profile, with a maximum speed of $573$~\si{ \km \,\, \s^{-1} } 
and a speed difference of $80$~\si{ \km \,\, \s^{-1} } between the front and the back.
The presence of a north-to-south rotation in the MC $B_z$ component led to a moderate geomagnetic storm, as indicated by the $Dst$ index reaching a minimum value of -86~nT on 17 June.

From a visual inspection of the magnetic field, we observe that the $B_z$ component rotates from north to south, while $B_y$ is positive at the cloud center, implying this MC is compatible with a right-handed NES flux-rope type at Earth.
At the same time, fitting the in-situ flux-rope with the MVA analysis, \citet{palmerio:2018} found an orientation of the ICME axis equal to $(\theta_\mathrm{MVA}, \phi_\mathrm{MVA})=(-28^\circ, 99^\circ)$, confirming that this was a low flux-rope axis inclination at Earth.
The flux-rope tilt angle at Earth is almost identical to the tilt angle of the PIL at the Sun, indicating that the structure underwent little rotation as it propagated in the corona and heliosphere.
They also suggested that the flux-rope impacted on Earth from the very center, as indicated by the small location angle, $L \simeq -8^\circ$.

%========================================================================
%========================================================================
\section{EUHFORIA results and comparison with observations} 
\label{sec:results}
%========================================================================
%========================================================================

In this Section we discuss the results of the simulations performed with EUHFORIA and compare them to remote-sensing and in-situ observations at Earth and other planetary locations.

%===========================================
\subsection{Simulation set up} 
\label{subsec:simulation_setup}
%===========================================

We simulate the heliospheric propagation of both CME events discussed in Section~\ref{sec:case_studies} using EUHFORIA.
For each event we run the semi-empirical coronal model in the same set up described by \citet{pomoell:2018},
using as input conditions GONG standard synoptic maps generated on the day of the CME eruptions.
The computational domain of the heliospheric model used in this work extends from 0.1~AU to 2~AU in the radial direction, over the range $\pm 60^\circ$ in latitudinal direction, and over the full angular extent of $360^\circ$ in longitude. 
We use a $2^\circ$ angular resolution in longitude and latitude, and $512$ cells in the radial direction with the heliospheric inner boundary at 0.1~AU and its outer boundary at 2.0~AU.
% corresponding to a radial resolution of $0.0037$~AU ($\sim 0.8$~$R_s$) per cell.
% 
To initialise the CMEs in the simulations we use the observation-based input parameters derived in Section~\ref{sec:parameters}.
For each event, we run EUHFORIA using the cone CME model, and we then present the results obtained using the spheromak CME model, discussing its use and limitations in the two specific cases.
The detailed input parameters used in each simulation are presented below.
All results in this work are obtained using EUHFORIA version 1.0.4.

Simulation outputs include 3D outputs of the whole heliospheric domain, and 1D text files containing the time series for the whole set of MHD variables at given positions in space.
Default outputs are given at planetary locations and notable spacecraft locations such as the STEREO mission, 
and additional virtual spacecraft can be put by the user at any other position of interest in the heliosphere.
To track the CME as it propagates in the simulations, 
in this work we place a set of virtual spacecraft between 0.1~AU and 1.0~AU along the Sun-Earth line. The spacecraft are distributed more densely near the Sun, i.e. with a separation of 0.05~AU between 0.1~AU and 0.4~AU, and with a separation of 0.2~AU between 0.4~AU and 1.0~AU. 
We put a second set of virtual spacecraft located at 1.0~AU, at $ 5^\circ$ and $ 10^\circ$ separation in longitude and/or latitude from Earth, in order to assess the spatial variability of the results in the vicinity of Earth. 

%===========================================
\subsection{Event 1: CME on 12 July 2012} 
\label{subsec:20120712_results}
%===========================================

\begin{figure}
\centering
\subfloat[Radial speed $v_r$]
{\includegraphics[width=\hsize,trim={20mm 0mm 0mm 0mm},clip]{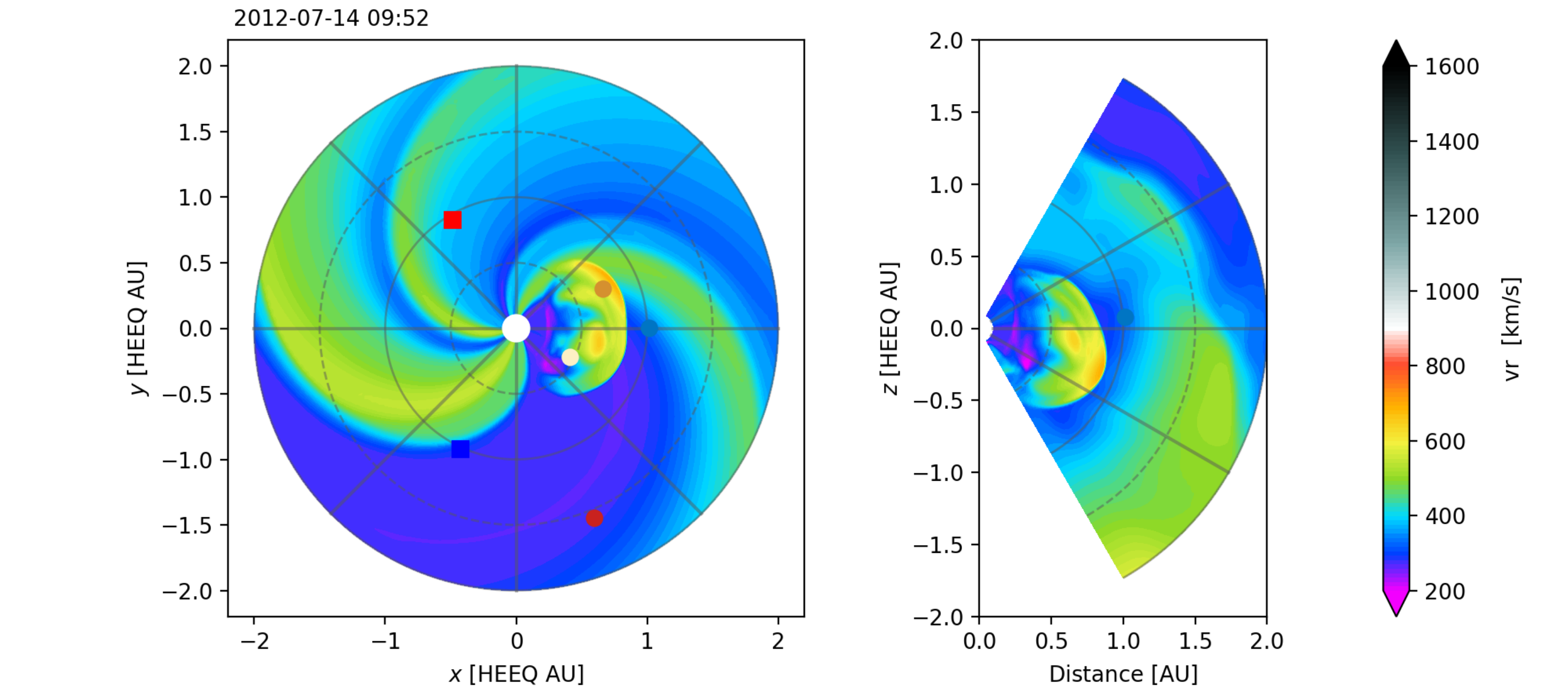} } \\
\subfloat[Scaled number density $n \big (\frac{r}{1 \,\, \mathrm{AU}} \big )^2 $ ]
{ \includegraphics[width=\hsize,trim={20mm 0mm 0mm 0mm}, clip]{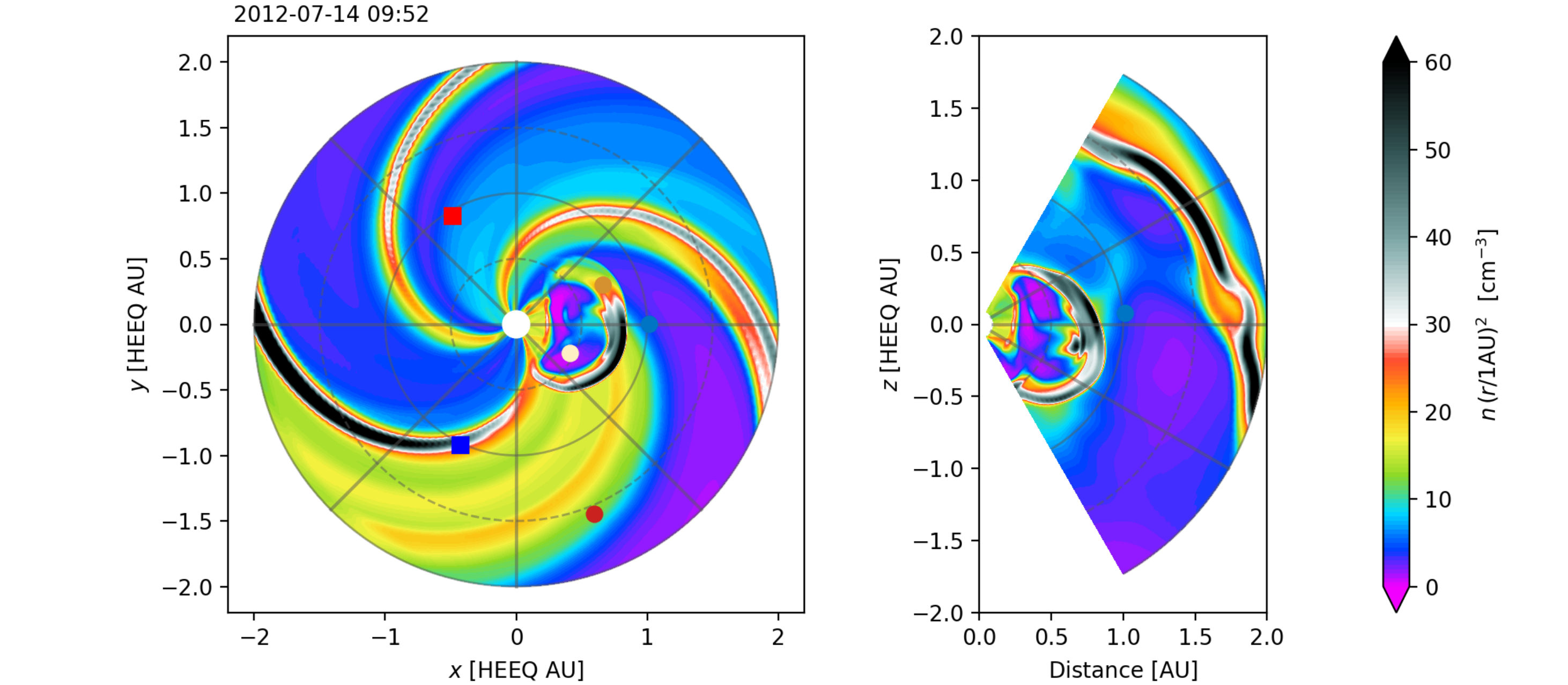} } \\
\subfloat[Co-latitudinal magnetic field $B_{clt}$]
{ \includegraphics[width=\hsize,trim={20mm 0mm 0mm 0mm}, clip]{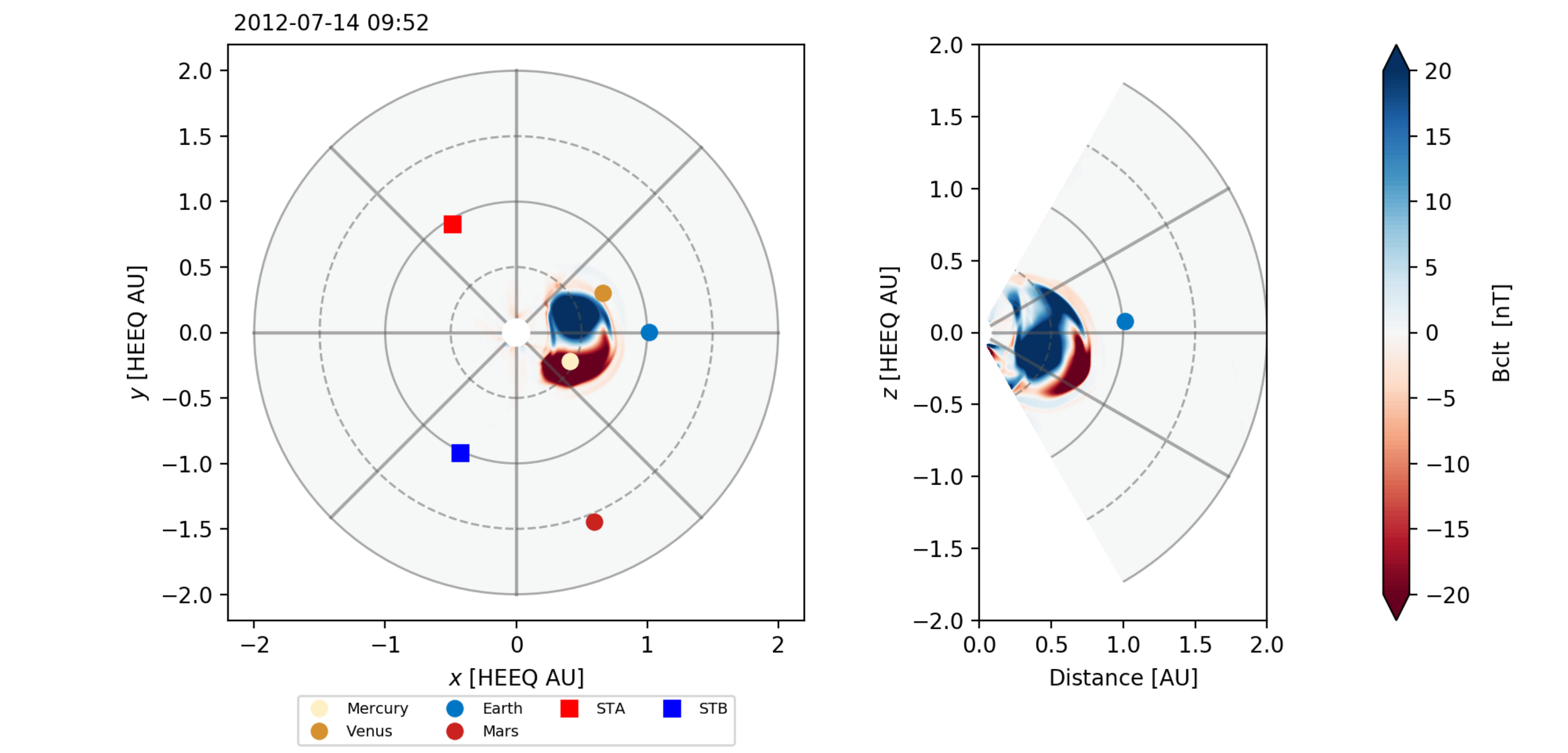}} 
\caption{Event 1: snapshot of the EUHFORIA Run~03 (spheromak CME with reduced speed $v_\mathrm{CME}=v_{rad}$) 
on 14 July 2012 at 09:52~UT in the heliographic equatorial plane (left) and in the meridional plane that includes the Earth (right). }
\label{fig:20120712_euhforia} 
\end{figure}

We first simulate the CME using the cone model (Run~01), employing the parameters determined by the GCS reconstruction as input.
We then perform a second simulation run of the same CME using the spheromak model, keeping the kinematic/geometric CME parameters as in Run~01, and adding the three magnetic parameters as determined in Section~\ref{subsec:magnetic_parameters} (Run~02). 
In a third simulation, we initialise the CME using the spheromak model using a reduced speed calculated as $v_\mathrm{CME} =  v_{3D} - v_{exp} = v_{rad}$ (Run~03).
In all three cases, as input for the coronal model we use the synoptic standard GONG map on 12 July 2012 at 11:54~UT.
Table \ref{tab:event_1} lists the CME input parameters used to simulate the CME with the cone and spheromak models.
The mass density and temperature are set to be homogeneous within the CME.
Using the default values listed in Table \ref{tab:event_1}, the density ratio in the CME body is approximately 1 with respect to the surrounding solar wind, while the pressure ratio is about 3.8.
% Table
\begin{table*}
\centering
\begin{tabular}{lll}
\hline
\hline
Parameter & Run~01 & Run~02 (Run~03)\\
 \hline
 CME model           & cone & spheromak \\ 
 Insertion time      & 2012-07-12T19:24 &  2012-07-12T19:24 \\  
 $v_\mathrm{CME}$                & 1266 \si{ \km \,\, \s^{-1} } & 1266 \si{ \km \,\, \s^{-1} } (763 \si{ \km \,\, \s^{-1} })\\ 
 $\phi$             & $-4^\circ$    &  $-4^\circ$\\   
 $\theta$           & $-8^\circ$    &  $-8^\circ$ \\   
 $\omega/2$         & $38^\circ$    & - \\   
 $r_0$                & -             &  $16.8 \,\, R_s$ \\  
 $\rho$             & $1 \cdot 10^{-18} \,\, \si{ \kg \,\, \m^{-3} }$   & $1 \cdot 10^{-18} \,\, \si{ \kg \,\, \m^{-3} }$  \\ 
 $T$                & $0.8 \cdot 10^{6} \,\, \si{\kelvin}$              & $0.8 \cdot 10^{6} \,\, \si{\kelvin}$  \\ 
 $H$                & -         & +1 \\ 
 Tilt               & -         & $-135^\circ$  \\ 
 $\phi_t$           & -         & $1.0 \cdot 10^{14}$~Wb  \\
 \hline
 Predicted ToA at Earth       &  2012-07-14T20:52 &  2012-07-14T07:03 (T22:33) \\
 \hline
\end{tabular}
\caption{CME input parameters used in the EUHFORIA simulations of the 12 July 2012 CME, 
and its predicted arrival times at Earth.}
\label{tab:event_1}
\end{table*}
An example of simulation results for Run~03 is provided in Figure~\ref{fig:20120712_euhforia}, which shows a snapshot in the ecliptic and  meridional planes containing the Earth, of the radial speed, scaled number density and $B_{clt}$ component of the magnetic field (see supplementary material for movies of the dynamics).

\begin{figure}[t]
\centering
{\includegraphics[width=\hsize,trim={0mm 61mm 0mm 0mm},clip]{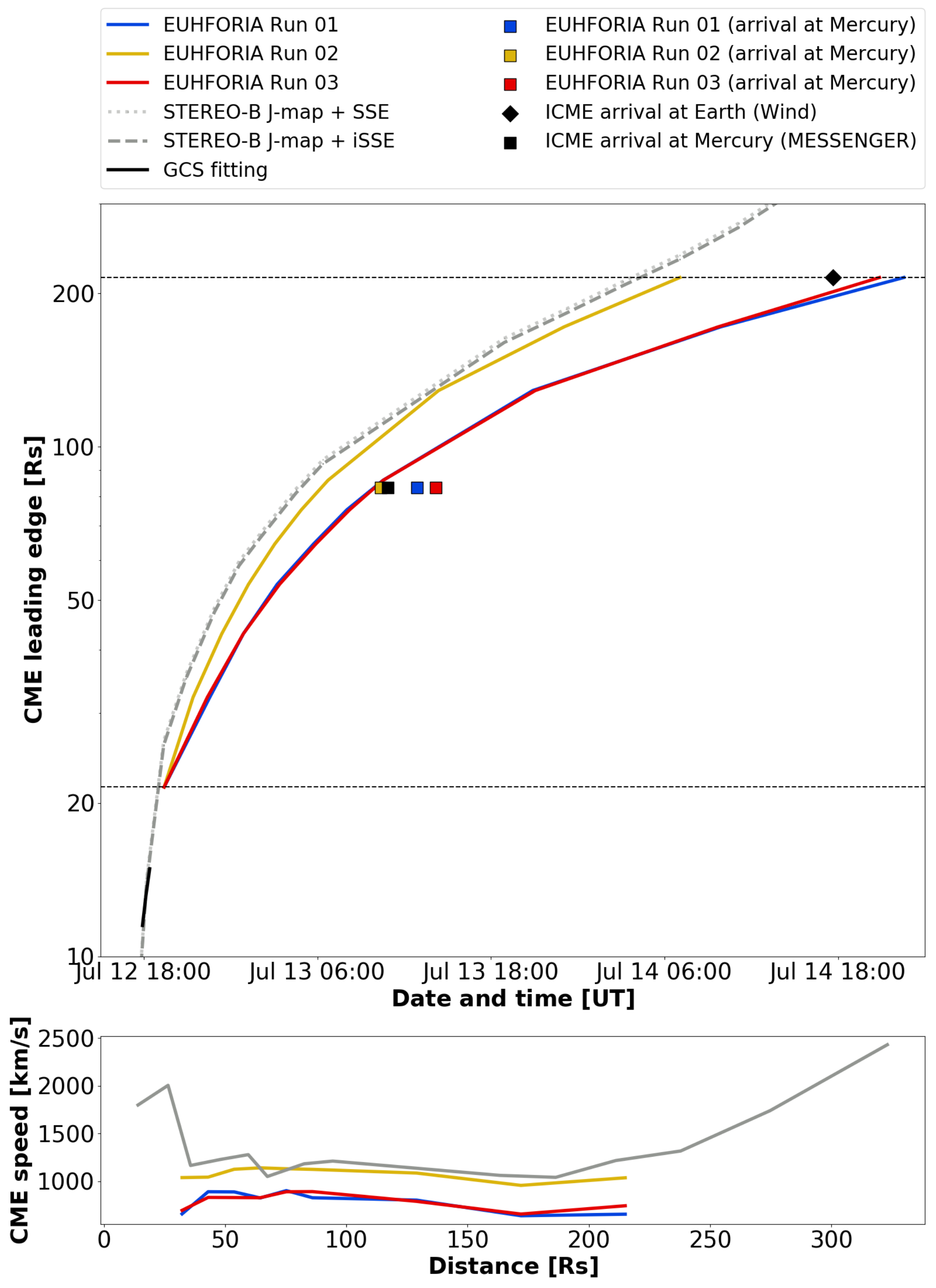} } 
\caption{Observed and modelled CME propagation in the heliosphere for the 12 July 2012 CME (Event 1).
%Top: position of the CME leading edge in time. 
The black dashed lines mark the EUHFORIA heliospheric inner boundary at 0.1~AU, and 1.0~AU.
%Bottom: speed of the CME leading edge as function of the heliocentric distance. 
}
\label{fig:20120712_propagation} 
\end{figure}

\medskip
% EUHFORIA results in the heliosphere
\textit{CME propagation in the heliosphere.}
Using the time series at the virtual spacecraft, we extract the time of arrival of the CME-driven shock at each one, and construct time-height profiles of the front along the Sun-Earth line. 
%From the arrival times of the CME front shock at different spacecraft, we can also construct the profile of the CME speed at different distances from the Sun.
Figure~\ref{fig:20120712_propagation} shows the result of the computation compared with the
time-height maps determined from the J-maps extracted at the PA corresponding to the direction to Earth.
%, together with the evolution of the CME speed as function of the heliocentric distance.
%
In EUHFORIA Run~01 (cone with $v_\mathrm{CME}=v_{3D}$, blue curve) and Run~03 (spheromak with $v_\mathrm{CME}=v_{rad}$, red curve) the propagation of the CME-driven shock along the Sun-Earth line is very similar all the way up to 1 AU.
On the other hand, Run~02 (spheromak with $v_\mathrm{CME}=v_{3D}$, yellow curve) shows that the front of the CME propagates faster already very early in the simulation. 
The difference between the time-height profile from Run~01 and Run~02 is entirely due to including an internal magnetic field in the CME, and therefore, it provides an estimate of the importance of the Lorentz force (and particularly of the magnetic pressure) on the propagation of the CME itself.
The difference between the time-height profiles in Run~02 and Run~03 is entirely due to the different initial speeds given to the CME in the model. 
The fact that the propagation of the CME in Run~03 is similar to the one observed for Run~01, 
shows that the differences in the CME propagation resulting from inclusion of the magnetic field can be mitigated by initialising the magnetised CME with a reduced speed.
Instead of choosing this speed based on some \textit{ad hoc} number, 
we computed it through a direct observational estimation of the expansion of the CME in the corona. 
%where we emphasised how CME expansion can be used as indication of dominant pressure terms acting within CMEs in the corona. 
A detailed discussion on the interpretation of the simulation results in terms of the Lorentz force acting on cone and spheromak CMEs is presented in the next paragraphs.

Time-height profiles based on STEREO-B J-maps and the SSE and iSSE techniques model the CME leading edge propagation along the Sun-Earth line similarly to EUHFORIA Run~02, predicting the CME arrival time at Earth to occur around 04:00~UT on 14 July, i.e. about 15 hours earlier than observed in-situ.

At Mercury/MESSENGER, EUHFORIA Run~01 predicts the CME ToA about 30 minutes earlier than the one reported by \cite{winslow:2015} from MESSENGER data, while Run~02 and Run~03 are 2 hours and 3 hours late respectively. 
Mercury was located about $30^\circ$ away from the Sun-Earth line (see Figure~\ref{fig:20120712_ecliptic}).
As the CME main direction of propagation was almost coincident with the Sun-Earth line, 
the CME hit Mercury with its western flank.
By comparing the arrival time of the CME leading edge at Mercury with that at the same radial distance but along the Sun-Earth line in EUHFORIA, we conclude that the CME would have been observed 4 hours earlier if Mercury would have been on the Sun-Earth line, i.e. the CME flank propagates with about 4 hours of delay with respect to the CME center.  

\begin{figure*}
\centering
\subfloat[Arrows indicating the direction of the Lorentz force 
$\vec{j} \times \vec{B} = + \frac{( \vec{B} \cdot \nabla) \vec{B}}{\mu_0} -\nabla P_{mag} $ 
at the $\beta=0.5$ surface. The arrows colour code is based on the magnitude of the Lorentz force (in Pa AU$^{-1}$).
\label{fig:20120712_3d_forces_a}]
{\includegraphics[width=0.48\textwidth]{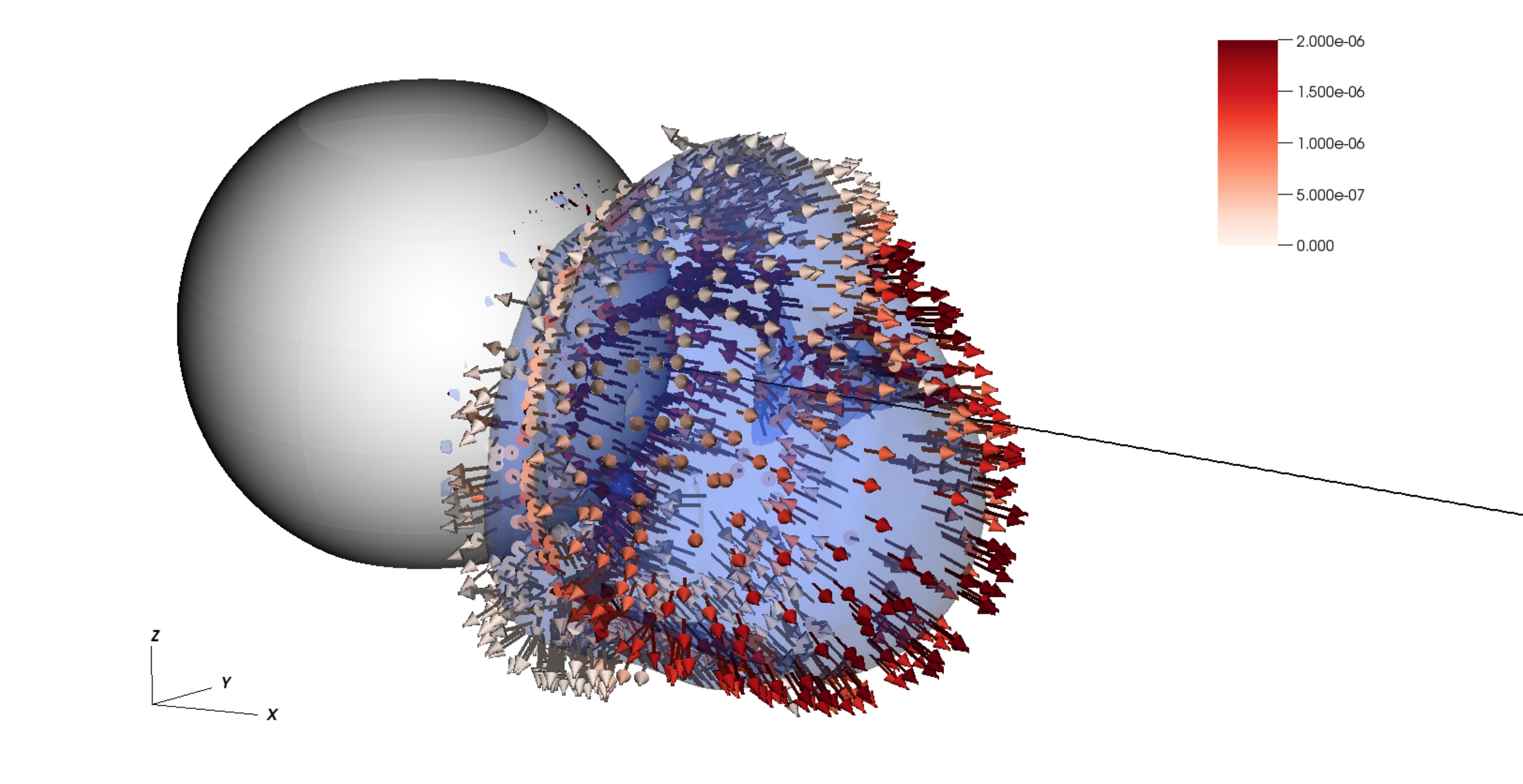}} \quad
\subfloat[Magnetic field lines in the CME, coloured based on the angle between the current density $\vec{j}$ and the magnetic field $\vec{B}$ (in rad, colour scale between 0 an $\pi/4$).
\label{fig:20120712_3d_forces_b}]
{\includegraphics[width=0.48\textwidth]{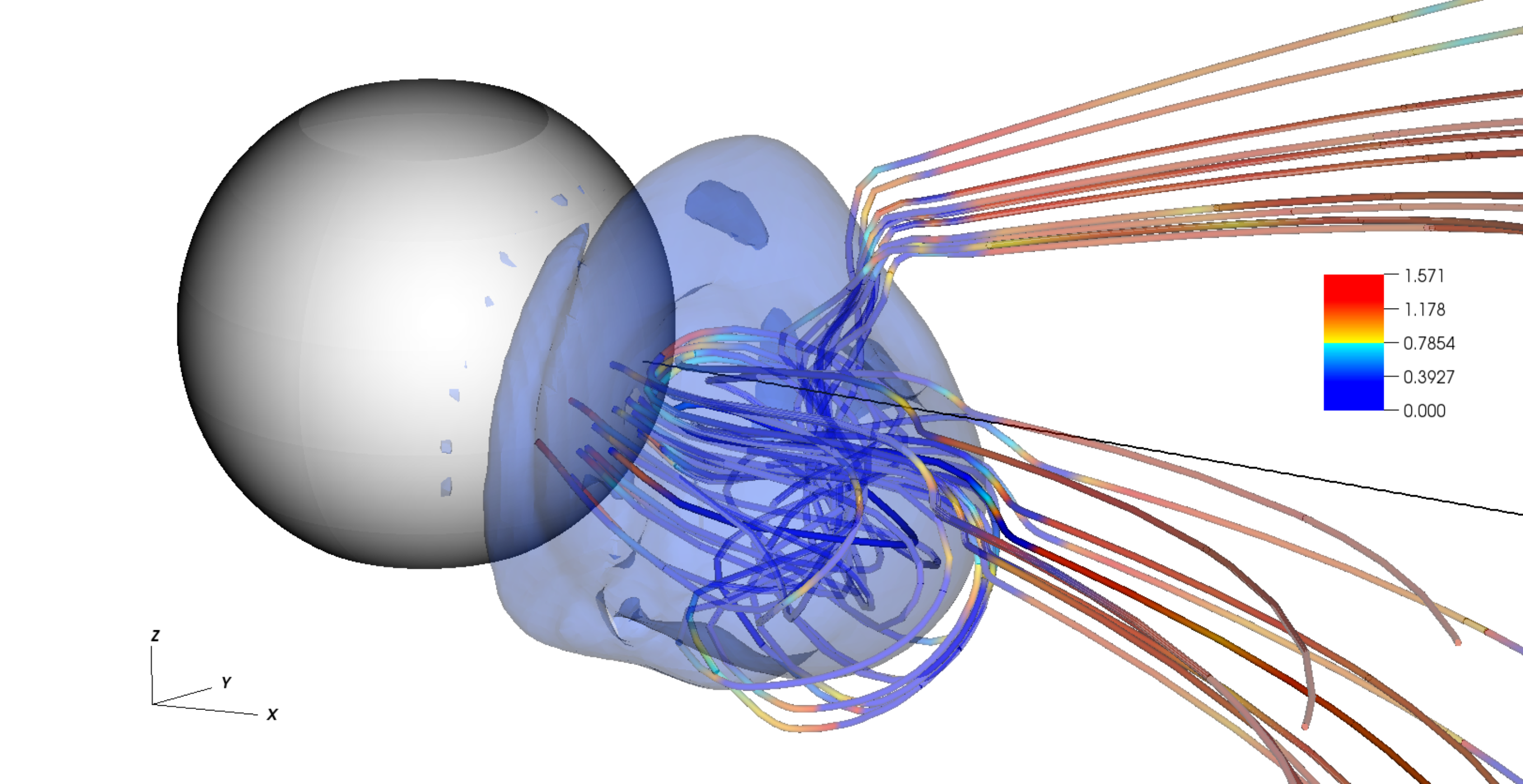}}
\caption{ 
Event 1: visualisation of the forces acting on the CME in Run~03 on 13 July 2012 at 03:53~UT, when the CME leading edge was close to 0.3~AU. The {blue} 3D surface marks the contour of the $\beta=0.5$ surface.
The black line indicates the direction to Earth.
The spherical 3D surface represents the heliospheric inner boundary at 0.1~AU.
}
\label{fig:20120712_3d_forces} 
\end{figure*}

\medskip
% Lorentz force
\textit{CME magnetic structure and Lorentz force.}
To further investigate the role of thermal pressure, magnetic pressure, and magnetic tension on the CME propagation, 
in Figure~\ref{fig:20120712_3d_forces_a} we plot the direction of the Lorentz force (coloured arrows) in the CME on the $\beta=0.5$ surface (blue isocontour), after the CME in Run~03 has fully entered the computational domain. 
The figure shows that the magnetic field of the spheromak CME, originally defined as force-free ($\vec{j} \times \vec{B} = 0$), loses this characteristics after insertion in the heliosphere, reasonably as consequence of its non-equilibrium with the surrounding solar wind. 
The Lorentz force at the front of the $\beta=0.5$ surface is stronger than at the flanks, and it is predominantly parallel to the surface normal and pointing away from the center of the CME. This indicates that the magnetic pressure gradient $-\nabla P_{mag}$ is dominating over the tension force $\frac{( \vec{B} \cdot \nabla) \vec{B}}{\mu_0}$. This force imbalance leads to the expansion of the CME.
As depicted by the $\beta=0.5$ surface, the bulk of the interior of the CME is characterised by a magnetically-dominated plasma.
This suggests that the CME expansion is caused by an over-pressure in the CME as compared to the ambient solar wind, and that this over-pressure is predominantly due to the magnetic pressure.
This force imbalance, on the other hand, is not present in cone CMEs (e.g. Run~01), where the Lorentz force is negligible as the magnetic field inside CMEs is just the one of the background solar wind.
On the CME flanks, the Lorentz force is weaker and it is almost tangential to the $\beta=0.5$ surface (i.e. perpendicular to the surface normal), so that the CME propagates in the heliosphere retaining its angular width, i.e. self-similarly.

Figure~\ref{fig:20120712_3d_forces_b} gives an indication of the curvature of magnetic field lines within the flux-rope structure 
and provides insights about the nature of the Lorentz force within the CME body. 
A twisted magnetic field configuration that has partly reconnected with the surrounding solar wind (as indicated by open field lines) is clearly visible. 
The colour code used for the field lines reflects the misalignment between the current density and the magnetic field inside the CME body.
The regions where this misalignment is higher correspond to regions of higher Lorentz force.
As misaligned currents and magnetic fields are not present within the CME (the angle between $\vec{j}$ and $\vec{B}$ is close to zero), 
the figure provides evidence that the originally force-free spheromak configuration preserves this characteristic even after insertion in the heliosphere, and that the expansion of the CME observed in simulations is mainly due to the Lorentz force acting at the CME-solar wind interface (due to the magnetic pressure gradient). Its net result is an expansion of the CME body.

In view of the results discussed in the previous paragraph and from the consideration of Figure~\ref{fig:20120712_3d_forces_a}, our interpretation of the three simulations of this event is the following: 
\begin{itemize}
\item 
Run~01: the (cone) CME is initialised with a speed $v_{CME} = v_{3D}$ that accounts for both the translational/radial motion of the CME center of mass and for the self-similar expansion of the CME nose as it propagates outwards in the corona. As cone CMEs are characterised by an over-pressure with respect to the surrounding solar wind but have no significant internal magnetic field, force imbalances at the CME-solar wind interaction surface are expected to be mostly due to gradients in the (thermal) pressure distribution at the interface and not due to Lorentz forces. Under these conditions we can see that the evolution of the CME front is such that the CME leading edge is predicted to arrive at Earth at a time consistent with in-situ observations.
\item 
Run~02: the (spheromak) CME is initialised with a speed $v_{CME} = v_{3D}$ that accounts for both the translational/radial motion of the CME center of mass and for the self-similar expansion of the CME nose as it propagates outwards in the corona. As in this case we have a spheromak CME that is characterised by an over-pressure with respect to the surrounding solar wind but also by strong internal magnetic fields, force imbalances at the CME-solar wind interaction surface are significantly stronger due to presence of strong Lorentz forces. As a result, the CME leading edge propagates faster than in Run~01, and the CME arrives at Earth about 14 hours earlier than indicated in in-situ observations.
\item 
Run~03: the (spheromak) CME is initialised with a reduced speed $v_{CME} = v_{rad}$ that only accounts for the translational/radial motion of the CME center of mass in the corona. In this case, the presence of strong Lorentz forces inducing an expansion of the CME front (as visible from Figure~\ref{fig:20120712_3d_forces_a}) compensates for the lower translational speed used to initialise the CME body in the simulation, so that the CME leading edge propagates in the heliosphere similarly to the original cone CME run (Run~01).
\end{itemize}
The fact that the propagation of the CME leading edge is similar between the cone model simulation (Run~01) and the simulation where the spheromak is initialised using a reduced speed corresponding to the translational/radial CME speed only (Run~03), provides evidence that the otherwise faster evolution of spheromak CMEs compared to cone CMEs is mostly due to Lorentz forces leading to an expansion of the CME front.

\begin{figure}[t]
\centering
\includegraphics[width=.70\hsize]{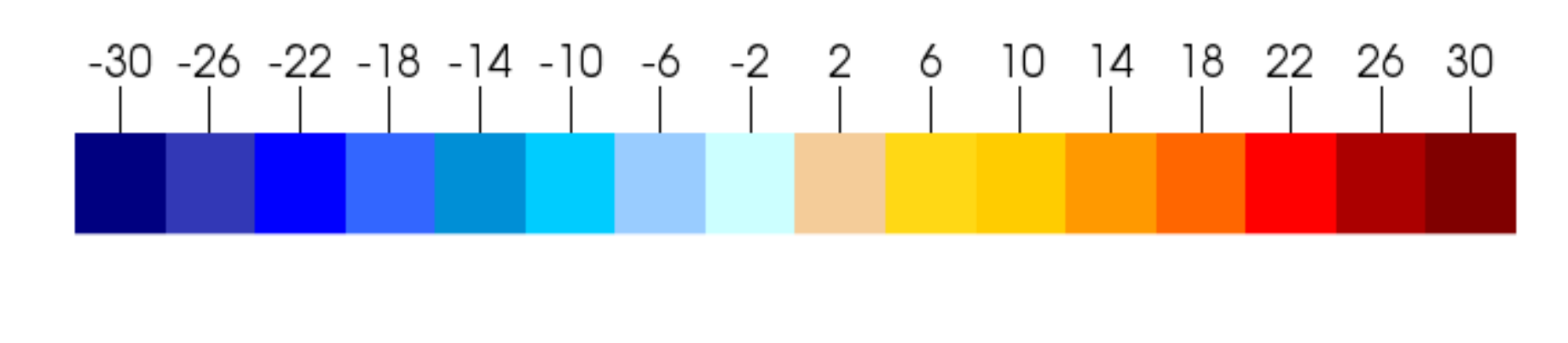} \\
\subfloat[13 July 2012 at 04:53~UT]
{   \includegraphics[width=.50\hsize,trim={20mm 40mm 60mm 40mm},clip,valign=c]{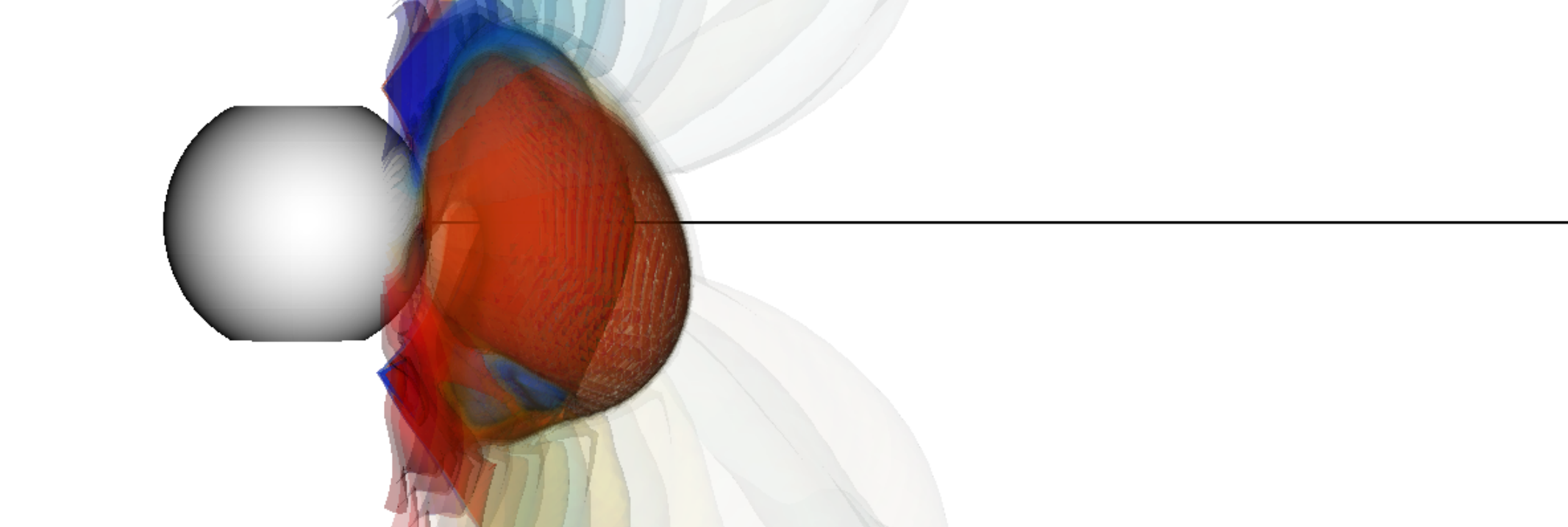} 
   \includegraphics[width=.48\hsize,trim={60mm 40mm 60mm 40mm},clip,valign=c]{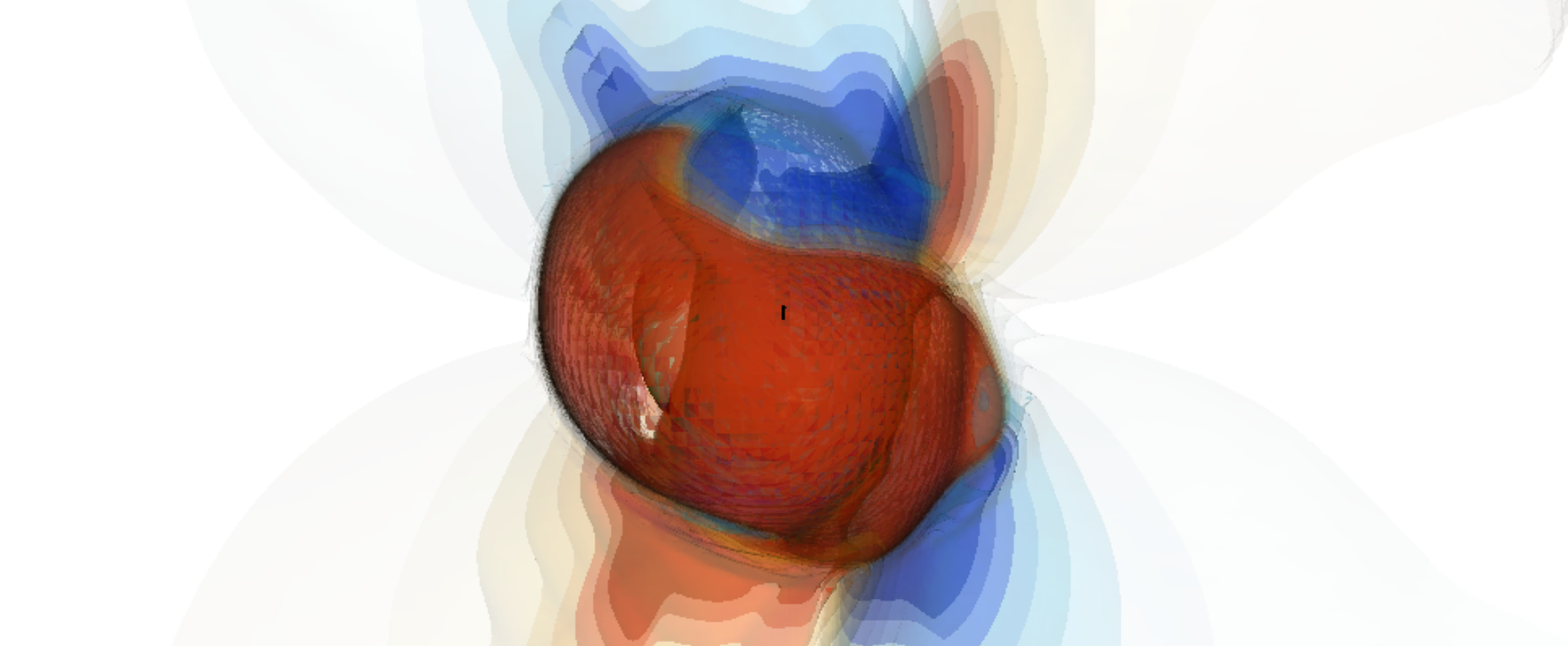} } \\
\subfloat[14 July 2012 at 00:53~UT]
{  \includegraphics[width=.50\hsize,trim={20mm 10mm 60mm 10mm},clip,valign=c]{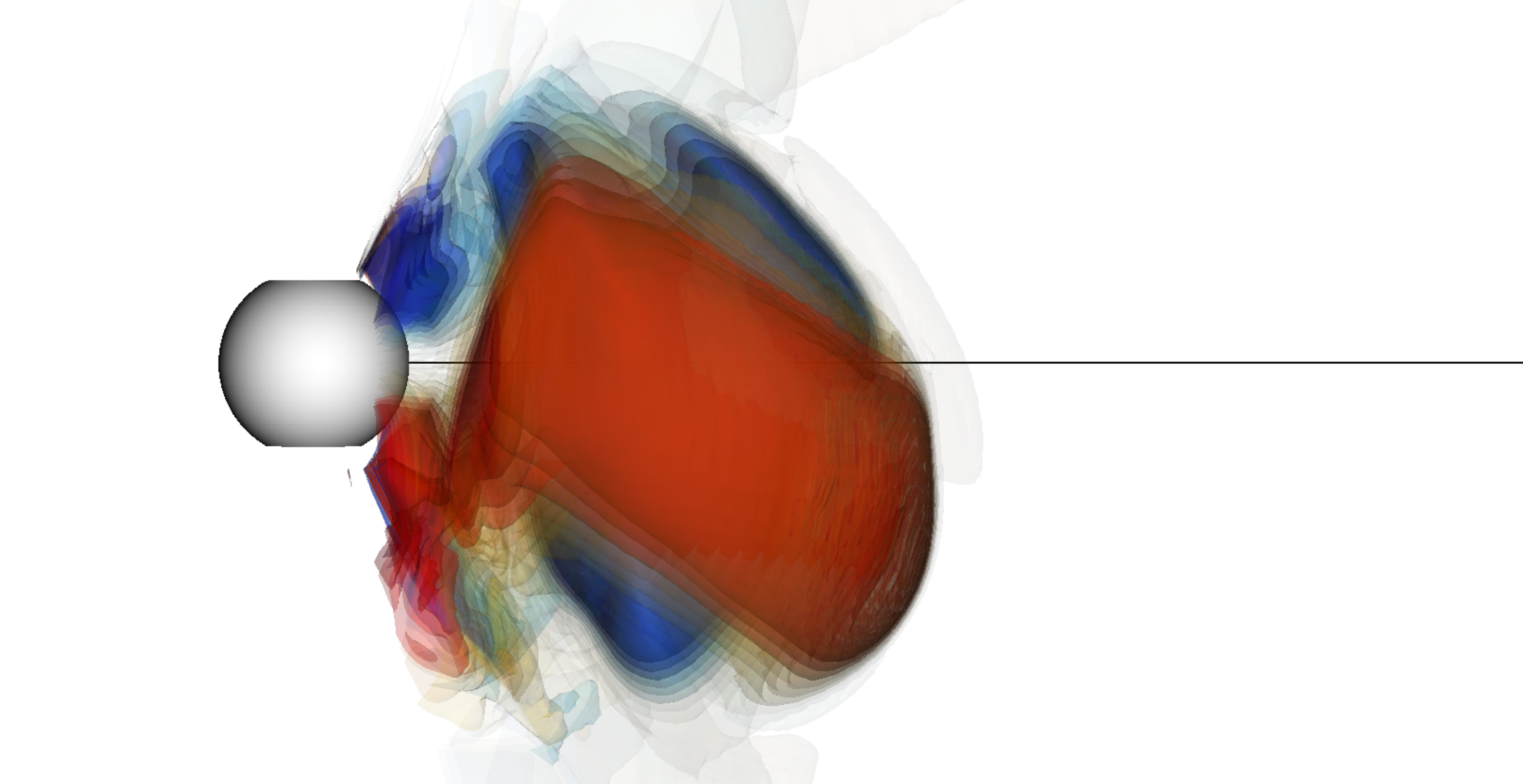} 
\includegraphics[width=.48\hsize,trim={60mm 20mm 60mm 20mm},clip,valign=c]{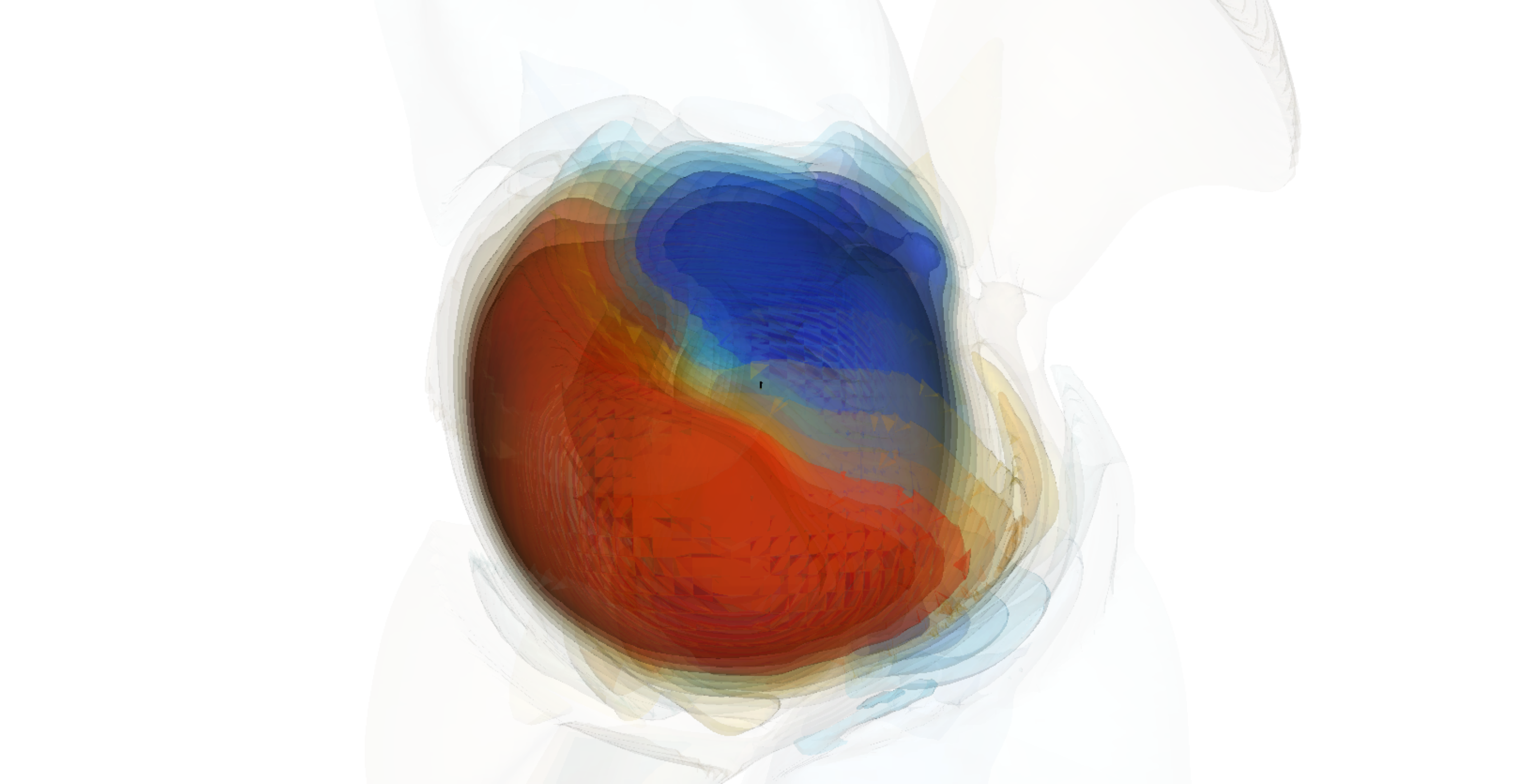} } \\
\subfloat[14 July 2012 at 20:53~UT]
{    \includegraphics[width=.50\hsize,trim={20mm 0mm 60mm 0mm},clip,valign=c]{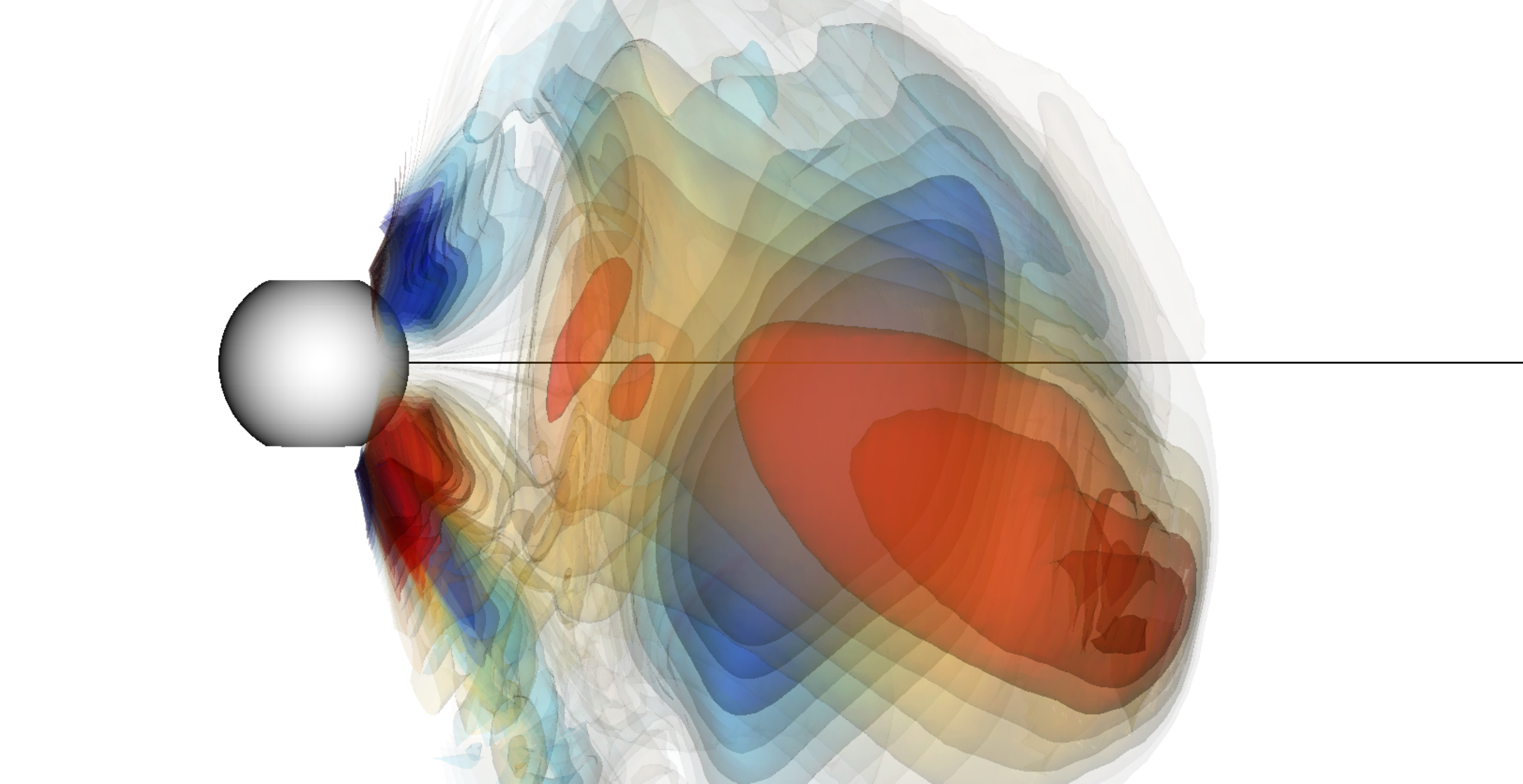} 
   \includegraphics[width=.48\hsize,trim={60mm 0mm 40mm 0mm},clip,valign=c]{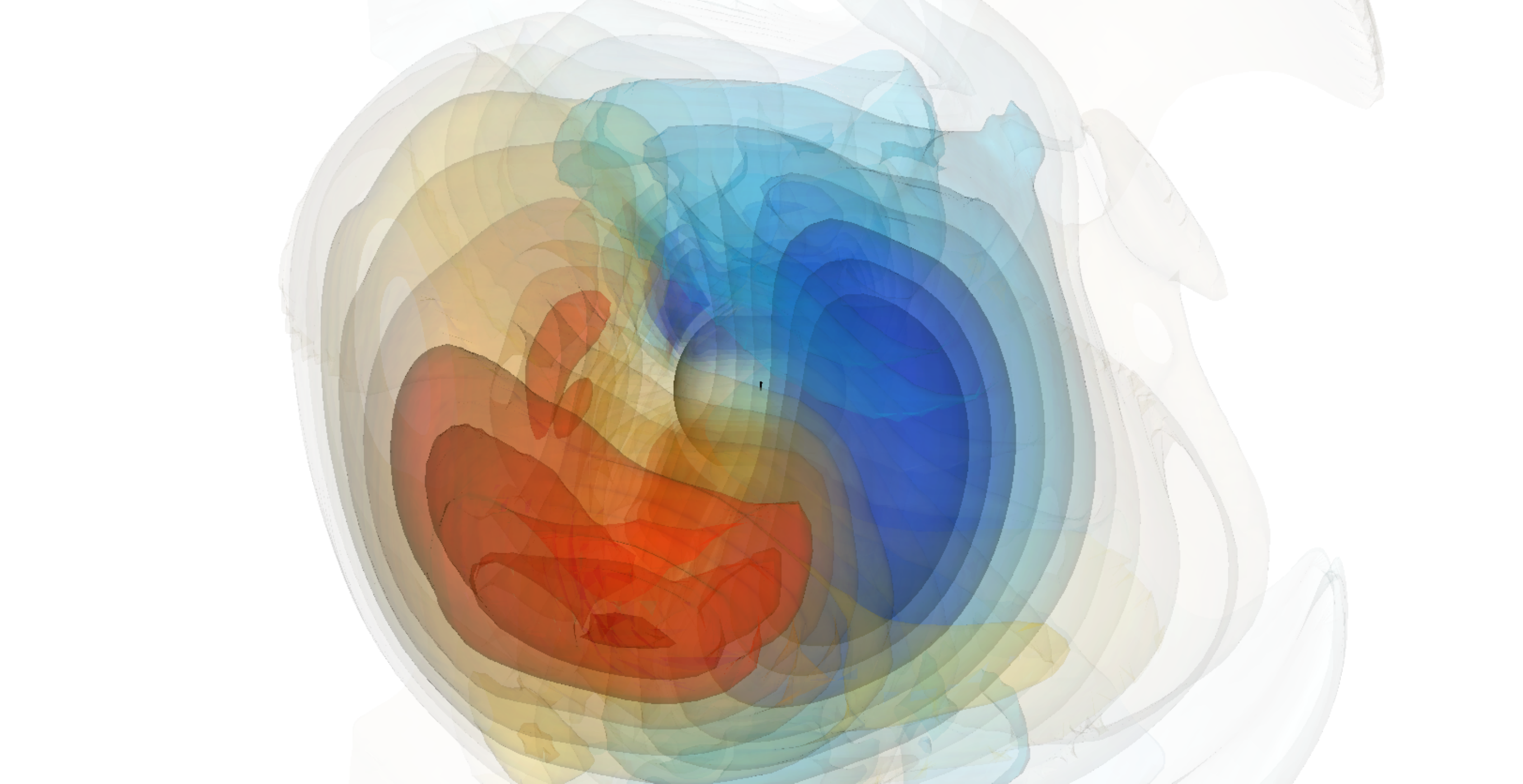} }
\caption{ 
Event 1: visualisation of the CME $B_{z}$ magnetic polarity [in nT] in Run~03, at three different times. Contour plot for $B_{z}$ as viewed from the side (left) and from Earth (right): negative and positive $B_{z}$ regions are denoted in reds and blues respectively. 
The 3D grey sphere corresponds to the heliospheric inner boundary at 0.1~AU.
The black line indicates the direction to Earth.
}
\label{fig:20120712_3d} 
\end{figure}
Figure~\ref{fig:20120712_3d} shows a 3D contour map of the different flux-rope $B_{z}$ polarity regions (northward and southward) at three different times in the simulation (from Run~03). 
Right after launch, the CME front is uniformly characterised by a positive $B_z$, 
while by the time it reaches Earth the positive $B_z$ region in the north-west part of the CME front has moved southward, so that the Earth is eventually predicted to cross a negative $B_z$ (geo-effective) region only. This is therefore a case where the use of the spheromak CME model driven by observation-based flux-rope parameters successfully predict the sign of $B_z$ at Earth.

\begin{figure}[t]
\centering
{
   \includegraphics[width=\hsize]{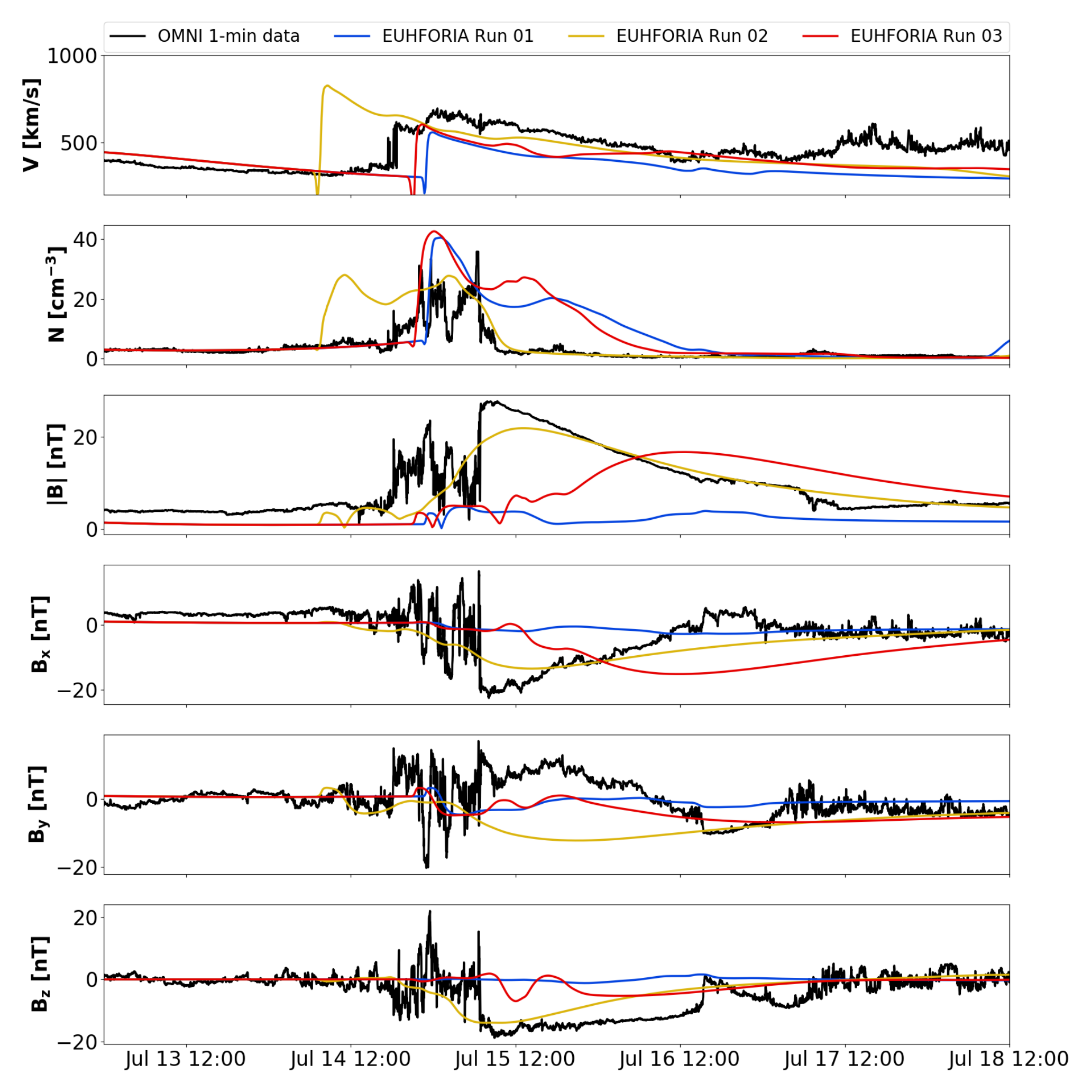} } 
\caption{Event 1: EUHFORIA time series at Earth compared to in-situ data from 1-min OMNI data (black). 
From top to bottom: speed, number density, magnetic field strength, $B_x$, $B_y$, $B_z$ components in GSE coordinates.}
\label{fig:20120712_earth} 
\end{figure}

\medskip
% EUHFORIA results at Earth
\textit{EUHFORIA predictions at Earth.}
 Figure~\ref{fig:20120712_earth} shows the simulation result at Earth, compared to in-situ measurements of the solar wind properties provided by the OMNI database.
 In the following discussion, we provide a first quantification of the prediction improvements associated to the use of the spheromak model focusing on CME ToA and ICME peak values of the magnetic field components in time only.
We leave out from the discussion other relevant metrics recently identified by the community \citep{owens:2018, verbeke:2019a}, 
as a detailed comparison of the different metrics used in operational forecasts goes beyond the scope of this work.
However, we point out that the use of such metrics could certainly provide a more complete quantification of the prediction improvements associated to the spheromak model, and highlight additional strengths and limitations that could be valuable for operational uses. We plan to further address the topic in future publications.

% arrival times
The CME arrival times at Earth for different runs are listed in Table~\ref{tab:event_1}.
Comparing the CME ToA at Earth from Run~01 with that from Run~02, the impact of the Lorentz force on the CME propagation to 1.0~AU is immediately clear, resulting in a difference of about 14 hours.
After reducing the speed of the spheromak CME (Run~03) so to account for the internal pressure introduced by the internal magnetic field, the predicted CME ToA is in good agreement with observations of the ICME-driven shock arrival time from OMNI data ($\sim 3$ hours from the observed shock time)
as well as with the CME ToA prediction from the cone model ($\sim 1$ hour difference in the ToA).
For comparison, the current typical error on the prediction of CME ToAs at Earth is $\pm 6$ hours, as recently reported by \citet{riley:2018} considering 32 different models.

% time series shapes
Comparing results from Run~01 and Run~03 with observations, it is evident that the EUHFORIA model is most successful in the speed and number density profiles, while the prediction of the magnetic field signatures is significantly more challenging.
Looking at the time series shapes, we observe that while the arrival time of the shock is well predicted by the spheromak model, there is a time delay in the peak of the magnetic field compared to the observations. This is possibly due to an overestimation of the CME radial size in the simulation, e.g. the model does not account for possible flattening or "pancaking" effects occurring already at distances comparable to the heliospheric inner boundary \citep{riley:2004,savani:2011,isavnin:2016}.
Extending the model to allow for some CME shape deformations to be specified at the simulation inner boundary, such as an elongation in the longitudinal and/or latitudinal directions in order to make the spheromak elliptical, could improve the time series for the magnetic field components, i.e. by compressing the signal in time.

% magnetic field peak values
Investigating the prediction of the IMF components (${B}$, $B_x$, $B_y$ and $B_z$) associated with the ICME, 
the cone model (Run~01), as expected, is unable to predict any magnetic signature associated with the magnetic could simply due to the lack of an internal magnetic structure in the modelled CME.
On the other hand, the use of a spheromak model improves significantly the prediction of the ICME magnetic field properties in terms of peak values.
While the observed maximum $B$ during the passage of the MC was 27~nT, 
the prediction of the cone CME model (Run 01) was $4$~nT (corresponding to $\sim 4\%$ of the observed maximum $B$),
and the one of the spheromak CME model (Run~03) was $16$~nT (corresponding to $\sim 60\%$ of the observed maximum $B$).
The prediction of the negative $B_z$ signature is considered to be the most important for prediction of the CME impact on Earth. 
In this case the observed minimum $B_z$ during the passage of the MC was -18~nT, while the minimum $B_z$ predicted from the cone model was negligible ($\sim -1$~nT), and the one of the spheromak model was $-7$~nT (corresponding to $\sim 40\%$ of the observed minimum $B_z$).
As peak values for $B$ and $B_z$ in simulated in-situ time series do not necessarily occur at the same time of the observed ones, to extract the predicted peak values we considered the whole time series presented in Figures \ref{fig:20120712_euhforia}.

\begin{figure}[h]
\centering
{
   \includegraphics[width=\hsize]{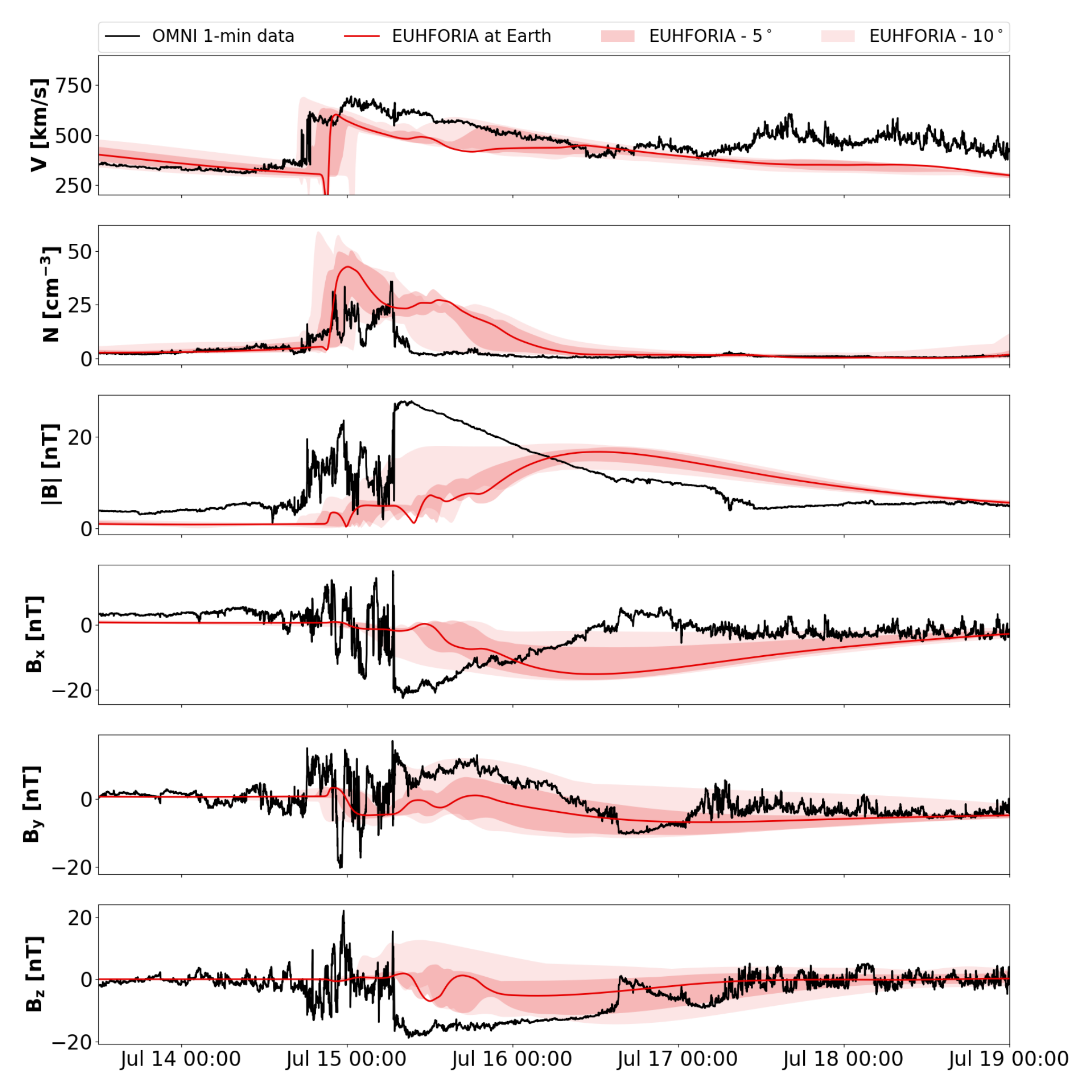}  } 
\caption{Event 1: EUHFORIA time series from Run~03 at Earth (red), compared to in-situ data from 1-min OMNI data (black).
The dark red and light red shaded areas show the maximum variation of EUHFORIA predictions 
at positions separated by $5^\circ$ and $10^\circ$ in longitude and/or latitude from Earth.
From top to bottom: speed, number density, magnetic field strength, $B_x$, $B_y$, $B_z$ components in GSE coordinates.
}
\label{fig:20120712_around_earth} 
\end{figure}

\medskip
% points around the Earth
\textit{Location sensitivity.}
In Figure~\ref{fig:20120712_around_earth} we show the EUHFORIA time series at Earth from Run~03, 
together with shaded areas indicating the variability of the plasma parameters in the vicinity of Earth, namely at virtual spacecraft located at $\pm 5^\circ$ and $\pm 10^\circ$ in longitude and/or latitude from Earth. 
As $\pm 10^\circ$ is the typical uncertainty associated to the CME direction of propagation (longitude/latitude) as reconstructed from the GCS model \citep{thernisien:2009}, considering outputs at those specific virtual spacecraft is a way to account for the uncertainty associated to the CME direction of propagation (in longitude/latitude).
Considering the range of predictions in this area around Earth, the CME ToA spans of $\pm 2$ hours around the ToA at Earth for the closer spacecraft ($\pm 5^\circ$ separation from Earth), and of $\pm 5$ hours around the ToA at Earth for the outer spacecraft ($\pm 10^\circ$ separation). 
The prediction for the maximum $B$ in the MC ranges between $15$~nT and $18$~nT at spacecraft located at $\pm 5^\circ$ from Earth,
and between $13$~nT and $19$~nT at spacecraft located at $\pm 10^\circ$.
The minimum $B_z$ in the MC ranges between $-4$~nT and $-12$~nT at spacecraft located at $\pm 5^\circ$ from Earth, and between $-2$~nT and $-14$~nT at spacecraft located at $\pm 10^\circ$.
In summary, considering the prediction at spacecraft located around the Earth at angular separations within the uncertainty of the CME direction of propagation derived from the GCS model, the best prediction accounts for $\sim 70\%$ of the maximum $B$ measured in the MC, 
and for $\sim 78\%$ of the minimum $B_z$ measured in the MC. 
The figure provides an indication of the sensitivity of the model to uncertainties related to the propagation direction of the CME, and similar results have previously been reported for EUHFORIA and other heliospheric MHD models \citep{verbeke:2019,torok:2018}.

%===========================================
\subsection{Event 2: CME on 14 June 2012} 
\label{subsec:20120614_results}
%===========================================

As presented in Section~\ref{subsec:20120614}, the second event studied in this work was composed of two CMEs that erupted from the same AR on two consecutive days (13 and 14 June 2012).

From the GCS reconstruction, CME1 had a direction of $\theta = -35^\circ$ and $\phi = -20^\circ$, with a half width of $26^\circ$.
In-situ observations show that at Earth it was associated with a shock followed by a long-lasting sheath region preceding the arrival of a second shock associated with CME2. 
Based on these arguments, this CME appears to have propagated significantly away from the Sun-Earth line. Therefore, we decide to simulate CME1 by means of a cone CME model only, as no flux-rope signature was observed at Earth.

\begin{figure*}[t]
\centering
\subfloat[]
{\includegraphics[width=.50\hsize, trim={20mm 47mm 0mm 0mm},clip]{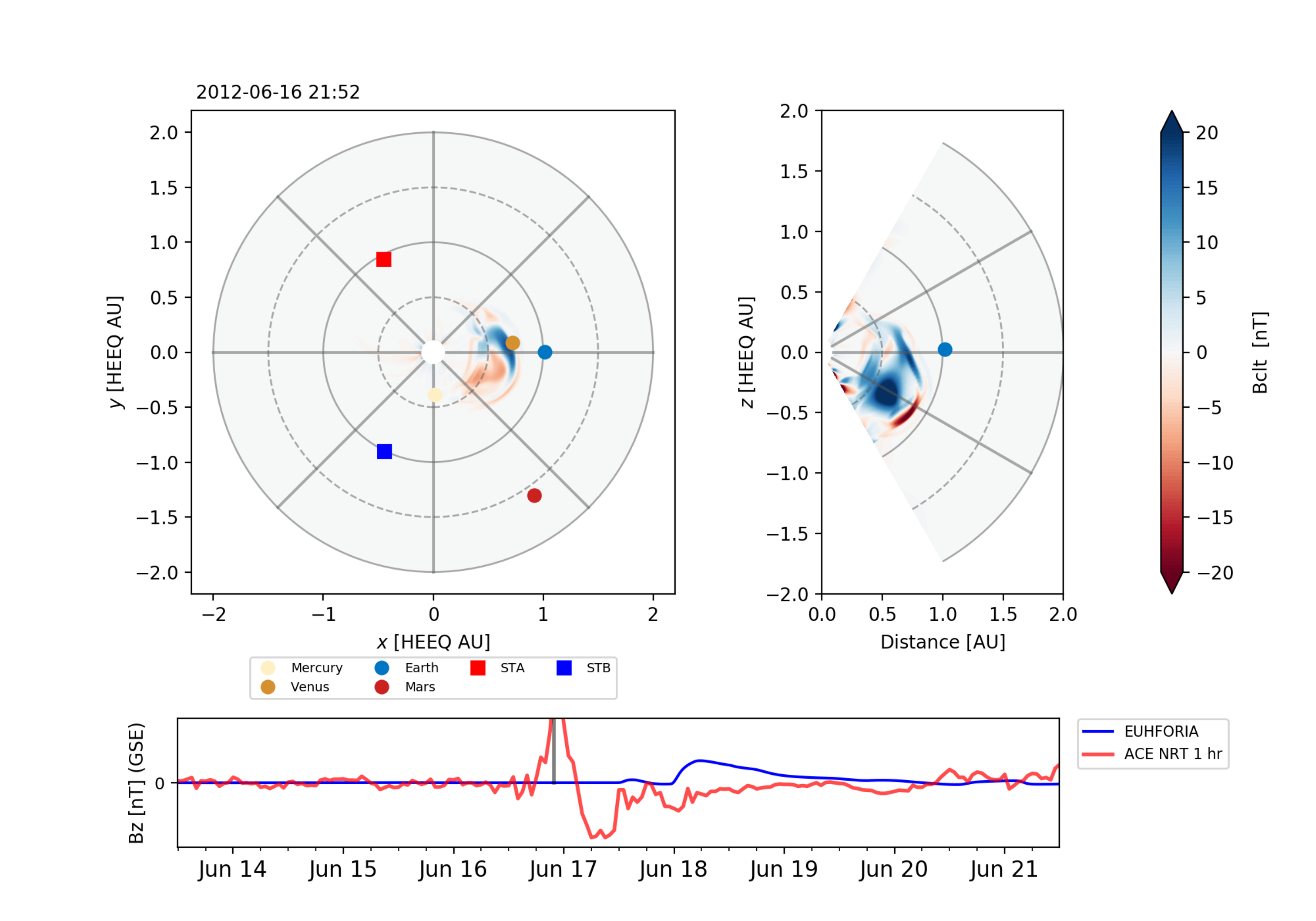} } 
\subfloat[]
{\includegraphics[width=.50\hsize, trim={20mm 47mm 0mm 0mm},clip]{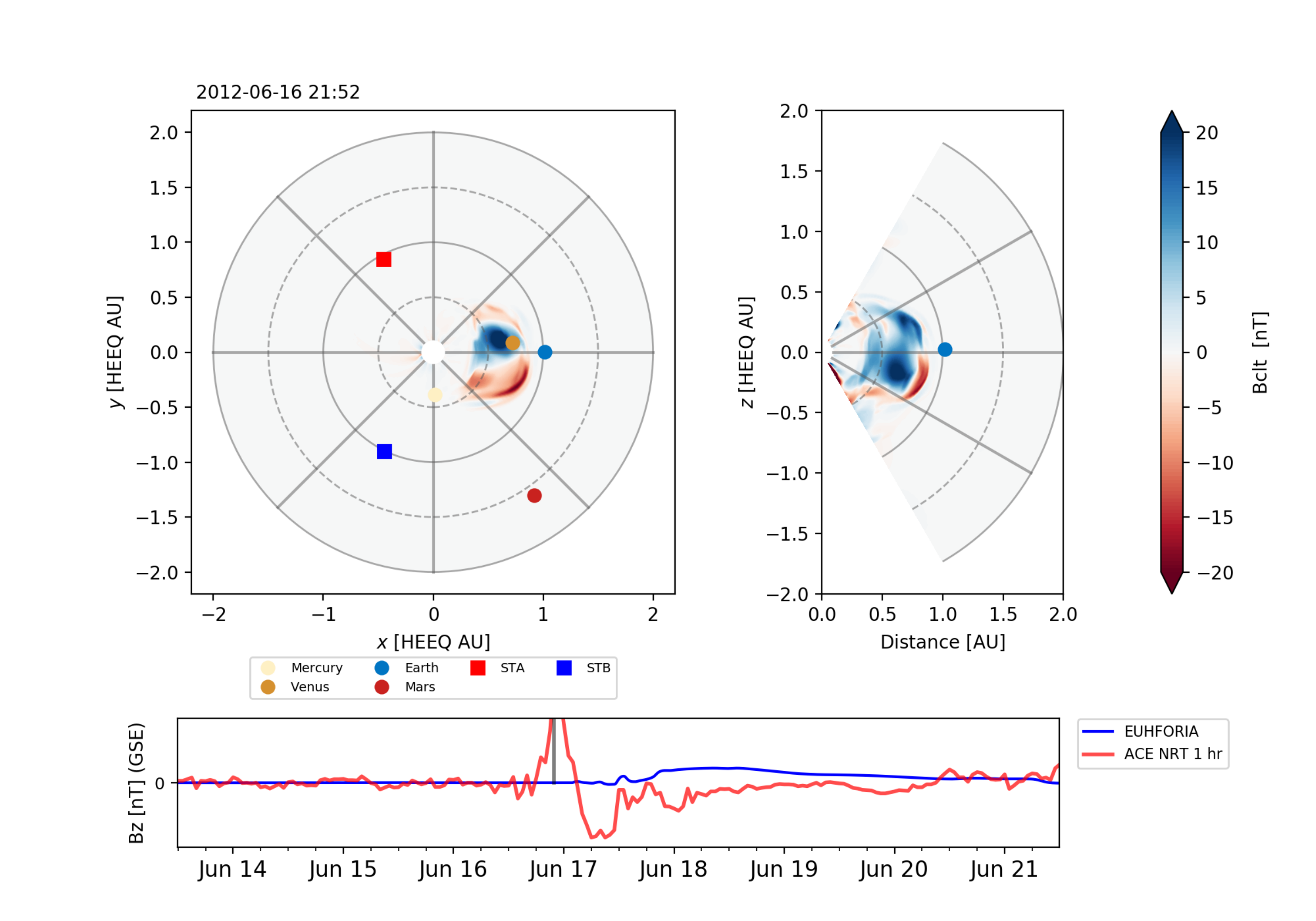} } \\
\subfloat[]
{\includegraphics[width=.45\hsize, trim={40mm 10mm 40mm 16mm},clip]{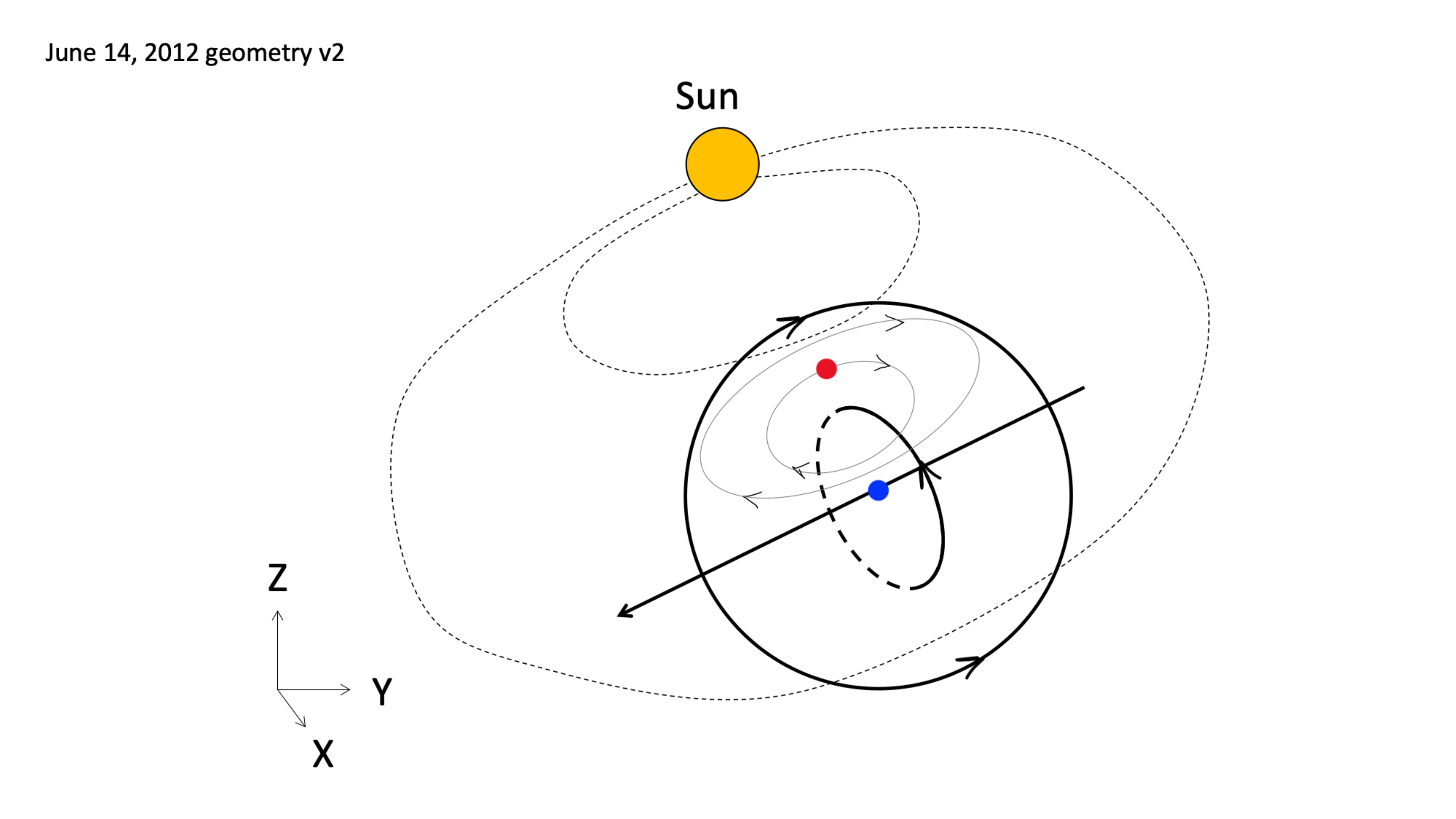} } \qquad
\subfloat[]
{\includegraphics[width=.45\hsize, trim={40mm 10mm 40mm 16mm},clip]{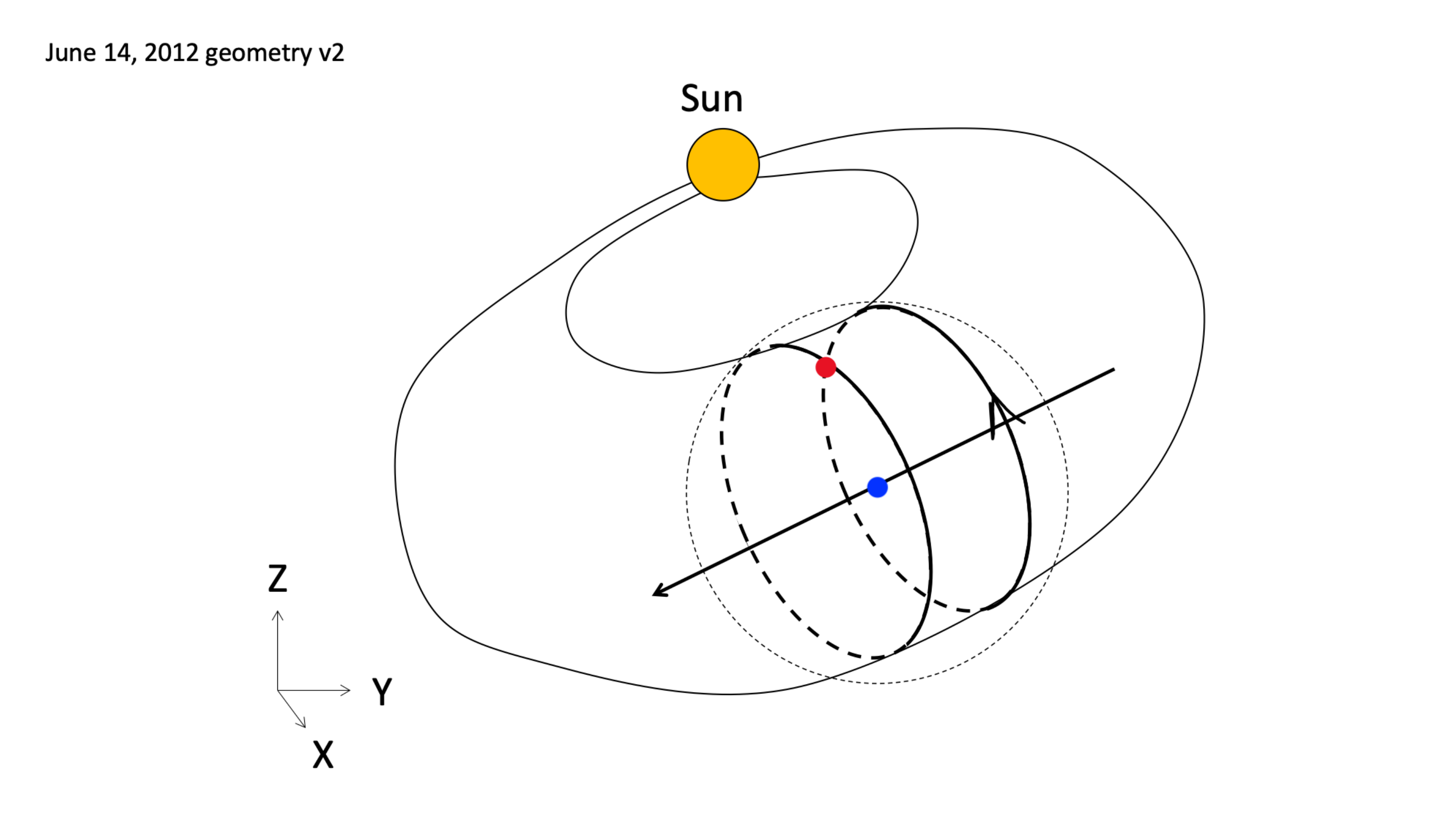} } 
\caption{Event 2: 
EUHFORIA snapshots of Run~03 (a) and Run~05 (b):
$B_{clt}$ in the heliographic equatorial plane and in the meridional plane that includes the Earth.   
(c): sketch showing the trajectory of the Earth through the spheromak magnetic structure (continuous black line) in Run~03 (red dot) and Run~05 (blue dot). The dashed black line shows the flux-rope structure as an elliptical object connected to the Sun. 
(d): sketch showing the magnetic structure of an elliptical flux-rope (continuous black line), in relation to the Earth trajectory in the two runs. Here the dashed black line marks the spheromak structure.
The actual trajectory of Earth is consistent with that marked by the red dot.}
\label{fig:20120614_euhforia_bz} 
\end{figure*}

\begin{figure}
\centering
\subfloat[Radial speed $v_r$]
{\includegraphics[width=\hsize,trim={20mm 0mm 0mm 0mm},clip]{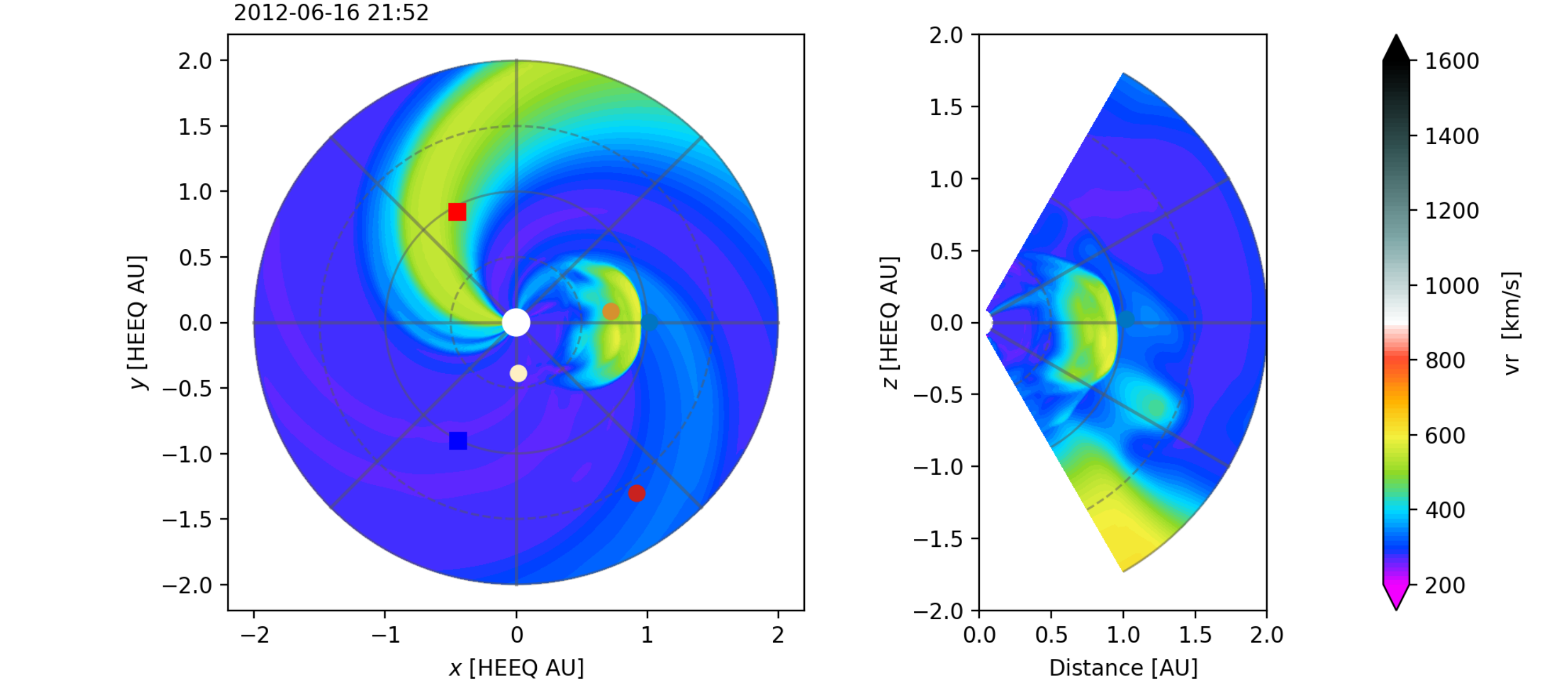} } \\
\subfloat[Scaled number density $n \big (\frac{r}{1 \,\, \mathrm{AU}} \big )^2 $ ]
{ \includegraphics[width=\hsize,trim={20mm 0mm 0mm 0mm},clip ]{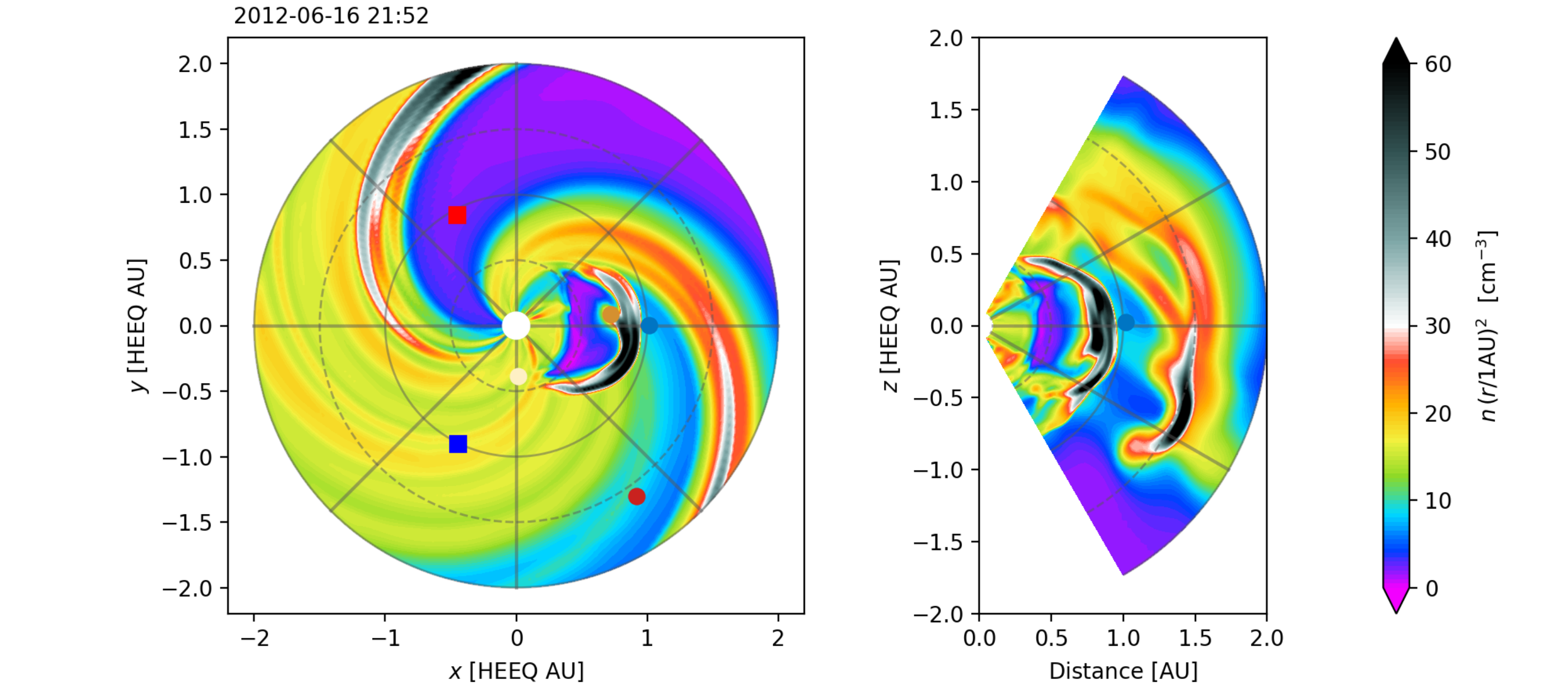} } \\
\subfloat[Co-latitudinal magnetic field $B_{clt}$]
{ \includegraphics[width=\hsize,trim={20mm 0mm 0mm 0mm},clip]{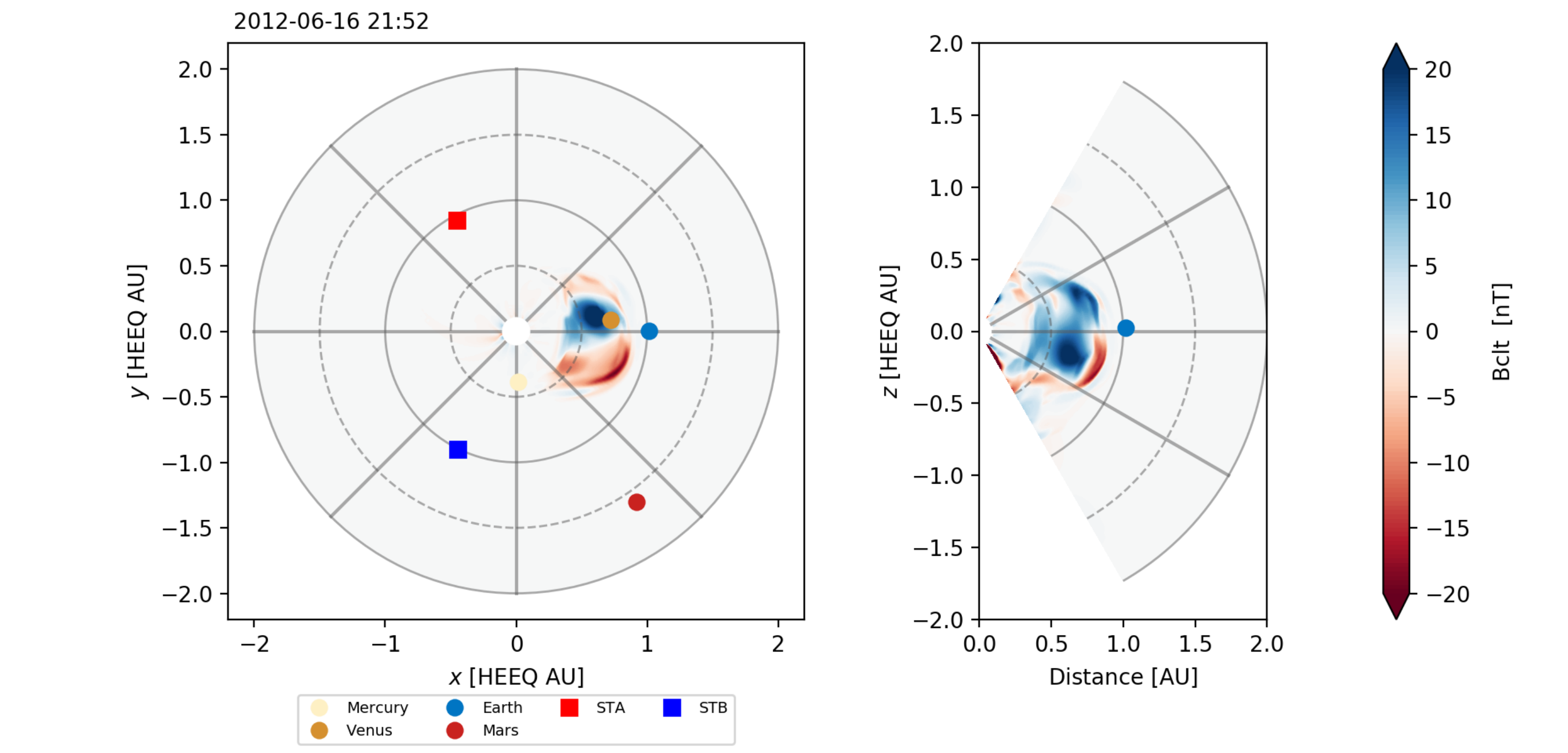}} 
\caption{Event 2: snapshot of the EUHFORIA Run~05 (spheromak CME with reduced speed $v_\mathrm{CME}=v_{rad}$, launched at $\theta = 0^\circ$) on June  16,  2012  at  21:52~UT  in the heliographic equatorial plane (left) and in the meridional plane that includes the Earth (right). }
\label{fig:20120614_euhforia} 
\end{figure}

On the other hand, as CME2 exhibited clear flux-rope signatures at Earth, we simulate it by means of both a cone and a spheromak CME model.
As for Event 1, we first simulate CME2 using the cone model (Run~01), 
using as input parameters the results of the GCS reconstruction.
We then perform a second simulation (Run~02) modelling CME2 as a spheromak CME
(using the three magnetic parameters as determined in Section~\ref{subsec:magnetic_parameters}), 
and keeping the kinematic/geometric CME parameters as in Run~01.
Finally, we perform a third simulation of CME2 using the spheromak model, 
but imposing a reduced speed determined as $v_\mathrm{CME} =  v_{3D} - v_{exp} = v_{rad}$ (Run~03).
Based on the coronagraph data, CME2 was observed to propagate south of the ecliptic plane, at a latitude of about $-25^\circ$ below the solar equatorial plane, and it was characterised by a half width of about $40^\circ$.
As shown in Figure~\ref{fig:20120614_euhforia_bz} (a), due to the spherical shape of the spheromak model, 
launching CME2 at the latitude derived from the GCS reconstruction ($\theta=-25^\circ$) the bulk of the ejecta magnetic field is modelled to propagate south of the equatorial plane, e.g. the model predicts a flux-rope tangential encounter with Earth, which does not correspond to reconstructions of the flux-rope configuration based on in-situ data (see discussion in Section~\ref{subsec:20120614}).
In order to reproduce the in-situ flux-rope configuration, we run a fourth simulation, using the spheromak CME model and launching the CME along the Sun-Earth line ($\theta= 0^\circ$) with a speed $v_{CME} = v_{3D}$ (Run~04). 
In this case, the bulk of the magnetic field propagates in the equatorial plane, reproducing the observed ICME properties in a more realistic way (Figure~\ref{fig:20120614_euhforia_bz}, d). 
A sketch of the trajectories of Earth across the spheromak CME in the two runs is given in Figure~\ref{fig:20120614_euhforia_bz} (c). 
A comparison with Figure~\ref{fig:20120614_euhforia_bz} (d) shows that the spheromak magnetic field is similar to one of an elliptical flux-rope only when a spacecraft crosses the CME near its center, while the two are very different in all their components in the case of off-axis encounters.
While the observed trajectory of Earth through the ICME is consistent with the red dot in Figure~\ref{fig:20120614_euhforia_bz} (d), 
the need to artificially launch the CME at $\theta= 0^\circ$ is justified by the fact that the spheromak model is incapable of reproducing the magnetic structure of an elliptical flux-rope away from its axis.
In a final simulation (Run~05), we simulate the spheromak CME as in Run~04, e.g. launched directly on the ecliptic plane, using a reduced speed determined as $v_\mathrm{CME} =  v_{3D} - v_{exp} = v_{rad}$.
Table \ref{tab:event_2} lists the input parameters used to simulate CME1 and CME2 with the cone and spheromak models.
In all five cases, as input for the coronal model we use the GONG synoptic standard map observed on 14 June 2012 at 11:54~UT.
In this case, the density ratio within the CME bodies is approximately 0.8 with respect to the surrounding solar wind values, while the pressure ratio is about 3.9.
%
% Table
\begin{table*}
\centering
\begin{tabular}{l|l|lll}
\hline
\hline
                        & \multicolumn{1}{c|}{CME1} & \multicolumn{3}{c}{CME2}\\
\hline
                        & All runs & Run~01 & Run~02 (Run~03) & Run~04 (Run~05) \\
\hline
CME model                  & {cone} & cone  & spheromak & spheromak\\
Insertion time         & {2012-06-13T19:38} & 2012-06-14T16:55 & 2012-06-14T16:55 & 2012-06-14T16:55  \\
$v_\mathrm{CME}$       & {719 \si{ \km \,\, \s^{-1} }}  
                       & 1213 \si{ \km \,\, \s^{-1} } 
                       & 1213 \si{ \km \,\, \s^{-1} } (713 \si{ \km \,\, \s^{-1} })
                       & 1213 \si{ \km \,\, \s^{-1} } (713 \si{ \km \,\, \s^{-1} })\\ 
$\phi$                 & {$-20^\circ$ } & $-5^\circ$ & $-5^\circ$ & $-5^\circ$ \\   
$\theta$               & {$-35^\circ$ } & $-25^\circ$ & $-25^\circ$ & $0^\circ$  \\      
$\omega/2$             & {$26^\circ$} & $40^\circ$ & - & - \\   
$r_0$                    & - & - &  $18.0 \,\, R_s$ &  $18.0 \,\, R_s$ \\  
$\rho$                 & {$1 \cdot 10^{-18} \,\, \si{ \kg \,\, \m^{-3} }$ }
                       & $1 \cdot 10^{-18} \,\, \si{ \kg \,\, \m^{-3} }$ 
                                & $1 \cdot 10^{-18} \,\, \si{ \kg \,\, \m^{-3} }$ 
                                & $1 \cdot 10^{-18} \,\, \si{ \kg \,\, \m^{-3} }$\\ 
$T$                    & {$0.8 \cdot 10^{6} \,\, \si{\kelvin}$ }
                       & $0.8 \cdot 10^{6} \,\, \si{\kelvin}$ 
                        & $0.8 \cdot 10^{6} \,\, \si{\kelvin}$ 
                        & $0.8 \cdot 10^{6} \,\,  \si{\kelvin}$\\ 
$H$                    & - & - & +1 & +1\\ 
Tilt                   & - & - & $-120^\circ$ & $-120^\circ$ \\ 
$\phi_t$               & - & -  & $4.0 \cdot 10^{13}$~Wb & $4.0 \cdot 10^{13}$~Wb  \\ 
\hline
Predicted ToA at Earth   & - &   2012-06-16T23:32   &  2012-06-16T12:53 & 2012-06-16T04:02\\
                        &   &   &   (2012-06-17T12:32) & (2012-06-17T01:32)\\
 \hline
\end{tabular}
\caption{CME input parameters used in the EUHFORIA simulations of the 13-14 June 2012 CMEs, 
and their predicted arrival times at Earth.}
\label{tab:event_2}
\end{table*}
An example of simulation results from Run~05 is provided in Figure~\ref{fig:20120614_euhforia}, which shows a snapshot on the ecliptic and meridional planes containing the Earth, of the radial speed, scaled number density and $B_{clt}$ component of the magnetic field (see supplementary material for movies of the dynamics).

\medskip
% EUHFORIA results in the heliosphere
\textit{CME propagation in the heliosphere.}
\begin{figure}[t]
\centering
{\includegraphics[width=\hsize,trim={0mm 59.5mm 0mm 0mm},clip]{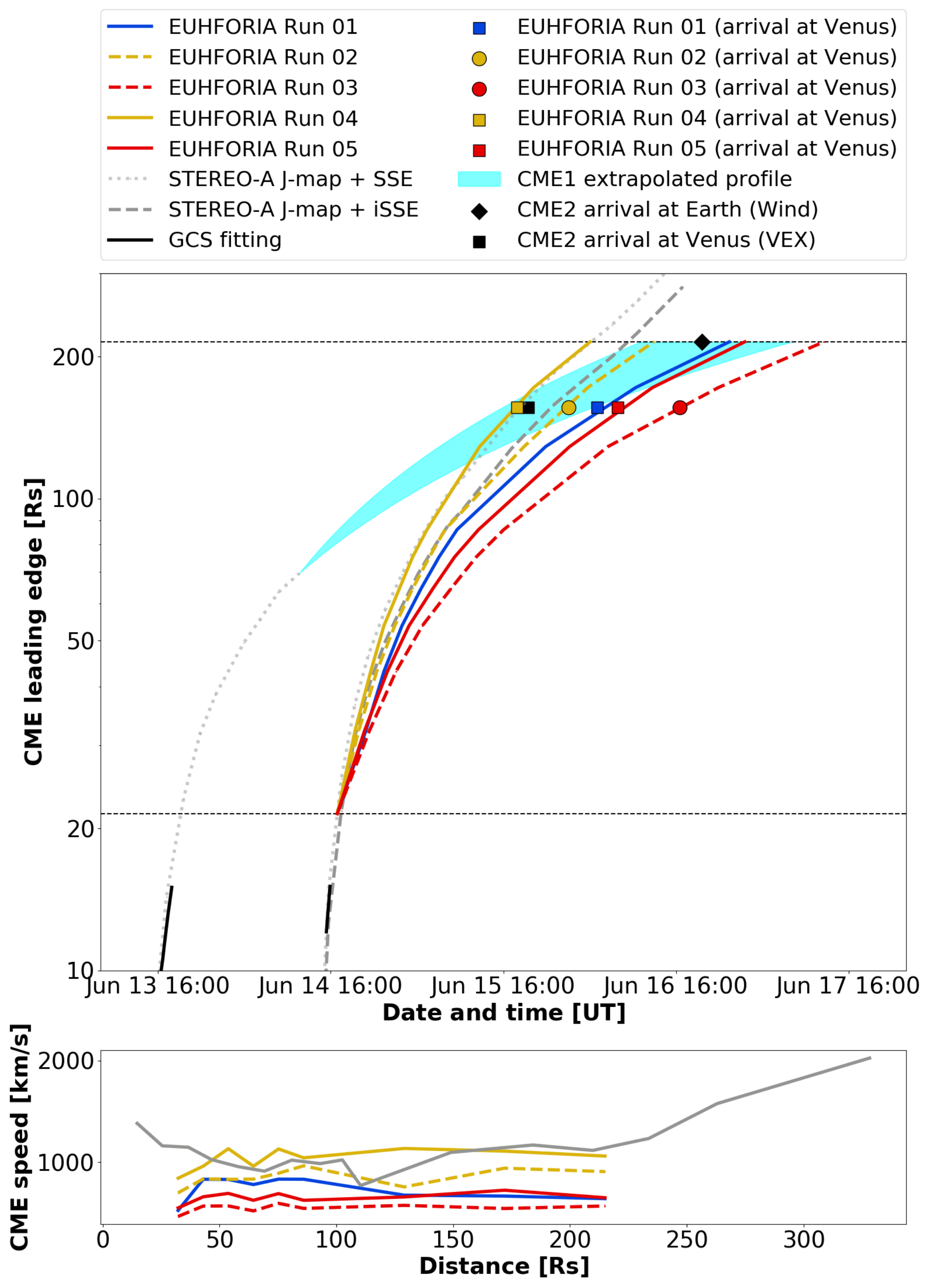} } 
\caption{Event 2: comparison of the observed and modelled propagation of CME1 and CME2 in the heliosphere.
%Top: position of the CME leading edges in time. 
The black dashed lines mark the EUHFORIA heliospheric inner boundary at 0.1~AU and 1.0~AU.
%Bottom: speed of the CME1 leading edge as function of the heliocentric distance.
}
\label{fig:20120614_propagation} 
\end{figure}
Similarly to Section~\ref{subsec:20120712_results}, in Figure~\ref{fig:20120614_propagation} 
we compare height-time profiles of the CME leading edge from simulations, to that derived from STEREO J-maps.
For this event, regardless of the CME latitude used, the spheromak CMEs initialised using the full 3D speed ($v_\mathrm{CME}=v_{3D}$; Run~02 and 04, yellow curves) propagate faster than the cone CME (Run~01, blue curve), while the spheromak CMEs initialised using the reduced speed $v_\mathrm{CME}=v_{rad}$ (Run~03 and 05, red curves) propagate slower.
The propagation of the CME-driven shock along the Sun-Earth line is very similar
between Run~01 (cone with $v_\mathrm{CME}=v_{3D}$, blue curve), 
and Run~05 (spheromak with $v_\mathrm{CME}=v_{rad}$ launched on the ecliptic plane, continuous red curve).
On the other hand, in simulations where the spheromak CME is initialised using $v_\mathrm{CME}=v_{3D}$ the propagation is faster already very early in the simulation (i.e. Run~02, spheromak with $v_\mathrm{CME}=v_{3D}$, dashed yellow curve, and Run~04, spheromak with $v_\mathrm{CME}=v_{3D}$ launched on the ecliptic plane, continuous yellow curve).
The only difference between Run~01 and Run~02 is the internal magnetic field used to initialise the CME in the model, hence the differences in the CME propagations in these two simulations reflect the impact of the Lorentz force on the propagation of the CME.
Run~02 and Run~03 (and similarly Run~04 and Run~05) differ in the speeds used to initialise the flux-rope CME in the model. 
As for Event 1, the fact that the propagation of the CME in Run~05 is similar to the one observed for Run~01 indicates that 
(a) the faster propagation of Run~04 is due to the internal magnetic pressure acting within the spheromak CME as an expansion force,
and that (b) the separation between the radial and expansion speed contributions from coronagraph observations proposed in Section~\ref{subsec:gcs} can be used as observational proxy to assess the CME expansion in the corona.

Time-height profiles based on STEREO-A J-maps and the SSE technique allow to estimate the propagation of the CME2 leading edge along the Sun-Earth line as if it was corresponding to the apex of the CME.
In this case, results for the time-height profile appear similar to EUHFORIA Run~04.
This is consistent with the fact that in Run~04 the CME was launched at $\theta = 0^\circ$, so that in the simulation the CME was in fact directed along the Sun-Earth line.
As for Event 1, we notice that this observation-based method estimates the CME arrival time to occur
around 04:00~UT on 16 June, i.e. about 14 hours earlier than observed in situ.
However, CME2 was actually propagating about $25^\circ$ away from the Sun-Earth line, as recovered by the GCS reconstruction. In this case, the iSSE technique is considered to provide a more accurate approximation of the propagation of the portion of the CME leading edge that actually propagated towards the Earth.
As visible from Figure~\ref{fig:20120614_propagation}, 
the iSSE reconstruction matches with Run~02, when the CME was actually launched at $\theta=-25^\circ$ south of the Sun-Earth line.
In this case, the CME arrival time is estimated to occur
around 07:00~UT on 16 June, i.e. about 12 hours earlier than observed in situ.

According to simulations, the interaction of CME1 and CME2 already took place before the time when the CMEs arrive at Venus/VEX (see supplementary material for movies of the dynamics), and as the interaction was limited to a region well below the ecliptic, only CME2 was predicted to arrive at Venus.
These results are in agreement with the observational analysis conducted by \citet{kubicka:2016}.
EUHFORIA Run~01 predicts the CME2 ToA to occur on 16 June 2012 at 05:03,
about 9 hours and a half later than reported by \cite{good:2016} based on VEX data.
All the other runs predict CME2 to arrive even later.
In detail, Run~02 and Run~04 predict CME2 to arrive on 16 June 2012 at 01:03 and on 15 June 2012 at 17:53 respectively,
while Run~03 and Run~05 predict CME2 to arrive on 16 June 2012 at 16:32 and on 16 June 2012 at 07:03 respectively.
Although a complete understanding of the exact reasons for such unsatisfactory prediction of the CME ToA at VEX (more than 9 hours off compared to in-situ observations) would require further investigations, as discussed below we note that this result may depend on limitations in the modelling of the CME-CME interactions using a cone-spheromak model combination for the two CMEs under consideration.

\medskip
% CME-CME interactions
\textit{CME1-CME2 interactions.}
Run~01 and Run~05 best reproduce the arrival time at Earth of CME2.
In contrast to Run~05, in Run~01 the CME is launched in the direction reconstructed from the GCS model. Looking at the global evolution of CME1 and CME2 in that simulation, 
we define their interaction as marked by the moment when the leading edge of CME2 catches up with the leading edge of CME1. 
This happens on 15 June 2012 around 17:52~UT, when the leading edge of CME1 is at about 0.65~AU from the Sun ($\simeq 140 \,\, R_s$). 
Extrapolating the propagation of the CME1 leading edge obtained from the J-maps tracking and the SSE technique, we also conclude that the interaction between CME1 and CME2 most probably occurred between $\sim 120$~$R_s$ and $\sim 160$~$R_s$ (Figure~\ref{fig:20120614_propagation}).  
These results are approximately in agreement with the result obtained by \citet{srivastava:2018}, who report the interaction to occur around $ 100 \,\, R_s$ based on observations of the CME leading edges in STEREO-A J-maps on the ecliptic plane. 

The fact that we modelled CME2 only as a flux-rope structure may be at the origin of the differences between the interaction distance found in our simulations with that reported by \citet{srivastava:2018}, while a more rigorous investigation of the CME-CME interaction would require the modelling of both CME1 and CME2 using a flux-rope model. A detailed study of the CME-CME interaction with EUHFORIA will be addressed in future studies.

\begin{figure}[t]
\centering
{
   \includegraphics[width=\hsize]{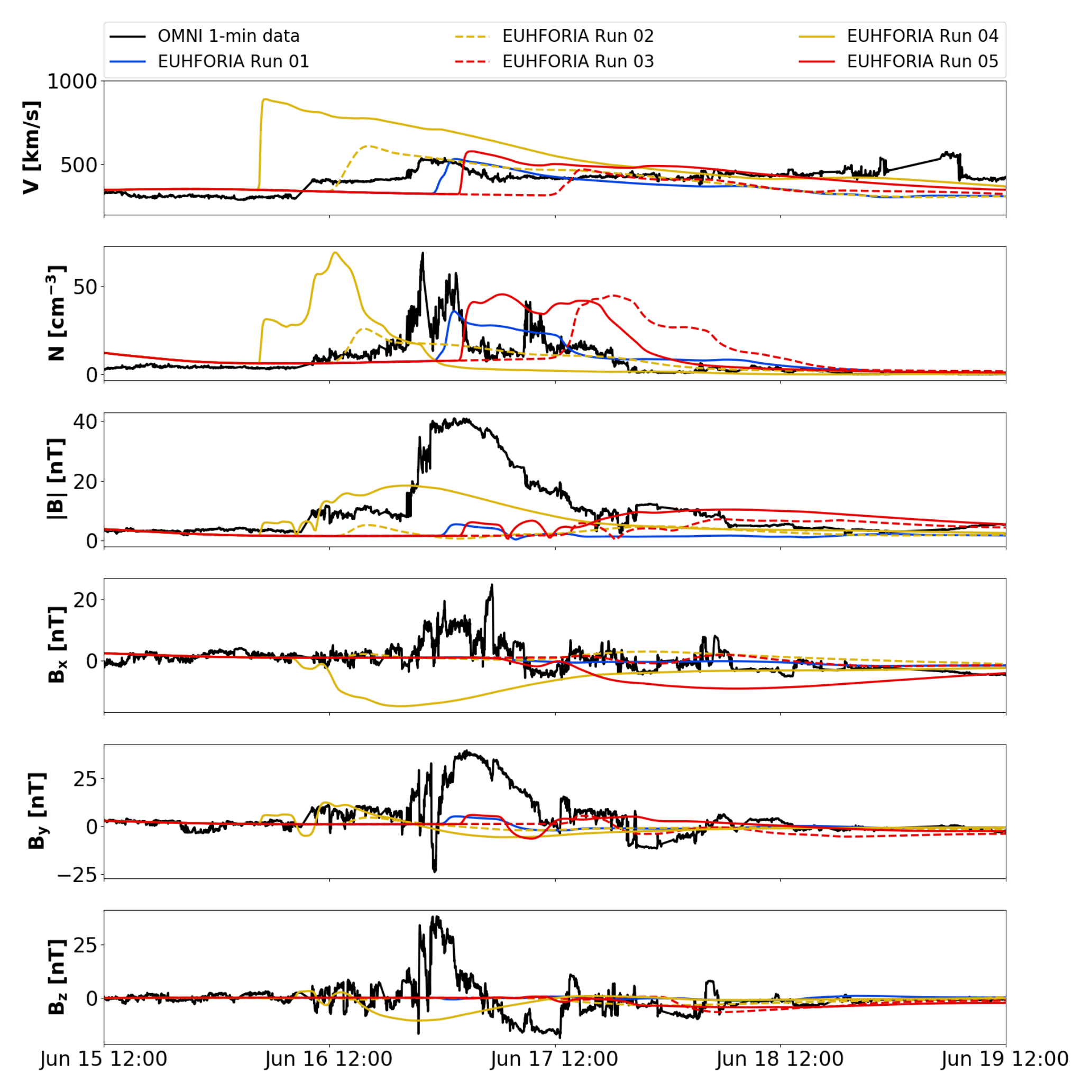} } 
\caption{ Event 2: EUHFORIA time series at Earth compared to in-situ data from 1-min OMNI data (black). 
From top to bottom: speed, number density, magnetic field strength, $B_x$, $B_y$, $B_z$ components in GSE coordinates.
}
\label{fig:20120614_earth} 
\end{figure}

\medskip
% EUHFORIA results at Earth
\textit{EUHFORIA predictions at Earth.}
 Figure~\ref{fig:20120614_earth} shows the results from the simulations (listed in Table \ref{tab:event_2}) at Earth, compared to in-situ measurements of the solar wind properties from the OMNI database.
%
% CME arrival times
CME1 is not predicted to arrive at Earth at all.
The arrival times of CME2 at Earth, for all the different runs, are listed in Table~\ref{tab:event_2}.
In this case, from the comparison of the CME ToA at Earth in Run~01 and Run~02 we observe that, at the CME flank, the impact of the magnetic pressure on the CME propagation is about 11 hours at 1.0~AU.
Launching the CME on the ecliptic plane (Run~04), the predicted CME ToA shifts about 9 hours earlier.
Reducing the speed of the spheromak CME (Run~05), the predicted CME ToA returns in good agreement with observations ($\sim 6$ hours from the observed shock time) as well as with the CME ToA prediction from the cone model (Run~01, $\sim 2$ hour difference in the ToA).

% time series profiles
As for Event~1, predicted ICME magnetic field profile in Run~05 appears to be elongated in the radial/temporal duration, indicating that the model tends to over-estimate the radial extension of the ICME at 1~AU.
% peak magnetic field value
The cone model (Run~01) is unable to predict the magnetic signatures observed in association to the MC.
For this particular event, observations of the source region lead to an estimation of the toroidal magnetic flux $\phi_t$ that is a factor 0.4 of the one used in the case of Event 1.
However, OMNI data show that the MC appears to be characterised by a very strong $B$, with a peak $B$ that is a factor 1.5 higher than the one observed in association to Event 1. 
While the reason for this unusual behaviour is something that still needs to be clarified, 
from the simulation point of view this results in an under-prediction of the peak values of all the IMF components. 
In fact, while the observed maximum $B$ during the passage of the MC was 40~nT, 
the prediction of the cone CME model (Run~01) is $5$~nT (corresponding to $\sim 13\%$ of the observed maximum $B$),
while the one of the spheromak CME model predicting the best CME arrival time and ICME magnetic signature (Run~05) is $10$~nT (corresponding to $\sim 25\%$ of the observed maximum $B$).
In this case the observed minimum $B_z$ during the passage of the MC was $-19$~nT, while the minimum $B_z$ predicted from the cone model is negligible ($\sim -1$~nT), and the one of the spheromak in Run~05 is $-4$~nT (corresponding to $22\%$ of the observed minimum $B_z$).

In this case, all the scores assessing the quality of the prediction considered are significantly lower than in the case of Event 1. This is possibly due to a combination of reasons, including the fact that the analysis of the solar, coronal, and heliospheric signatures of this CME-CME interaction event were more complex than the 12 July 2012 CME event.

\begin{figure}[t]
\centering
{\includegraphics[width=\hsize]{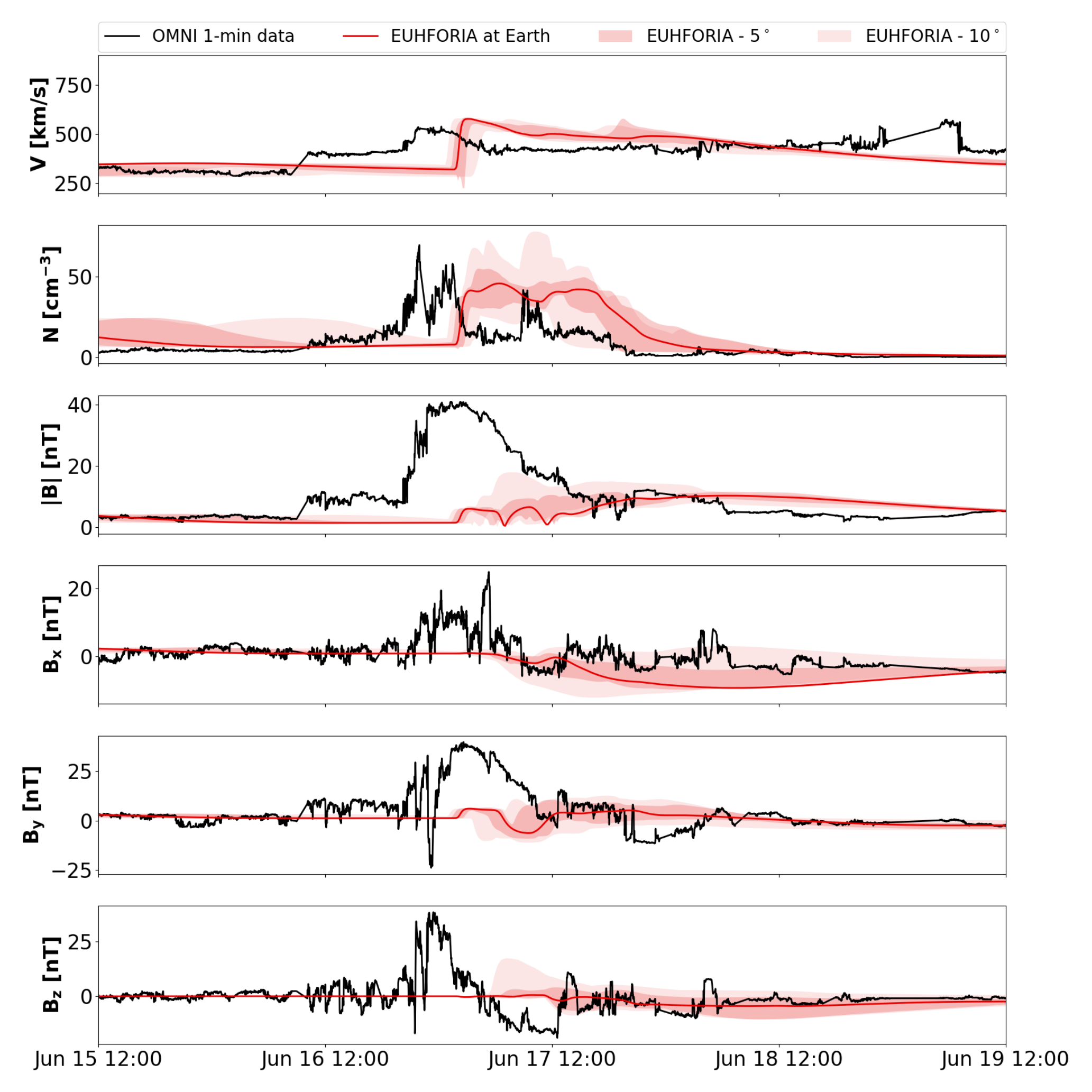}  } 
\caption{Event 2: EUHFORIA time series from Run~05 at Earth, compared to in-situ data from 1-min OMNI data (black).
The dark red and light red shaded areas show the maximum variation of EUHFORIA predictions 
at positions separated by $5^\circ$ and $10^\circ$ in longitude and/or latitude from Earth.
From top to bottom: speed, number density, magnetic field strength, $B_x$, $B_y$, $B_z$ components in GSE coordinates.
}
\label{fig:20120614_around_earth} 
\end{figure}

\medskip
% points around the Earth
\textit{Location sensitivity.}
Similarly to what discussed for Event 1, the predictions above are meaningful only when discussed in relation to prediction at surrounding points, i.e. at the location of virtual spacecraft around the Earth.
Figure~\ref{fig:20120614_around_earth} shows the EUHFORIA time series at Earth from Run~05, 
together with shaded areas indicating the variability in the predictions at virtual spacecraft located at $\pm 5^\circ$ and $\pm 10^\circ$ in longitude and/or latitude from Earth.
Considering the range of predictions in this area around Earth, the CME ToA spans of $\pm 2$ hours around the ToA at Earth for the closer spacecraft ($\pm 5^\circ$ separation from Earth), and of $\pm 5$ hours around the ToA at Earth for the outer spacecraft ($\pm 10^\circ$ separation). 
The prediction for the maximum $B$ in the MC ranges between $10$~nT and $11$~nT at spacecraft located at $\pm 5^\circ$ from Earth,
and between $9$~nT and $18$~nT at spacecraft located at $\pm 10^\circ$.
The minimum $B_z$ in the MC ranges between $-2$~nT and $-11$~nT at spacecraft located at $\pm 5^\circ$ from Earth,
and between $-1$~nT and $-11$~nT at spacecraft located at $\pm 10^\circ$.
In conclusion, considering the prediction at spacecraft located around the Earth at angular separations within the uncertainty of the CME direction of propagation derived from the GCS model, 
the best prediction accounts for $\sim 45\%$ of the maximum $B$ measured in the MC, 
and for $\sim 58\%$ of the minimum $B_z$ measured in the MC. 

%============================================================
%============================================================
\section{Discussion and summary}
\label{sec:conclusions}
%============================================================
%============================================================

In this work we have studied two Earth-directed CME events using the EUHFORIA helispheric model. Our main aim was assessing the capabilities of the new spheromak CME model in predicting the ICME magnetic properties at Earth, when initialised using CME parameters derived directly from remote-sensing observations. 
In Section~\ref{sec:euhforia} we presented the cone and spheromak CME models currently implemented in EUHFORIA, and discussed their differences in terms of the dominating forces acting on the CMEs in the MHD description.
In Section~\ref{sec:parameters} we presented in detail the approaches used to derive the input parameters from remote-sensing observations of the CMEs and their source regions, and in Section~\ref{sec:case_studies} we discussed the application of those methods to two case study CME events, one occurring on 12 July 2012 (Event 1) and one on 14 June 2012 (Event 2).
In Section~\ref{sec:results} we presented the results of the simulations of the CME events above, discussing their propagation and arrival at Earth and comparing simulation results with remote-sensing and in-situ observations.

For each event, we simulated the CMEs using both the cone model and the spheromak model.
Our analysis indicates that the use of a spheromak model initialised with observations-based CME input parameters significantly improves the prediction of the ICME internal magnetic field intensity and orientation at Earth.
These results also make us expect a net improvement in the prediction of the CME geo-effectiveness in terms of impact on the terrestrial environment.
The prediction of the CME arrival time at Earth was found to be highly dependent on the CME model and CME input parameters used, and a detailed investigation of the forces acting on the CMEs was needed in order to understand its dependence on the CME initialisation.
The key findings of this work can be summarised as follows.
\begin{enumerate}
\item
% input parameters
The determination of the CME parameters at 0.1~AU from remote-sensing observations is an extremely challenging issue. 
It can be performed only in the case of well-observed events, and even in such cases it still needs to be complemented by a number of assumptions and approximations. 
Being aware of such limitations, we performed this first analysis for two CMEs that were observed by the full constellation of spacecraft monitoring the Sun and its atmosphere from different view points on the ecliptic plane.
We focused on relatively simple CME events, characterised by almost no rotation/deflection in their magnetic structure after the eruption, so that their evolution from the eruption site up to 0.1~AU could be considered approximately radial and self-similar. Both events were also observed as clear MC signatures at L1, making the comparison with simulation outputs easier as magnetic field rotations were smooth and easy to identify.

\item
% speed separation from observations
In Section~\ref{subsec:gcs} (see Appendix~\ref{app:appendix_b} for the full calculations) we have proposed a new observational method, based on GCS fitting outputs, to separate the radial and expansion speeds of CMEs. 
Testing this method against empirical relations in literature, we observed that 
single-spacecraft observations and multi-spacecraft observations provide quite different estimations of the CME expansion and radial speeds. 
On one side, the geometrical approach based on multi-spacecraft observations is based on 3D geometrical relations, and hence it is in principle more consistent with the geometry used in the coronagraphic reconstruction of the CMEs and in heliospheric simulations. 
On the other hand, it can only be applied to CME events that were observed from more than one view point, and it has not been tested yet on a large set of events, as it is the case of the empirical relations considered here.
Although additional testing would be needed in order to assess the performance of this approach in the case of a statistical set of CME events, in our opinion the geometrical approach has two major additional values compared to the empirical ones: (1) it allows to go beyond empirical relations that may work on large sets of event but may fail in single cases, and (2) it allows to quantify the contribution of the expansion and radial speeds for any CME, and not only to those there were observed as full halos.

\item
% implications on CME propagation
As presented and discussed in Section~\ref{sec:results}, the separation of the radial and expansion speeds is critical in order to model the propagation of spheromak CMEs in the heliosphere and to predict their arrival time at Earth. 
In fact, the simulation results show that spheromak CMEs propagate significantly faster than cone CMEs when initialised with the same kinematic parameters. For both case studies we have shown that those differences in the propagation can be mitigated by initialising spheromak CMEs at 0.1~AU using a reduced speed ($v_{rad}$) that considers only the translational motion of the CME center of mass, instead of the combination of the translational and of the expansion motion of the CME apex ($v_{rad}+v_{exp}$).
Based on further analysis of the simulation outputs, we interpreted these differences as the result of the different Lorentz force acting on cone and spheromak CMEs (particularly at the CME-solar wind interface due to magnetic pressure gradients), which in turn leads to different CME expansions in the heliosphere. 

\item
% predictions at L1
Considering predictions of the peak ICME magnetic field parameters at Earth, results for these first case studies shown that using the spheromak CME model improves the predictions of $B$ ($B_z$) at Earth up to 60 (40) percentage points for Event 1, and 12 (22) percentage points for Event~2, compared to the cone model.
At the same time, the model predictions appear to be sensitive to the exact position sampled in the heliosphere. Considering virtual spacecraft separated by $5^\circ$ and $10^\circ$ in longitude/latitude from Earth, $B$ ($B_z$) predictions improved significantly, reaching up to 70$\%$ (78$\%$) of the observed peak value for Event 1, and 45$\%$ (58$\%$) for Event 2.
This provides an indication of the spatial variability of the predictions at 1~AU.
As such separations are consistent with the typical uncertainties in the reconstruction of the CME direction of propagation from the GCS model, in the case of a background solar wind that is uniform at angles up to $10^\circ$ around the Sun-Earth line, results obtained from virtual spacecraft around Earth can also be used as efficient alternative to otherwise time-consuming ensemble simulations of CMEs performed varying the CME direction of propagation.

\item
% radial/temporal extent and spheromak shape
In both events, the predicted magnetic field time series show very extended MC radial/temporal signatures, suggesting that the spheromak model tends to over-estimate the radial size of the CME at 1~AU, not fully accounting for global shape deformation effects such as pancaking. This was also visible in the time delay of the magnetic field peaks compared to observations.
Extending the current CME model to include the possibility to introduce ellipsoidal spheromak CMEs could mitigate this effect, and is something that will have to be quantified in future work.

\item
% limitation of spheromak
Event 1 (12 July 2012 CME) represents a well-understood CME event in terms of its solar, coronal, and heliospheric evolution, and it was associated with a successful prediction of the ICME magnetic properties using the spheromak CME model.
Event 2 (14 June 2012 CME) on the other hand turned out to be a more complicated event than expected.
In fact, it appeared to be a single non-interacting CME event considering its evolution along the Sun-Earth line (including its in-situ signature characterised by a nicely-rotating magnetic field structure), but it was actually associated with a CME-CME interaction occurring between the main CME and a previous CME that erupted from the same AR on 13 June 2012.
The analysis and modelling of this event was \textit{per se} already more complicated than that of Event~1.
Moreover, Event 2 also provided an example of the limitations of the spheromak CME model in reproducing the global shape of flux-rope structures in the heliosphere. 
In fact, simulations of CME2 using the latitude and half width derived from the GCS reconstruction predicted the CME to propagate almost completely south of the ecliptic, so that almost no magnetic field signatures associated to the ICME were predicted at Earth (Run~02 and 03, Figure~\ref{fig:20120614_euhforia_bz}). On the other hand, in-situ observations of the event show magnetic signatures compatible with an almost-central crossing of Earth through the flux-rope structure.
In Section~\ref{subsec:20120614_results}, we have shown that a way to cope with such limitations and have a correct prediction of the magnetic fields rotations observed at Earth is that of launching the spheromak CME directly on the equatorial plane (Run~04 \& 05, Figure~\ref{fig:20120614_earth}). 

%At the same time, this issue underlined the need to incorporate in EUHFORIA a flux-rope CME models able to reproduce the large-scale magnetic field structure of magnetic clouds.   
\end{enumerate}

% future works
In conclusion, in this work we focused on Earth-directed CME events that were associated with clear MC signatures at Earth, with the aim of benchmarking the current EUHFORIA prediction capabilities in the case of well-observed CME events. 
Initial results indicate significant improvements in the predictions of the ICME magnetic field structures at Earth when using a more realistic spheromak CME model compared to a traditional cone CME model.
However, most CME events are more complex than the case studies presented in this work, either due to their interaction with other structures in the solar wind - including other CMEs, or due to the limited observations available.
This particularly will become a critical issue in the next years as observations from the STEREO spacecraft will reduce as the mission goes towards its end, eventually preventing 3D reconstruction of CME events until the launch of new missions.
Therefore, several efforts to assess the predictive capabilities of the EUHFORIA model in more complex scenarios are also needed. 
In particular, a detailed analysis of the EUHFORIA capabilities in modelling CME-CME interactions will be addressed in future studies. 
On the observational side, an important issue regards the quantification of the uncertainties associated to the CME input parameters determined from observations.
Finally, we stress the fact that in this work we focused on the modelling of the CME propagation in the heliosphere, and on the predictions of the CME arrival time and its magnetic signatures at Earth. 
However, ultimate predictions of the CME geo-effectiveness intrinsically imply the need to go beyond predictions of the solar wind and ICME properties at L1. In this regard, a detailed analysis of the predicted CME impacts on the magnetospheric-ionospheric-ring current systems in terms of induced geomagnetic activity will be addressed in a companion paper. 

%-------------------------------------------------------------------

\begin{acknowledgements}
The authors thank the anonymous referee for the constructive comments and suggestions for improvements.
CS was funded by the Research Foundation - Flanders (FWO) SB PhD fellowship no. 1S42817N.
LR acknowledges funding from the CCSOM (Constraining CMEs and Shocks by Observations and Modelling) project.
MM was funded by the SWAP and LYRA projects of the Centre Spatial de Li\`{e}ge and the Royal Observatory of Belgium, funded by the Belgian Federal Science Policy Office (BELSPO).
% Jens
JP acknowledges funding from the University of Helsinki three-year grant project 490162 and the SolMAG project (4100103) funded by the European Research Council (ERC) in the framework of the Horizon 2020 Research and Innovation Programme. 
% Stefaan
These results were obtained in the framework of the projects GOA/2015-014 (KU Leuven), 
G.0A23.16N (Research Foundation - Flanders, FWO), and C90347 (ESA Prodex).
% EUHFORIA
EUHFORIA is developed as a joint effort between the University of Helsinki and KU Leuven.
The full validation of solar wind and CME modeling is being performed within the CCSOM project (\url{http://www.sidc.be/ccsom/}). The EUHFORIA website, including a repository for simulation results and the possibility for users to access the code, is currently under construction; please contact the authors for full 3D simulation outputs and model details.
The simulations were carried out at the VSC -- Flemish Supercomputer Center, funded by the Hercules foundation and the Flemish Government -- Department EWI. 
% observational data
We acknowledge the use of data contained in the CME catalog generated and maintained at the CDAW Data Center by NASA and The Catholic University of America in cooperation with the Naval Research Laboratory, 
SOHO/LASCO images (courtesy of the SOHO consortium),
SDO/AIA and SDO/HMI images (courtesy of NASA/SDO and the AIA and HMI science teams),
STEREO/SECCHI images, and data from the Venus Express MAG instrument and the MESSENGER MAG instrument.
The authors thank J. Davies and D. Barnes for providing the programs to analyse STEREO J-maps.
This paper uses data from the Heliospheric Shock Database, generated and maintained at the University of Helsinki.
\end{acknowledgements}

%-------------------------------------------------------------------
% BIBLIOGRAPHY
%\bibliographystyle{aa} % style aa.bst
%\bibliography{Bibliography} % your references Yourfile.bib

%\begin{thebibliography}{}

%\end{thebibliography}

%-------------------------------------------------------------------

% Activate the appendix
% from now on sections are numerated with capital letters
\begin{appendix}

% ========================================
\section{Derivation of $v_{rad}$ and $v_{exp}$ from the Graduated Cylindrical Shell model}
\label{app:appendix_b}
% ========================================

Using the same notation as \citet{thernisien:2011}, 
the heliocentric distance of the CME front at its apex, $h_{front}$, is defined as 
\begin{equation}
h_{front} = OH = \frac{b+\rho}{1-\kappa} = \frac{b+\rho}{1-\kappa^2} \,  (1+\kappa),
\label{eqn:gcs_hgt}
\end{equation}
and 
\begin{equation}
OH =  {OC_1} +  R(\beta=\pi/2),
\end{equation}
where $b = OB $ and $\rho = BD $.
From geometrical considerations, the total speed of the CME apex $v_{3D}$ is related to the variation over time of the parameter $h_{front}$, while the expansion speed $v_{exp}$ is related to the variation in time of $R(\beta=\pi/2)$, 
and the radial speed $v_{rad}$ is related to that of $ OC_1$.
Therefore, the radial and expansion speed can be calculated based on the standard GCS output parameters as:
\begin{equation}
\begin{cases} 
{v_{rad} = \frac{d OC_1}{d t} } \\
{v_{exp} = \frac{d R(\beta=\pi/2)}{d t}}.
\end{cases}
\end{equation}
The heliocentric distance of the apex centre, $OC_1$, and the cross section radius of the apex, $R(\beta=\pi/2)$,
are in turn related to the leading edge height $h_{front}$ by the following relations \citep{thernisien:2011}:
\begin{equation}
\begin{cases} 
OC_1 			= \frac{b+\rho}{1-\kappa^2} =  \frac{1}{1+\kappa} \, h_{front} \\
R(\beta=\pi/2) = \frac{b+\rho}{1-\kappa^2}  \,  \kappa  =  \frac{\kappa}{1+\kappa}  \, h_{front}.
\end{cases}
\end{equation}
so that $OC_1 + R(\beta = \pi/2) = OH$ (as shown in Figure~\ref{fig:gcs}).
Combining these results and remembering that all the GCS parameters are in principle time-dependent, one obtains
\begin{align}
v_{rad} &= \frac{d}{d t} \left( \frac{h_{front}}{1+\kappa} \right) 
                = \frac{1}{1+\kappa} \frac{d h_{front}}{dt} 
                - h_{front} \frac{1}{(1+\kappa)^2} \frac{d \kappa}{dt},
\label{eqn:gcs_vrad_full}
\end{align}
and 
\begin{align}
v_{exp} &= \frac{d}{d t} \left( \frac{\kappa}{1+\kappa}h_{front} \right) = \nonumber \\
        &= \frac{\kappa}{1+\kappa} \frac{d h_{front}}{dt} 
        + h_{front} \, \left( \frac{1}{1+\kappa} - \frac{\kappa}{(1+\kappa)^2} \right) \frac{d \kappa}{dt}.
\label{eqn:gcs_vexp_full}
\end{align}
For CMEs where $\kappa$ can be kept fixed in time, the above equations simplify to 
\begin{equation}
\boxed{v_{rad} = \frac{1}{1+\kappa} \frac{d h_{front}}{dt},}
\label{eqn:gcs_vrad_full_02}
\end{equation}
and 
\begin{equation}
\boxed{v_{exp} = \frac{\kappa}{1+\kappa} \frac{d h_{front}}{dt}.}
\label{eqn:gcs_vexp_full_02}
\end{equation}

% ========================================
\section{From $\phi_{p}$ to spheromak parameters}
\label{app:appendix_a}
% ========================================
The linear force-free spheromak solution is \citep{chandrasekhar:1957, shiota:2016,verbeke:2019}: 
\begin{equation}
\begin{cases} 
B_r = 2 B_0 \frac{J_1(\alpha r)}{\alpha r} \cos \theta \\ 
B_\theta = - \frac{B_0}{\alpha r} ( J_1(\alpha r) + \alpha r  \, J_1'(\alpha r) ) \sin \theta  \\ 
B_\phi=  B_0 \, J_1(\alpha r) \sin \theta, \\
\end{cases} 
\end{equation}
where $x_{01} = \alpha r_0 = 4.4934$ (1st zero of $J_1$) and $r_0$ is the spheromak radius. 
The poloidal magnetic flux (function of $r$) can be calculated as
\begin{align}
\phi_p(r)  & = \iint B_r r^2 \sin \theta \, d \theta \,  d \phi = \nonumber \\
        %& = \int_0^{\pi/2} \sin \theta d \theta \int_0^{2 \pi} B_r r^2 d \phi = \nonumber \\
        %& = \int_0^{\pi/2} \sin \theta d \theta \int_0^{2 \pi} \Big [ 2 B_0 \frac{J_1(\alpha r)}{\alpha r} \cos \theta \Big ] r^2 d \phi = \nonumber \\
        & = \frac{2 B_0}{ \alpha r} r^2 J_1(\alpha r) \int_0^{\pi/2} \cos \theta \sin \theta \, d \theta \int_0^{2 \pi} \, d \phi = \nonumber \\
        %& = \frac{2 B_0}{ \alpha}  
        %\cdot \Big[ - \frac{1}{2} \cos^2 \theta \Big|^{\pi/2}_0 \Big] 
        %\cdot J_1(\alpha r) r \cdot  \int_0^{2 \pi}  d \phi = \nonumber \\
        & 
        %= \frac{4 \pi B_0}{ \alpha} \cdot \frac{1}{2} \cdot J_1(\alpha r) r 
        =  \frac{2 \pi B_0}{ \alpha}  J_1(\alpha r) \,  r.
\end{align}
To compute the actual poloidal magnetic flux value, 
one needs first to determine at which distance $r_*$ from the center of the spheromak, 
on the plane $\theta = \pi/2$, the magnetic field becomes completely axial ($B_r = 0, B_\theta = 0$).
This distance can be calculated as:
\begin{align}
B_\theta = - \frac{B_0}{\alpha r} ( J_1(\alpha r) + \alpha r  \,  J_1'(\alpha r) ) \sin \theta & = 0 \nonumber \\
J_1(\alpha r) + \alpha r \,  J_1'(\alpha r) & = 0 \nonumber \\
\frac{\sin x - x \cos x + 2 x^2 \cos x + x (x^2-2) \sin x}{x^2} & = 0,
\label{eqn:r_star}
\end{align}
for $x = \alpha r$ and  $0 \le x \le x_{01}$.
The only acceptable solution to Equation \ref{eqn:r_star} is $ x_* = \alpha r_* = 2.4048$, 
corresponding to $(r_*/r_0) = 2.4048 / x_{01} = 0.5352 $. 
The actual poloidal magnetic flux can then be calculated as
\begin{align}
\phi_p (r_*)    & = \frac{2 \pi B_0}{\alpha} 
                    \,   J_1(\alpha r_* )  \, r_* = \nonumber \\
                & = \frac{2 \pi B_0}{\alpha} \,  r_*
                    \,  \frac{\sin(\alpha r_* ) - \alpha r_*  \cos(\alpha r_* )}{(\alpha r_*)^2} = \nonumber \\
                & = \frac{2 \pi B_0}{\alpha^3} \,  \frac{1}{r_*}
                    \,  \Big( \sin(\alpha r_* ) - \alpha r_*  \cos(\alpha r_* ) \Big).
\label{eqn:phip_sph}
\end{align}
Inverting Equation \ref{eqn:phip_sph}, knowing the poloidal magnetic flux from observations, the axial magnetic field can then be calculated as
\begin{equation}
\boxed{
B_0 =   \frac{\alpha^3}{ 2 \pi}
        \,  \frac{ \phi_p (r_*) \, r_*}{\Big(\sin(\alpha r_* ) - \alpha r_*  \cos(\alpha r_* ) \Big)}.
}
\label{eqn:B0_sph}
\end{equation}
The toroidal magnetic flux can then be derived as 
\begin{align*}
\phi_t  & = \iint B_\phi r \, dr \, d \theta = \\
        & = \int_0^{r_0} r \, dr \int_0^\pi B_\phi \, d \theta = \\
        & = \int_0^{r_0} r \, dr \int_0^\pi [ B_0 \, J_1(\alpha r) \sin \theta ] \, d \theta  = \\
%        & = B_0 \int_0^{\pi} \sin \theta d \theta \int_{0}^{r_0} J_1(\alpha r) r dr = \\
%        & = B_0 \cdot 2\cdot \int_{0}^{r_0} J_1(\alpha r) r dr = \\
%       & = 2 B_0 \cdot \int_{0}^{x_0 = \alpha r_0} J_1(x) x dx \frac{1}{\alpha^2} = \\
%        & = \frac{2 B_0}{\alpha^2} \int_{0}^{x_0} x \frac{\sin x - x \cos x}{x^2} dx = \\
%        & = \frac{2 B_0}{\alpha^2} \Big ( \int_{0}^{x_0} \frac{\sin x}{x} dx - \int_{0}^{x_0} \cos x dx \Big ) = \\
        & = \frac{2 B_0}{\alpha^2} \Big ( - \sin x_{01} + \int_{0}^{x_{01}} \frac{\sin x}{x} dx  \Big ),
\end{align*}
hence
\begin{equation}
\boxed{\phi_t = \frac{2 B_0}{\alpha^2} \Big ( - \sin x_{01} + \int_{0}^{x_{01}} \frac{\sin x}{x} dx  \Big ). }
\label{eqn:phit_sph}
\end{equation}
% Figure~\ref{fig:spheromak} shows a sketch of a spheromak FR and its relevant parameters.

\end{appendix}
% ========================================
\end{document}